%% file: main_full.tex
\documentclass[11pt]{extarticle}

\usepackage{color}
\usepackage{threeparttable}
\usepackage[english]{babel}
\usepackage{graphicx}
\usepackage{framed}
\usepackage[normalem]{ulem}
\usepackage{amsmath}
\usepackage{amsthm}
\usepackage{amssymb}
\usepackage{amsfonts}
\usepackage{enumerate}
\usepackage[utf8]{inputenc}
\usepackage[top=0.8 in,bottom=0.8in, left=1 in, right=1 in]{geometry}

\theoremstyle{definition}

\graphicspath{{./Figure/}}

\usepackage{graphicx,kantlipsum,setspace}
\usepackage[font=small,labelfont=bf]{caption}
\usepackage{newfloat}
\DeclareFloatingEnvironment[name={Supplementary Figure},fileext=lof]{suppfigure}

\DeclareFloatingEnvironment[name={Supplementary Table},fileext=lof]{supptable}

  \DeclareSymbolFont{numbers}{T1}{ptm}{m}{n}
  \SetSymbolFont{numbers}{bold}{T1}{ptm}{bx}{n}
  \DeclareMathSymbol{0}\mathalpha{numbers}{"30}
  \DeclareMathSymbol{1}\mathalpha{numbers}{"31}
  \DeclareMathSymbol{2}\mathalpha{numbers}{"32}
  \DeclareMathSymbol{3}\mathalpha{numbers}{"33}
  \DeclareMathSymbol{4}\mathalpha{numbers}{"34}
  \DeclareMathSymbol{5}\mathalpha{numbers}{"35}
  \DeclareMathSymbol{6}\mathalpha{numbers}{"36}
  \DeclareMathSymbol{7}\mathalpha{numbers}{"37}
  \DeclareMathSymbol{8}\mathalpha{numbers}{"38}
  \DeclareMathSymbol{9}\mathalpha{numbers}{"39}

\usepackage[misc]{ifsym}

\usepackage{authblk}
\usepackage{indentfirst}
\usepackage{amsthm,amsmath,amssymb}
\usepackage{mathrsfs}
\usepackage[]{hyperref}

\usepackage{titlesec}

\titleformat*{\section}{\normalsize \bfseries}
\titleformat*{\subsection}{\normalsize \bfseries}

\title{\large \textbf{Demonstration of fully integrated parity-time-symmetric electronics}}

\author[1,7]{\normalsize  Weidong Cao}
\author[1]{Changqing Wang}
\author[1,2,3]{Weijian Chen}
\author[4,6]{Song Hu}
\author[4,5]{Hua Wang}
\author[1,7]{Lan Yang}
\author[1,7]{Xuan Zhang}

\begin{scriptsize}

\affil[1]{Department of Electrical and Systems Engineering, Washington University, St. Louis, MO, USA.}
\affil[2]{Department of Physics, Washington University, St Louis, MO, USA.}
\affil[3]{Center for Quantum Sensors, Washington University, St Louis, MO, USA.}
\affil[4]{School of Electrical and Computer Engineering, Georgia Institute of Technology, Atlanta, GA, USA.}
\affil[5]{Department of Information Technology and Electrical Engineering, Swiss Federal Institute of Technology Zurich, Zurich, Switzerland.}
\affil[6]{Present address: Apple Inc, 1 Apple Park Way, Cupertino, CA, USA.}
 \affil[7]{Corresponding authors: weidong.cao@wustl.edu; yang@seas.wustl.edu; xuan.zhang@wustl.edu.}
\end{scriptsize}

\date{}                 
\begin{document}

\maketitle

\newpage

\input{main_text.tex}


\pagebreak





\appendix
\renewcommand{\thesection}{Supplementary Information \arabic{section}} 

\setcounter{equation}{0}
\setcounter{figure}{0}
\setcounter{table}{0}

\renewcommand{\theequation}{S\arabic{equation}}

\input{Design_analysis_PT}
\input{Scattering_analysis}
\input{Measurement_scat}

\input{VCO_theory}

\input{Exper_Set}

\input{Su_re}
\input{Versatile_implemen}

\input{Extended_dis}

\newpage
\bibliographystyle{IEEEtran}
\bibliography{main_full}

\end{document}

%% file: main_text.tex
\begin{abstract}
\textbf{Harnessing parity-time (PT) symmetry with balanced gain and loss profiles has created a variety of opportunities in electronics from wireless energy transfer to telemetry sensing and topological defect engineering. 
However, existing implementations often employ ad-hoc approaches at low operating frequencies and are unable to accommodate large-scale integration. Here, we report a fully integrated realization of PT symmetry in a standard complementary metal-oxide semiconductor technology. 
Our work demonstrates salient PT symmetry features such as phase transition as well as the ability to manipulate broadband microwave generation and propagation beyond the limitations encountered by exiting schemes. 
The system shows 2.1 times bandwidth and 30 percentage noise reduction compared to conventional microwave generation in oscillatory mode and displays large non-reciprocal microwave transport from 2.75 to 3.10 gigahertz in non-oscillatory mode due to enhanced nonlinearities.
This approach could enrich integrated circuit (IC) design methodology beyond well-established performance limits and enable the use of scalable IC technology to study topological effects in high-dimensional non-Hermitian systems. }
\end{abstract}

\section{Introduction}
Symmetry is one of the most essential notions to influence the fundamental properties of physical systems.
Quantum systems whose Hamiltonians commute with a joint parity-time ({PT}) operator ($PT\hat{H}=\hat{H}PT$), possess a special kind of symmetry, known as {PT} symmetry.
In general, open quantum systems interacting with environments can be described by non-Hermitian Hamiltonians which preserve complex eigenvalues. 
However, a system with {PT} symmetry possesses purely real eigenspectra in certain regimes whereas the eigenstates are non-orthogonal to each other.
Over the past years, PT-symmetric systems featured with balanced gain and loss profiles have been studied in optics~\cite{wang_EIT,wang_EIT_1,CPA1,CPA2,CPA3,CPA4,topo_r_0, single_mode1,single_mode2,single_mode3,sensor1,sensor2,weijian_sensor,nonre2,nonre4}, optomechanics~\cite{opmech1,opmech2},
optoelectronics~\cite{oeo1}, and acoustics~\cite{aco1,aco2,NE_nonreci}, and initiated a number of exotic effects and applications including electromagnetically induced transparency~\cite{wang_EIT,wang_EIT_1}, coherent perfect absorption-lasing~\cite{CPA1,CPA2,CPA3,CPA4}, topological light steering~\cite{topo_r_0}, single-mode lasing~\cite{single_mode1,single_mode2,single_mode3}, ultrasensitive sensors~\cite{sensor1,sensor2,weijian_sensor}, opto-electronic microwave generation~\cite{oeo1}, and non-reciprocal photon~\cite{nonre2,nonre4} and phonon~\cite{NE_nonreci} transmission.

Electronics has recently emerged as a promising field to study PT symmetry due to the flexibility and reliability of controlling active and passive electronic resonators~\cite{PT_E, LRC_1, Dual_beh, wireless_p, NE_wireless, NE_sensor,topo_ele,topo_ele_1,topo_ele_2,PTX}. Experiments have been reported on printed circuit boards~\cite{PT_E,LRC_1,Dual_beh, wireless_p,NE_wireless,NE_sensor,topo_ele,topo_ele_1,topo_ele_2} and in microelectromechanical systems~\cite{PTX}, showing robust wireless energy transfer~\cite{wireless_p,NE_wireless}, enhanced telemetry sensing~\cite{PTX,NE_sensor}, and topological effects~\cite{topo_ele,topo_ele_1,topo_ele_2}.
However, these electronic platforms are confined to low-frequency operation below a few hundred megahertz and are difficult
to scale to small physical dimensions and complex integrated structures.
To explore and unleash the full potential of PT symmetry in electronics, one must look beyond existing
ad-hoc implementation approaches.
Integrated circuit (IC) technology--the leading nanotechnology for electronics, provides a standard manufacturing process for flexible and customized designs that consist of millions of nanoscale integrated devices.
Its scalability in physical dimension enables ICs to be a powerful 
platform that covers a wide applied spectra from DC to terahertz.
It also supports integration of complex three-dimensional structures~\cite{3d_IC}, allowing one to extend a core electronic non-Hermitian unit into higher-dimensional structures to study topological electronics~\cite{topo_ele,topo_ele_1,topo_ele_2}.
Despite such intriguing properties, IC technology is yet to be employed to realize {PT} symmetry, though gain, loss, and their coupling effects do commonly exist there.

On the other hand, as has been shown in the field of optics and acoustics, PT symmetry can provide enhanced ability in wave generation~\cite{oeo1} and propagation~\cite{nonre2,nonre4,NE_nonreci}, making it especially attractive for IC technology.
Effective implementations of these functionalities in the microwave domain remain challenging in IC and the capability to exceed the conventional performance limits has long been sought after.
In particular, electrical non-reciprocal microwave transmission is highly desirable for diverse on-chip applications~\cite{con_non}, yet the existing approach that uses bulky and costly ferromagnetic devices suffers from a number of drawbacks and is  
incompatible with semiconductor fabrication process~\cite{mag_book}.
Advancing integrated magnetic-free non-reciprocal devices thus not only demands breakthroughs in materials and fabrication technologies but also relies on our ability to enrich the arsenal of IC design methodology.
PT-symmetric systems can break Lorentz reciprocity to produce enhanced non-reciprocity in the presence of nonlinearity~\cite{WJ}.
Such merit has been demonstrated in acoustic wave~\cite{NE_nonreci} and light transmissions~\cite{nonre2,nonre4}, but remains unexplored in electronics. 
Therefore, harnessing PT symmetry for broadband microwave generation and non-reciprocity with chip-scale implementation is particularly appealing and represents immense potential.

In this Article, we report a fully integrated implementation of PT symmetry in a 130-nanometer (nm) complementary metal-oxide-semiconductor (CMOS) technology, and
use it to create wideband high-quality microwave generation and broadband strong microwave isolation at gigahertz (GHz).
We demonstrate the PT symmetry phase transition feature on this scalable platform.
With the distinctive gain-loss tuning freedom, we show that in oscillatory mode, our system exhibits a wideband microwave generation from 2.63 {GH}z to 3.20 {GH}z with an average $-120$ dBc$/$Hz noise intensity, achieving 2.1 times bandwidth with 70 percentage of phase noise compared to a baseline conventional oscillator.
While in non-oscillatory mode, the intrinsic nonlinearity of the system is greatly enhanced by PT symmetry, leading to 7$\sim$21 dB non-reciprocal microwave transmission in a broad band of 2.75$\sim$3.10 GHz.
Our results show that the introduction of PT symmetry into IC technology could benefit a broad range of chip-based applications including waveform synthesis and generation~\cite{wave_gene}, frequency modulation~\cite{fre_modu}, and manipulation of microwave propagation~\cite{non_bro}.

\begin{figure}
\centering
\includegraphics[width=1\textwidth]{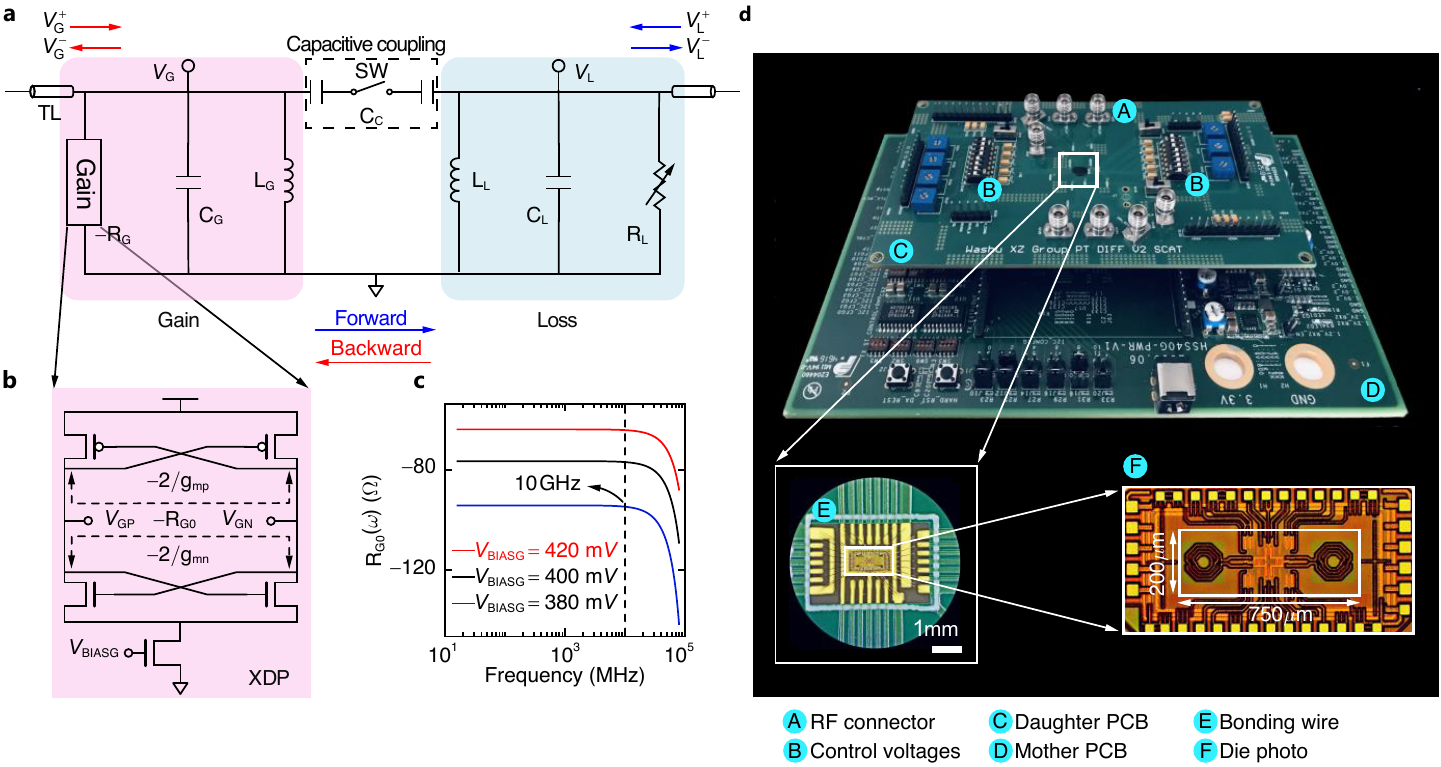} 
\caption{\textbf{{Illustration and characterization of a fully integrated PT-symmetric electronic system.}} \textbf{a}, The schematic diagram of the fully integrated {PT}-symmetric electronic system, where capacitive coupling ($C_C$) is used to connect two RLC resonators, one with gain $-R_G$ and the other one with equal loss ($R_L$). The two units can be coupled and decoupled by the switch SW. The forward transmission is defined from gain side to loss side. \textbf{b}, A cross-coupled differential pair (XDP) is used in our design to generate $-R_{G0}$ which is defined as the reciprocal of the total small signal transconductance of the PMOS differential pair and NMOS differential pair, i.e, $-R_{G0} = {-2}/{(g^{}_{mp}+g^{}_{mn})}$. \textbf{c}, The simulation results show that the negative resistance $-R_{G0}$ remains constant in a wide frequency range up to 10 GHz. \textbf{d}, Fully integrated PT-symmetric electronic system. The chip is wire-bonded on a daughter printed circuit board (PCB), which provides various control voltages and RF connectors for testing. A mother PCB supplies powers to the daughter PCB. The fully integrated {PT}-symmetric electronic system is fabricated in a 130 nm CMOS whose core area is 200$\times$750 $\mu${m}$^2$.}
\label{fig:design}
\end{figure}

\section{Fully integrated {PT-}symmetric electronic system}
Our fully integrated {PT}-symmetric electronic system consists of two capacitively coupled {resistor-inductor-capacitor} (RLC) resonators, one with gain and the other one with equivalent loss (Figure~\ref{fig:design}\textbf{a}).
A differential architecture (Supplementary Figure~\ref{fig:fig_1}) is implemented to obtain superior device matching, better robustness, and higher reliability against noise, interference, and failure.
The equivalent single-ended circuit schematic, shown in Figure~\ref{fig:design}\textbf{a} is used to simplify analysis.
We propose a cross-coupled differential pair~\cite{XDP} (XDP) as a negative resistance converter (NRC) in our system (Figure~\ref{fig:design}\textbf{b}), which generates a stable and tunable gain beyond 10 GHz (Figure~\ref{fig:design}\textbf{c}).
It can also better conserve energy than operational-amplifier~\cite{PT_E,LRC_1,Dual_beh,wireless_p,NE_wireless,NE_sensor,NE_nonreci} based NRCs and Colpitts-type~\cite{PTX} based NRCs, while generating the same amount of gain. 
These characteristics are ideal for broadband and energy-efficient microwave PT-symmetric electronic systems.

The active {RLC} resonator has a gain rate $-R_{G0}$ generated by the XDP, a variable loss rate $R_{G1}$, and an intrinsic loss rate $R_{G2}$, yielding a total gain of $-R_{G}=-R_{G0}\verb||||R_{G1}\verb||||R_{G2}$. The total loss rate $R_{L}=R_{L0}\verb||||R_{L1}$ in the passive {RLC} resonator is contributed by a variable loss $R_{L0}$ and an intrinsic loss $R_{L1}$.
The capacitance in each {RLC} resonator comes from a fixed metal-insulator-metal (MIM) capacitor and a varactor.
The coupling capacitor consists of two
equal MIM capacitors in serial connection via an on-chip switch (SW, Figure~\ref{fig:design}\textbf{a}). 
The inductance in each {RLC} resonator is provided by a symmetrical parallel inductor (Supplementary Figure~\ref{fig:fig_2}).
Both {RLC} resonators have the same natural frequency $\omega_0=3.20$ GHz.
The system is integrated on one monolithic chip with a core area of $200 \times750$ $\mu${m}$^2$ using a standard 130 nm CMOS technology  (Figure~\ref{fig:design}\textbf{d}).
Note that alternative fully integrated {PT}-symmetric structures
could also be implemented (\ref{sec: si_7}.
To show the functionalities and performances, we bonded the chip on a test board with golden wires  (Figure~\ref{fig:design}\textbf{d}).
The board provides a power supply, control voltages, and radio-frequency connectors for measurements.
The gain and loss rate, and varactor capacitance can be flexibly adjusted by the control voltages.
The {PT} symmetry condition is satisfied by setting $R_G \approx R_L=R$, $L_G \approx L_L=L$, and $C_G \approx C_L=C$.
The detailed circuit parameters are provided in Methods.

\section{{PT} symmetry phase transition}
We first numerically analyze the {PT} symmetry transition of our system based on the small signal model--a common methodology in analyzing analog circuits.
The system is considered as a linear time-invariant system, which is valid for small signal inputs.
By applying Kirchoff's law on the equivalent circuit (Figure~\ref{fig:design}\textbf{a}), four eigenfrequencies~\cite{PT_E} are found (Methods), i.e., 
\begin{equation}
\label{eq:resonance}
{\omega}_{1,2}=\pm \frac{\sqrt{{\gamma}^2_{EP} - {\gamma}^2} + \sqrt{{\gamma}^2_{UP} - {\gamma}^2}}{2\sqrt{1+2c}}\cdot \omega_0, ~\quad {\omega}_{3,4}=\pm \frac{\sqrt{{\gamma}^2_{EP}  - {\gamma}^2} - \sqrt{{\gamma}^2_{UP} - {\gamma}^2}}{2\sqrt{1+2c}}\cdot \omega_0.
\end{equation}
Here, $\gamma^{ }_{EP}=\sqrt{1+2c}-1$ denotes the exceptional point ({EP})~\cite{sensor1,sensor2} and ${\gamma}^{ }_{UP}=\sqrt{1+2c}+1$ is the upper critical point~\cite{LRC_1}; $c$ is defined as the capacitive coupling ratio between the coupling capacitance $C_C$ and the RLC resonator's capacitance $C$; and $\gamma$ is the normalized tuning parameter of gain (loss), defined as $\frac{{\sqrt{L/C}}}{R}$.
In the unbroken phase ($0< \frac{\gamma}{\gamma^{}_{EP}} < 1$), the system is characterized by four purely real eigenfrequencies, with two of them positive (${\omega}_{1},{\omega}_{3}$) and the other two negative (${\omega}_{2},{\omega}_{4}$).
In the broken phase ($\frac{\gamma}{\gamma^{}_{EP}} >1$, $\frac{\gamma}{\gamma^{}_{UP}} < 1$), the eigenfrequencies are complex conjugate pairs with non-vanishing real parts.
Above $\frac{\gamma}{\gamma^{}_{UP}} > 1$, the eigenfrequencies become two complex conjugate pairs with purely imaginary parts.

\begin{figure}
\centering
\includegraphics[width=1\textwidth]{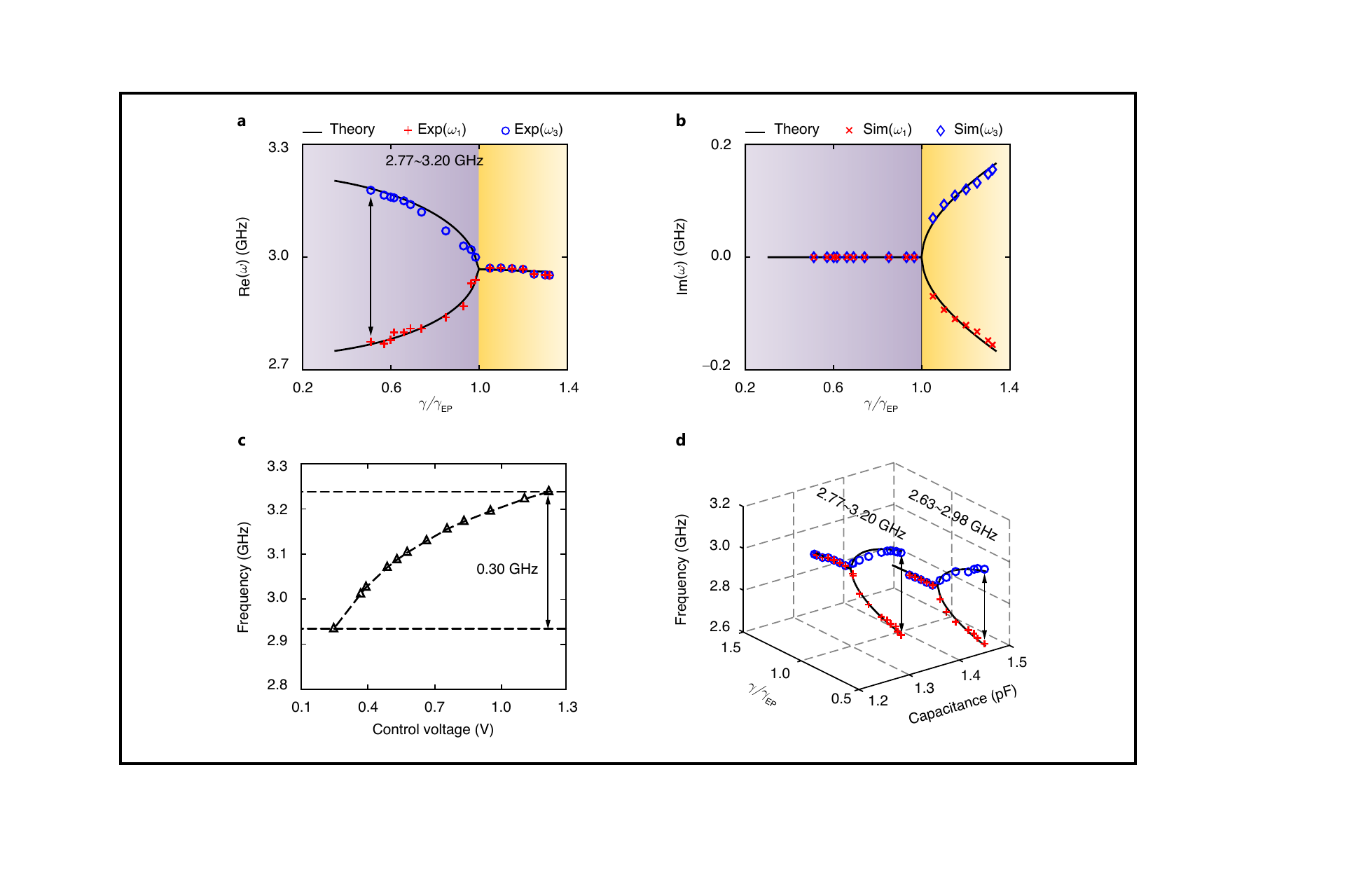} 
\caption{\textbf{{Eigenfrequencies and phase transition of the fully integrated PT-symmetric electronic system with the evolution of coupling factor $\gamma/\gamma^{ }_{EP}$}.} \textbf{a}-\textbf{b}, Real (\textbf{a}) and imaginary (\textbf{b}) parts of eigenfrequencies as a function of the coupling factor $\gamma/\gamma^{ }_{EP}$ when the resonators have a capacitance $C=1.30$ pF. In both \textbf{a} and \textbf{b}, symbols correspond to experimental (Exp) or simulated (Sim) data, while the curves show theoretical results obtained with the small signal model.
\textbf{c}, Frequency tuning range (FTR) of the baseline oscillator, which is achieved by adjusting the varactor in the gain resonator through external voltage bias. The tuning range of the capacitance $C$ is from 1.30 pF to 1.55 pF under different control voltages, corresponding to an FTR of $[2.93, 3.23]$ GHz. \textbf{d}, FTR of our system at two fixed capacitances of resonators, $C=1.30$ {pF} and $C=1.45$ {pF}. By tuning the capacitance from 1.30 {pF} to 1.45 {pF}, the FTR changes from the interval of $[2.77, 3.20]$ GHz to the interval of $[2.63, 2.98]$ GHz, enabling $[2.63, 3.20]$ GHz tuning range in total.}
\label{fig:phase_transition}
\end{figure}

Experimentally, we engineered $\gamma$ by adjusting the gain (loss) and maintaining other circuit parameters unchanged.
The bifurcation of the eigenfrequencies regards to the coupling factor $\frac{\gamma}{\gamma^{}_{EP}}$ was clearly demonstrated (Figure~\ref{fig:phase_transition}\textbf{a}).
In the unbroken phase, the system has two purely real eigenfrequencies.
We observed that the two {RLC} resonators had the same magnitude of voltage.
When $\gamma$ increases, the system undergoes a phase transition at the {EP}, where the real eigenfrequencies branch out into the complex plane.
In the broken phase, the system possesses two supermodes formed by the coupling of two RLC resonators.
Such supermodes have a single resonant frequency but with amplification and dissipation respectively.
The imaginary parts of the eigenfrequencies in the broken phase were obtained by Simulation Program with Integrated Circuits Emphasis (SPICE, simulation methods provided in \ref{sec: si_5_2}), since the oscilloscope used in our study could not capture the fast-changing dynamics of the exponentially oscillating amplitudes at the terminals ($V_{G}$ and $V_L$) of the resonators. 
In the simulation, we observed an exponential growth (ending up at a saturation level), corresponding to the supermode with gain; the decay rates of the other supermode are the mirror of those for amplification (Figure~\ref{fig:phase_transition}\textbf{b}). 
These results are in good agreement with the theoretical predictions.

\section{Wideband high-quality microwave generation}
The resonant behavior in the PT symmetry phase transition of our system provides a new strategy to generate microwave signals.
Conventional microwave generation uses gain to fully compensate for the intrinsic loss of LC resonators to generate a stable wave.
PT symmetry provides a new degree of freedom to modulate microwave generation: by manipulating gain-loss distribution in two coupled resonators, loss can play a role as important as that of gain.
This unique gain-loss tuning freedom can enhance the bandwidth of microwave generation beyond conventional microwave generators (i.e., oscillators based on a single-resonator structure or a coupled-resonator structure) with only capacitive~\cite{VCO} tuning or inductive~\cite{VCO_IND} tuning scheme.
We theoretically compared the eigenfrequencies of all these systems (\ref{sec: si_4}).
A PT-symmetric system inherently has a larger resonance tuning range given by Eq.~(\ref{eq:resonance}), enabled by the eigenfrequencies' bifurcation of tuning the gain-loss contrast $\gamma$.
\textcolor{black}{In comparison, the eigenfrequency of a conventional oscillator only depends on $\omega_0=\frac{1}{\sqrt{LC}}$, independent of gain-loss strength.}

To experimentally show the advantages of a PT-symmetric system for microwave generation, we then decoupled the two {RLC} resonators via an on-chip switch (SW, Figure~\ref{fig:design}\textbf{a}).
For a fair comparison, the active {RLC} resonator with a capacitive tuning was regarded as a baseline traditional oscillator (Methods).
In the experiments, the baseline yielded a 0.30 GHz (2.93$\sim$3.23 {GH}z) bandwidth tuning by adjusting the control voltage of varactor (Figure~\ref{fig:phase_transition}\textbf{c}).
In comparison, with the same amount of varactor tuning, the {PT}-symmetric system achieved a wider bandwidth tuning range of 0.57 {GH}z (2.63$\sim$3.20 {GH}z) by manipulating the gain-loss contrast (Figure~\ref{fig:phase_transition}\textbf{d}), effectively enabling 2.1 times frequency tuning range of the baseline.
We also observed that phase noise performances of our system were generally 1.5 dB better than the baseline across different oscillation frequencies (Supplementary Figure~\ref{fig:fig_9}), equivalent to 70 percentage of the baseline noise intensity. 
This quality improvement can be attributed to the enhanced intrinsic passive Q factor in a PT-symmetric system and the subsequently increased carrier power through gain-loss tuning (\ref{sec: si_4_2}).
These experiments demonstrate that {PT} symmetry can broaden the bandwidth tuning range of conventional microwave generation with improved quality.
Our further investigations also show that our system generates four-phase microwaves using the unique topology at the exceptional point (Supplementary Figure~\ref{fig:fig_8}).

\begin{figure}
\centering
\includegraphics[width=1\textwidth]{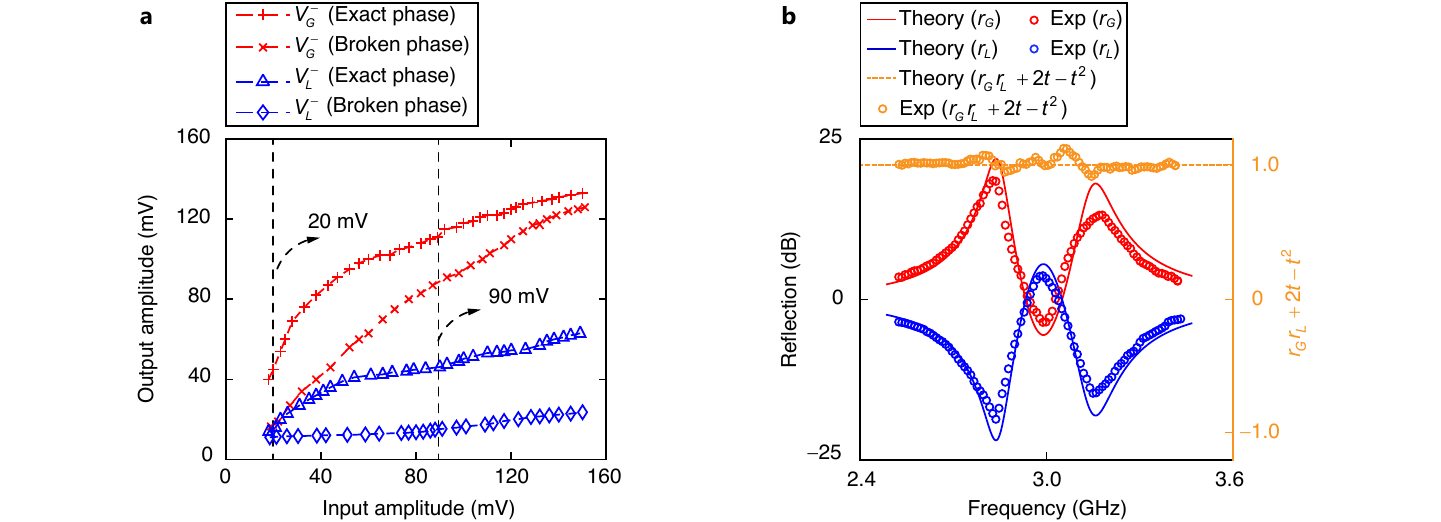} 
\caption{\textbf{{Nonlinearity characterization and generalized power conservation of the fully integrated PT-symmetric electronic system.}} \textbf{a}, Backward and forward input-output responses in both the exact phase and broken phase. Nonlinear responses occur in both regimes when the amplitudes of the input signals are high but are stronger in the broken phase with a steeper slope. The black dashed-dotted lines mark the input amplitudes we choose to investigate the phenomenon of nonreciprocal transmission.
\textbf{b}, Demonstrations of generalized unitary relationship (also known as generalized power conservation) of our system. We use two different ways to exhibit this property. First, we use decibel (dB, the left-side $y$-axis in the figure) to represent forward ($r_{G}$) and backward reflection coefficients ($r_{L}$) and show the comparison of $r_{G}$, and $r_{L}$ between experimental (Exp) measurements (empty circles) and {PT} symmetry theory (solid curves) in the linear region of the unbroken phase. The coefficients are symmetric about the $0$-dB axis.
Second, we demonstrate the quantity of $r_{G}\cdot r_{L}+2t-t^2$ (the right-side $y$-axis in the figure). It shows that the experimental results (Exp) are much close to the PT symmetry theory (dashed curve with a constant value 1) across the spectrum. Note that in this experiment we use a single-port~\cite{PT_E} set-up (\ref{sec: si_2_1}) for scattering measurement, where the transmission $t$ is 0. 
Both ways verify the property.}
\label{fig:scattering}
\end{figure}

\section{Nonlinearity characterization and scattering properties}

Although a small-signal model is used to simplify the analysis, nonlinearities are prevalent in CMOS integrated systems when they operate at high frequency and in a large-signal domain.
The XDP in our system is a nonlinear gain generator, which provides both a frequency-dependent~\cite{VCO} and a complex high-order amplitude-dependent~\cite{amp-depen} negative conductance (\ref{sec: si_1_2}), in contrast to the simple low-order amplitude-dependent gain intentionally introduced in previous {PT}-symmetric systems~\cite{Asym_t,NE_nonreci}.
To experimentally characterize the nonlinear response of our system, the equivalent of a transmission line (TL)~\cite{PT_E} with characteristic impedance $Z_0$ was attached to both sides of the system through an on-chip switch in the form of a resistor $R_0 = Z_0$.
We biased the system in either the unbroken or the broken phase and then monitored the output voltage $V^-_{G}$ ($V^-_{L}$) at the gain (loss) side by sending the amplitude-varying input $V^+_{L}$ ($V^+_{G}$) into the loss (gain) side (Figure~\ref{fig:design}\textbf{a}).
\textcolor{black}{For large input voltage (90 m$V$), a nonlinear response was clearly observed in both phases.
For small input voltage (20 m$V$), while a linear response is observed in the unbroken phase, a nonlinear response is clearly observed in the broken phase (Figure~\ref{fig:scattering}\textbf{a}).}
This enhanced nonlinearity can be attributed to the field localization in the active {RLC} resonator~\cite{NE_nonreci}.
Note that in this and the following experiments, we delicately tuned the gain and loss distribution to make our system operate in non-oscillatory mode while keeping the balanced PT symmetry condition.

Based on the characterization, we first studied the reflection of our system in the linear region.
The scattering theory~\cite{PT_E,gpc} of linear {PT}-symmetric systems has shown that their reflection fulfills generalized unitary relationships, i.e., the gain side reflection $r_{G}$ and the loss side reflection $r_{L}$ satisfy $r_{G}\cdot r_{L}=1$ across the spectrum (\ref{sec: si_2_1}).
To experimentally demonstrate this property, a TL was connected at the gain or loss side (Figure~\ref{fig:design}\textbf{a}), and the system was biased in the unbroken phase.  
Then, a sinusoidal signal with varied frequencies was applied to the system.
The incident wave, $V^+_{G}$ ($V^+_{L}$), and the reflected wave, $V^-_{G}$ ($V^-_{L}$), were extracted from the voltages at either side of the TL, from which the reflection coefficients $r_{G} ={V^-_{G}}/{V^+_{G}}$ and $r_{L} ={V^-_{L}}/{V^+_{L}}$ were calculated. 
The measured reflections in Figure~\ref{fig:scattering}\textbf{b} match well with the theoretical predictions.
This scattering property has been demonstrated before for telemetry sensing~\cite{PTX} by magnetically coupling two {RLC} resonators. 
Our system built upon capacitive coupling would also be promising as integrated chemical sensors if the capacitors of {RLC} resonators had been devised similarly to previous works~\cite{ic_sens}.
\textcolor{black}{We also investigated the two-port scattering property~\cite{CPA2,CPA3,CPA4} (\ref{sec: si_2_2}) and observed simultaneous existence of a coherent perfect absorption and lasing mode~\cite{CPA2,CPA3,CPA4} in this system (Supplementary Figure~\ref{fig:fig_14}).}
Moreover, extending our dimer system with more complex PT-symmetric structures can realize more advanced scattering phenomena, such as unidirectional invisibility~\cite{uni_d0,uni_d1} (\ref{sec: si_8_1}.

\section{Magnetic-free non-reciprocal microwave transmission}
PT-symmetric systems have demonstrated enhanced non-reciprocal acoustical and optical wave transmissions~\cite{nonre2,nonre4,NE_nonreci} with introduced nonlinear effects.
We then studied the non-reciprocal microwave transport of our system operating in different regions.
We measured the forward and backward transmissions of our system at different coupling factors ${\gamma}/{\gamma}^{}_{EP}$ by tuning gain-loss contrast.
In these experiments, the same experimental setup shown in Figure~\ref{fig:design}\textbf{a} was adopted.
The signal with variable frequencies was introduced into the gain (loss) side and captured at the loss (gain) side.
By measuring the incident wave $V^+_{G}$ ($V^+_L$) at the gain (loss) side and the transmitted wave $V^-_L$ ($V^-_G$) at the corresponding loss (gain) side, both the forward transmission $t_F={V^-_L}/{V^+_G}$ and backward transmission $t_B={V^-_G}/{V^+_L}$ were obtained.

\begin{figure}
\centering
\includegraphics[width=1\textwidth]{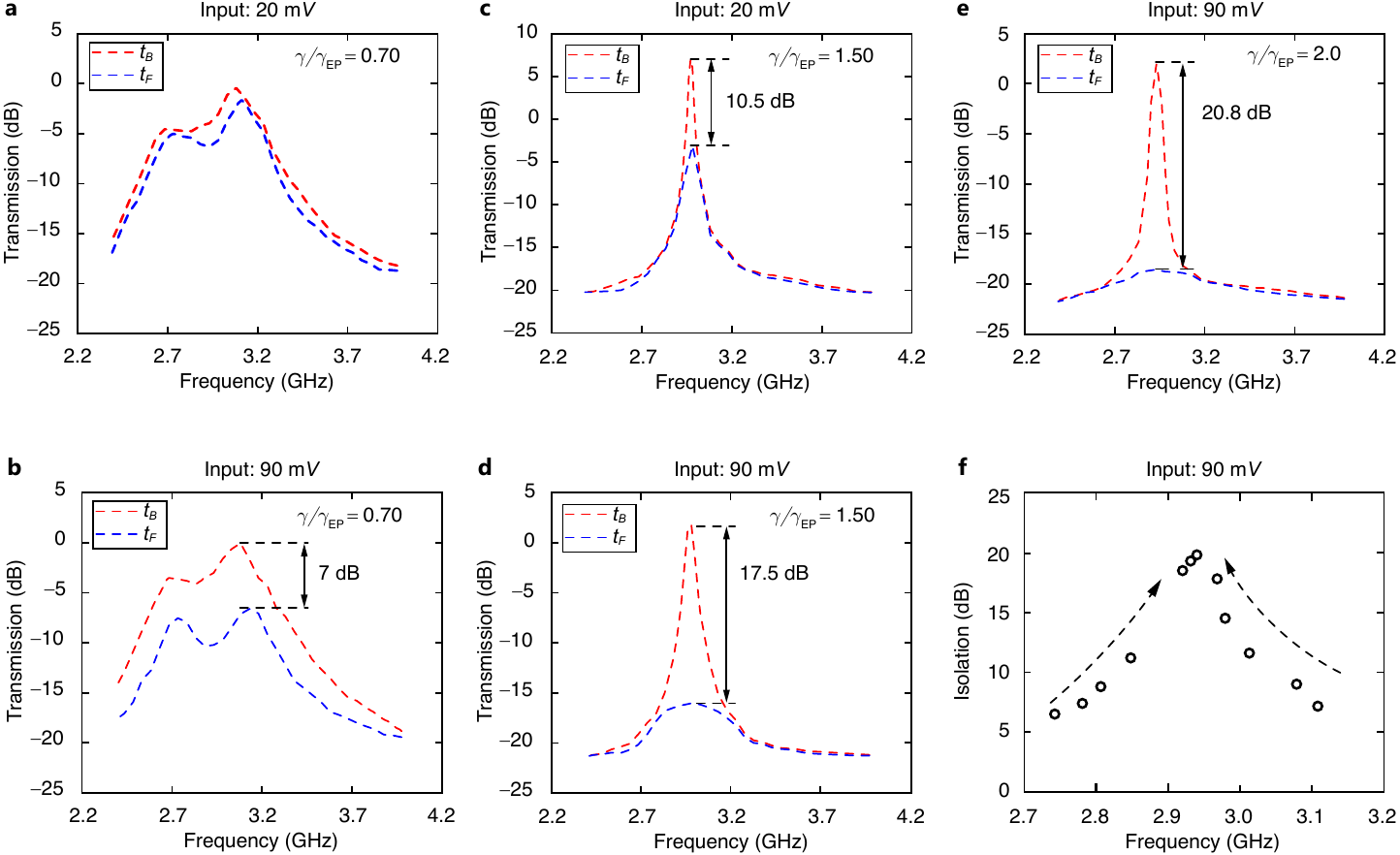} 
\caption{\textbf{{Non-reciprocal transmission of the fully integrated PT-symmetric electronic system.}} \textbf{a}, Reciprocal transmission in the unbroken phase (${\gamma}/{\gamma^{ }_{EP}}=0.70$) with a small input amplitude (20 m$V$). \textbf{b}, Non-reciprocal transmission in the unbroken phase (${\gamma}/{\gamma^{ }_{EP}}=0.70$) with increased input amplitude (90 m$V$). \textbf{c}, Non-reciprocal transmission in the broken phase (${\gamma}/{\gamma^{ }_{EP}}=1.50$) with a single peak at a small input amplitude (20 m$V$). \textbf{d}, Enhanced non-reciprocal transmission in the broken phase (${\gamma}/{\gamma^{ }_{EP}}=1.50$) with increased input amplitude (90 m$V$).
\textbf{e}, Further enhanced non-reciprocal transmission in the broken phase with increased coupling factor ($\gamma/\gamma^{ }_{EP}=2.0$) at the same input amplitude (90 m$V$). \textbf{f}, The isolation $t_B-t_F$ optimized by tuning the gain$/$loss parameter is shown at different probing frequencies in the microwave domain (2.75$\sim$3.10 GHz).}
\label{fig:non_reciprocal}
\end{figure}

In the unbroken phase (Figure~\ref{fig:non_reciprocal}\textbf{a},\textbf{b}), the transmission spectra in both directions show double resonances.
However, reciprocal transmission is observed with a 20 m$V$ input amplitude while the non-reciprocal transmission is observed with a 90 m$V$ input amplitude: the forward transmission goes up to $-$7 dB, and the backward transmission is 0.1 dB.
\textcolor{black}{In the low input power case, the system is linear and reaches an equilibrium between the two {RLC} resonators, giving rise to similar microwave transmissions in both directions.}
Nonlinearity appears with increasing input amplitudes.
The two resonators have different levels of nonlinearity, resulting in non-reciprocal transmission.
In the broken phase (Figure~\ref{fig:non_reciprocal}\textbf{c},\textbf{d}), the transmission spectra in both directions show single resonance.
Non-reciprocal transmissions are observed under both input cases with different input amplitudes, and the case with the larger input voltage shows more significant non-reciprocal transmission.
In addition, the comparison of the different phases under the same 90 m$V$ input amplitude (Figure~\ref{fig:non_reciprocal}\textbf{b},\textbf{d}) shows that the nonlinearity is greatly enhanced in the broken phase, owing to the field localization in the active {RLC} resonator, and therefore the non-reciprocity is also enhanced: $t_F$ reduces to $-$15.6 dB, and the $t_B$ slightly increases to 1.9 dB. 
Moreover, the non-reciprocity of the system is improved in the broken phase (Figure~\ref{fig:non_reciprocal}\textbf{e}) under the same 90 m$V$ input: $t_F$ decreases to $-$18.7 dB while the $t_B$ increases to 2.1 dB, due to the further enhanced nonlinearity by increasing coupling factor $\gamma/\gamma^{}_{EP}$.
By sweeping the coupling factor $\gamma/\gamma^{}_{EP}$ at a fixed 90 m$V$ input amplitude (Supplementary Figure~\ref{fig:fig_12}), this non-reciprocal behavior was observed over a broad bandwidth (2.75$\sim$3.10 GHz) at the resonance with the minimum isolation (Methods) over 7 dB (Figure~\ref{fig:non_reciprocal}\textbf{f}).

Recent device demonstrations have produced non-magnetic non-reciprocity in silicon based on temporal modulation~\cite{con_non}, but often exhibit narrow bandwidths and have significant area overheads because a number of passive devices are required to perform complex modulations.
Our system clearly demonstrates that PT symmetry with nonlinearity offers a new approach to achieving broadband non-magnetic non-reciprocal transmissions by tuning gain-loss contrast. 
Compared to state-of-the-art CMOS non-reciprocal devices~\cite{mag_1}, our fully integrated PT-symmetric electronic system shows strong isolation (7$\sim$21 dB) among a wider microwave (2.75$\sim$3.10 GHz) bandwidth without complex modulations that require huge and expensive on-chip area. 
Our system also shows strong non-reciprocity with a lower input power threshold ($-21$ dBm) compared to nonlinearity-induced non-reciprocal devices on other electronic platforms~\cite{non_bro,NE_nonreci} (Supplementary Table~\ref{tb:com_4}).
This excellent performance could lay the foundation for abundant application advancements in quantum computing~\cite{con_non}, device protection~\cite{non_bro,Wang2012}, and radar communication~\cite{nonlinearreci}.

\section{Conclusions}

We have reported a fully integrated electronic platform based on CMOS technology for non-Hermitian physics, validating the powerful role of IC to study PT symmetry in a scalable manner.
Fully integrated PT-symmetric electronics enables new capabilities in the microwave domain compared to the previous electronic platforms~\cite{PT_E,LRC_1,Dual_beh, wireless_p,NE_wireless,NE_sensor,topo_ele,topo_ele_1,topo_ele_2,PTX}.
With the unique gain/loss tuning mechanism of PT symmetry, our system shows extended broadband response and improved noise performance for microwave generation over conventional devices.
In particular, our chip demonstrates strong non-reciprocal microwave transmission with the enhanced intrinsic nonlinearity of IC, leading to a new generation of integrated non-magnetic non-reciprocal devices.
Our results shed light on PT symmetry as an innovative design approach to overcoming the limitations of IC performances and benefiting numerous applications.
In addition, more advanced IC technologies can be used to extend the functionality and performance benefit of PT-symmetric systems to the higher millimeter and terahertz frequency range.
The study is also expected to motivate further exploration
such as PT symmetry in opto-electronics~\cite{oeo1}, electro-acoustics~\cite{NE_nonreci}, and topological electronics (\ref{sec: si_8_2}) based on high-dimensional PT-symmetric structures~\cite{topo_ele, topo_ele_1,topo_ele_2}
with standard IC technology, enriching scientific discoveries of non-Hermitian physics.

\newpage
\noindent \textbf{Methods}

\noindent \textbf{Differential architecture.}
The fully integrated {PT}-symmetric electronic system was implemented in a differential topology (Supplementary Figure~\ref{fig:fig_1}).
The gain $-R_G$ is the parallel resistance of $-R_{G0}$, $-R_{G1}$ and $R_{G2}$.
$-R_{G0}$ is generated by the XDP.
$R_{G1}$ is realized by a voltage-controlled MOS resistor.
The loss $R_{L0}$ is realized in the same way as $R_{G1}$.
By fixing the bias voltage $V_{BIASG}$ of $-R_{G0}$ and controlling the bias voltage of gain (loss) side MOS resistor, $-R_G$ ($R_L$) can be continuously adjusted. 
The capacitor $C_G$ ($C_L$) in each {RLC} resonator is composed of a MIM capacitor $C_{G1}$ ($C_{L1}$) and a varactor $C_{G2}$ ($C_{L2}$).
The varactor only takes up a small proportion of the total capacitance in each RLC resonator and is used to compensate for the mismatch between the fixed MIM capacitor on both sides.
The coupling capacitance $C_C$ is designed by two equal MIM capacitors $C_{C1}$ and $C_{C2}$ which are serially connected through an on-chip switch (SW, Figure~\ref{fig:design}\textbf{a}).
The inductance $L_G$ ($L_L$) in both {RLC} resonators comes from symmetrical parallel inductors (symindp, Supplementary Figure~\ref{fig:fig_2}).
In the practical implementation, the equivalent of a transmission line (TL) with characteristic impedance $Z_0$ was attached to each side of the system through an on-chip switch in the form of an on-chip resistor $R_0=Z_0$.
By controlling the switch, the system could be flexibly configured to test one-port scattering or nonreciprocal transmission.

In the differential architecture, each signal is transmitted by a pair of differential wires where the signal is represented by the amplitude difference between the differential wires. 
For example, in our design, each voltage $V_G$ ($V_L$) at the terminal of {RLC} resonator is represented by a pair of differential signals ($V_{GP}$ and $V_{GN}$ for $V_G$, $V_{LP}$ and $V_{LN}$ for $V_L$).
The differential architecture is symmetric with respect to its virtual ground and can be divided into two equal parts (Supplementary Figure~\ref{fig:fig_3}).
Either of them is an equivalent single-ended representation of the differential one and can be used to derive the {PT} symmetry concept.
Note that in single-ended architecture, the gain $-R_G$, loss $R_L$, and inductor $L_G$ ($L_L$) will be half, and the capacitor $C_G$ ($C_L$) will be double.
The {PT} symmetry condition is satisfied by setting $R_G \approx R_L=R$, $L_G \approx L_L=L$, $C_G \approx C_L=C$.
In our system, $R\in[90, 380]$ $\Omega$, $L=1.85$ nH, $C\in[1300,1550]$ fF, $C_C=500$ fF, $Z_0=280$ $\Omega$.

\noindent \textbf{Phase transition.}
Applying Kirchoff’s law on the equivalent circuit representation in Figure~\ref{fig:design}\textbf{a} yields the following expression~\cite{PT_E}:
\begin{equation}
\label{VIrela1}
V_G=i{\omega}^{'}L/2\cdot I_1, ~\quad I_1 - V_G/(R/2)+i{\omega}^{'}2C\cdot V_G + i{\omega}^{'}C_C\cdot(V_G-V_L)=0,
\end{equation}
\begin{equation}
\label{VIrela2}
V_L=i{\omega}^{'}L/2\cdot I_2, ~\quad I_2 + V_L/(R/2)+i{\omega}^{'}2C\cdot V_L + i{\omega}^{'}C_C\cdot(V_L-V_G)=0.
\end{equation}
Here, ${\omega}^{'}$ is an angular frequency. 
Eliminating the current from the relations, scaling the frequency and time by ${\omega}_0=\frac{1}{\sqrt{LC}}$, and taking $c=C_C/2C$, $\gamma=\sqrt{L/C}/R$ gives the following matrix equation~\cite{PT_E}:
\begin{equation}
\label{eq:matrix}
\left[\begin{array}{cc}
{1}/{\omega}-j\gamma-{\omega}(1+c) & {\omega}c\\
{\omega}c & {1}/{\omega}+j\gamma-{\omega}(1+c) \\
\end{array}\right]\cdot \\
\left[\begin{array}{c}
 V_G\\
 V_L\\
\end{array}\right]=\left[\begin{array}{c}
 0\\
 0\\
\end{array}\right].
\end{equation}
Here, ${\omega}$ is the normalized frequency.
This linear, homogeneous system has four normal mode frequencies, as required to fulfill any arbitrary initial condition for voltage and current, given by~\cite{PT_E}
\begin{equation}
\label{eq:mode_main}
{\omega}_{1,2}=\pm \frac{\sqrt{{\gamma}^2_{EP} - {\gamma}^2} + \sqrt{{\gamma}^2_{UP} - {\gamma}^2}}{2\sqrt{1+2c}}, ~\quad {\omega}_{3,4}=\pm \frac{\sqrt{{\gamma}^2_{EP}  - {\gamma}^2} - \sqrt{{\gamma}^2_{UP} - {\gamma}^2}}{2\sqrt{1+2c}},
\end{equation}
where, the {PT}-symmetric breaking point (${\gamma}^{ }_{EP}$) and the upper critical point (${\gamma}^{}_{UP}$) are identified as ${\gamma}^{}_{EP}=|1 - \sqrt{1+2c}|$, ${\gamma}^{}_{UP}= 1 + \sqrt{1+2c}$.
The corresponding phase difference~\cite{PT_E} between the two RLC resonators can be expressed as
\begin{equation}
\label{eq:phi_main}
\phi_{1,3} = \frac{\pi}{2} - \tan^{-1}\left[\frac{1}{\gamma}\cdot(\frac{1}{{\omega}_{1,3}}-(1+c)\cdot{\omega}_{1,3})\right].
\end{equation}

\noindent \textbf{Frequency tuning range of microwave generation.} 
The frequency tuning range for microwave generation is defined as
\begin{equation}
\label{eq:ftr}
FTR=\frac{{\omega}_{\max}-{\omega}_{\min}}{({\omega}_{\max}+{\omega}_{\min})/2}\times 100\%.
\end{equation}
Here, ${\omega}_{\max}$ and ${\omega}_{\min}$ are the maximum and minimum frequency in the total tuning bandwidth.

\noindent \textbf{Isolation of non-reciprocal transmission.} 
The isolation of our system under a coupling factor ${\gamma}/{\gamma}^{}_{EP}$ is defined as 
\begin{equation}
\label{eq:iso}
I_{ISO}=\max(t_B-t_F).
\end{equation}
Here, $t_B$ and $t_F$ are the backward transmissions and forward transmissions of the system with the frequency sweeping.

\noindent \textbf{Chip implementation and fabrication.}
All CMOS devices were prepared in Cadence Virtuoso (an industry-standard design tool for frontend circuit design).
All designs satisfy standard CMOS manufacturing rules of IBM's commercial 130 nm CMOS8RF process, with physical verification performed using Mentor Graphics Calibre (an industry-standard design tool for backend layout design).
We resort to Metal Oxide Semiconductor Implementation Service (MOSIS) for fabrication.

\noindent \textbf{Measurement setups.}
 The chip was bonded on a 4-layer FR4 PCB (used as a daughter board in our experiments) by gold wires (Supplementary Figure~\ref{fig:fig_10}), forming all 38 electrical connections (including power and ground) from the chip to the PCB.
 The bonding wires were properly designed for use at frequencies above 50 {GH}z.
Our experimental setup comprised a bonded chip in a daughterboard, a motherboard, a power supply, a mixed signal oscilloscope (MSO, Agilent 9404A), an arbitrary wave generator (AWG, KEYSIGHT M8195A) and a personal computer (PC).
The daughter board provided control biases to the chip.
These biases had two main functions: 1) compensating the mismatch of CMOS components to minimize the unbalance between the two RLC resonators; 2) tuning the gain (loss) such that the system can evolve from exact phase to broken phase.
The mother board acted as a power board to supply all power/control voltages to the daughter board.
The benchtop power supply was the main power source and was used to power the mother board.
The MSO has four pairs of differential channels, and its highest sampling rate is 20 $GSa/s$.
The AWG has four pairs of differential channels, each pair of which can generate arbitrary waves up to 50 GHz with independently varying phases.
In the phase transition experiments, the outputs of two RLC resonators were connected to the MSO, where both the eigenfrequencies and phase differences could be directly observed on the panel. 
In the scattering experiments, the AWG sourced sinusoidal signals with varying frequencies or phases into the chip through TL.
Then signals on both terminals of the TL were sent into the MSO such that the incident wave and reflected wave could be captured. 
In the nonreciprocal experiments, the AWG fed sinusoidal signals with varying frequencies into the system through the gain (loss) side TL.
Then both the incident wave on the input terminal of the gain (loss) side TL and the reflected wave on the output terminal of loss (gain) side TL could be captured by MSO.
During the scattering experiments and nonreciprocal transport experiments, the AWG was controlled by software on a PC to generate frequency-varying signals.

\newpage

\noindent \textbf{Acknowledgements.} This work was supported in part by the National Science Foundation (NSF) grant no. CNS-1657562, no. CCF-1942900, and no. EFMA1641109.
W.D. C and X.Z. thank MOSIS educational program (MEP) for the chip fabrication. 
W.D. C, C.Q. W, W.J. C, L.Y., and X.Z. thank Tsampikos Kottos from Wesleyan University for the technical discussions.

\noindent \textbf{Author contributions.} L.Y. and X.Z. conceived the idea.
W.D. C and X.Z. designed the circuits. 
W.D. C performed the simulations and experiments with the aid of C.Q. W, W.J. C, and S.H. 
All authors contributed to formulating the analytical model, analyzing the data, and writing the manuscript.
L.Y. and X.Z. supervised the project.

\noindent \textbf{Competing interests.} The authors declare no competing interests.

\noindent \textbf{Data availability.} Source data are provided in this paper. The data that support the findings of this study are available within the article and its Supplementary Information. Additional data are available from the corresponding authors upon request.

%% file: Design_analysis_PT.tex


 This part is the Supplementary Information of ``Fully integrated parity-time-symmetric electronics''~\cite{cao2022fully} orginally published here \url{https://www.nature.com/articles/s41565-021-01038-4}.

\section{Implementation}
\subsection{Differential Architecture And Detailed Circuits}

The schematic overview of the proposed fully integrated parity-time- (PT-) symmetric electronic system is illustrated in Supplementary Figure~\ref{fig:fig_1}. 
As it shows, our system was implemented with a differential topology, in contrast with the single-ended architecture commonly used in board-level~\cite{S21} and MEMs-level~\cite{S22} PT-symmetric electronic systems. 
The differential architecture has the advantage to mitigate common-mode perturbations.

\begin{suppfigure}[h]
 \centering
    \includegraphics[width=0.70\linewidth]{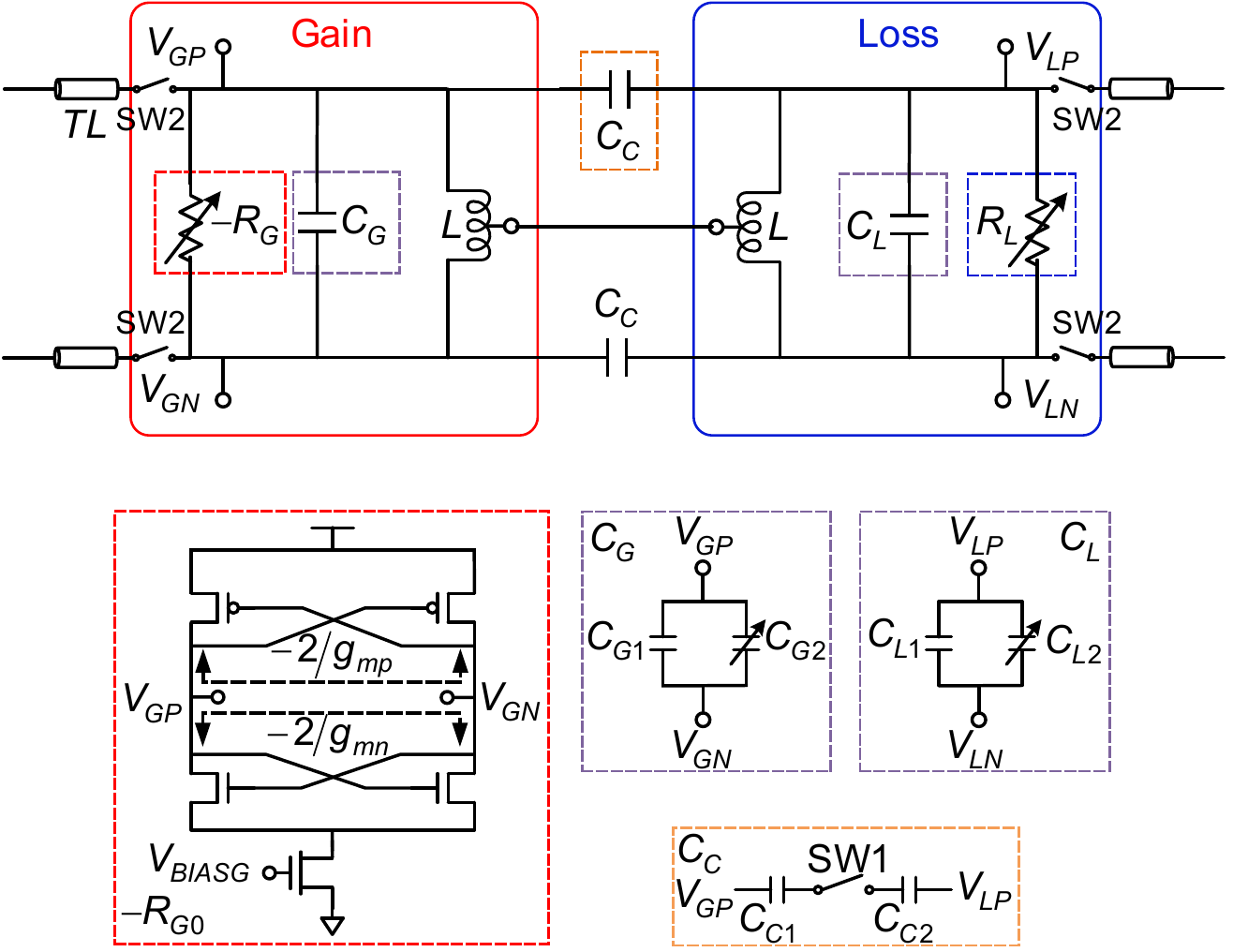}
    \caption{ \textbf{Schematic overview of fully integrated PT-symmetric electronic system with differential architecture.} It consists of two RLC resonators with balanced gain $-R_G$ and loss $R_L$. The gain $-R_{G0}$ is generated by the cross-coupled differential pair (XDP).  The capacitances $C_G$ ($C_L$) in both RLC resonators are comprised of an inherent parasitic capacitance $C_{G0}$ ($C_{L0}$), a fixed high-Q MIM capacitor $C_{G1}$ ($C_{L1}$) and a varactor $C_{G2}$ ($C_{L2}$). The varactor is used to compensate the mismatch between $C_{G1}$ and $C_{L1}$. The coupling capacitance ($C_C$) is made of two serially connected MIM capacitors ($C_{C1}$ and $C_{C2}$) with an on-chip switch (SW1). The two RLC resonators can be coupled (decoupled) by turning on (off) the SW1. TLs are attached to both sides of the system by SW2.}
    \label{fig:fig_1}
\end{suppfigure}

Our system consists of two RLC resonators, one with active gain $-R_G$ and the other one with
passive loss $R_L$.
The gain $-R_G$ is the parallel resistance of $-R_{G0}$, $R_{G1}$, and $R_{G2}$, namely, $-R_{G}=-R_{G0}||R_{G1}||R_{G2}$.
Here, $-R_{G0}$ is generated by the cross-coupled differential pair (XDP); $R_{G1}$ is a variable resistor realized by MOS transistors~\cite{mos_r1,mos_r2}; $R_{G2}$ is the inherent loss of the active RLC resonator. 
Similarly, the loss $R_L$ is the parallel resistance of $R_{L0}$, and $R_{L1}$, that is $R_{L}=R_{L0}||R_{L1}$.
$R_{L0}$ is a variable resistor realized by in the same way as $R_{G1}$;
$R_{L1}$ is the inherent loss of the passive RLC resonator. 
By controlling the bias voltage of gain (loss) side MOS transistors~\cite{mos_r1,mos_r2}, $-R_{G}$ ($R_{L}$) can be continuously adjusted. 
The capacitor $C_{G}$ ($C_{L}$) in each RLC resonator is composed of a parasitic capacitance $C_{G0}$ ($C_{L0}$), a fixed Metal-Insulator-Metal (MIM) capacitor $C_{G1}$ ($C_{L1}$) with high-quality factor (high-Q) and an adjustable varactor $C_{G2}$ ($C_{L2}$). 
The varactor takes up a small proportion of the total capacitance and is used to compensate for the fabricated mismatch between the fixed MIM capacitors of both sides. 
The coupling capacitance $C_{C}$ is designed by two
equal MIM capacitors $C_{C1}$ and ($C_{C2}$) in serial connection via an on-chip switch (SW1). 
Note that the two RLC resonators can also be coupled (decoupled) by turning on (off) of the SW1.

\begin{suppfigure}[h]
 \centering
    \includegraphics[width=0.4\linewidth]{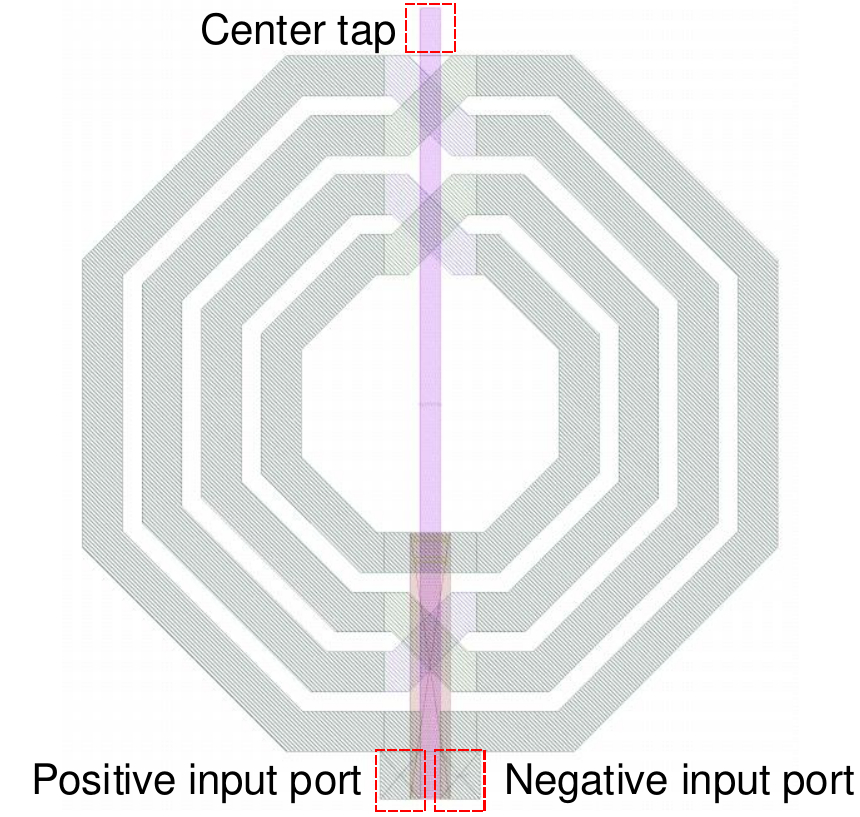}
    \caption{ \textbf{Symmetrical physical layout of symindp.} It has a pair of differential input ports and a center tap.}
    \label{fig:fig_2}
\end{suppfigure}

The inductor $L$ in each RLC resonator is a symmetrical parallel inductor (symindp) with three terminals: two input ports and one center tap. 
The center tap connection is provided such that by connecting the center taps of inductors in both RLC resonators, the passive RLC resonator shares the same common-mode voltage with the active one.
This symindp with cross-over connections to create a symmetrical layout (Supplementary Figure~\ref{fig:fig_2}) with low resistance and low parasitic capacitance, is ideally suited for differential resonators. 
In the practical implementation, the equivalent of a transmission line (TL) with characteristic impedance $Z_0$ was attached to both sides of the system through an on-chip switch (SW2) in the form of a resistor $R_0=Z_0$.

\begin{suppfigure}[!t]
 \centering
    \includegraphics[width=0.58\linewidth]{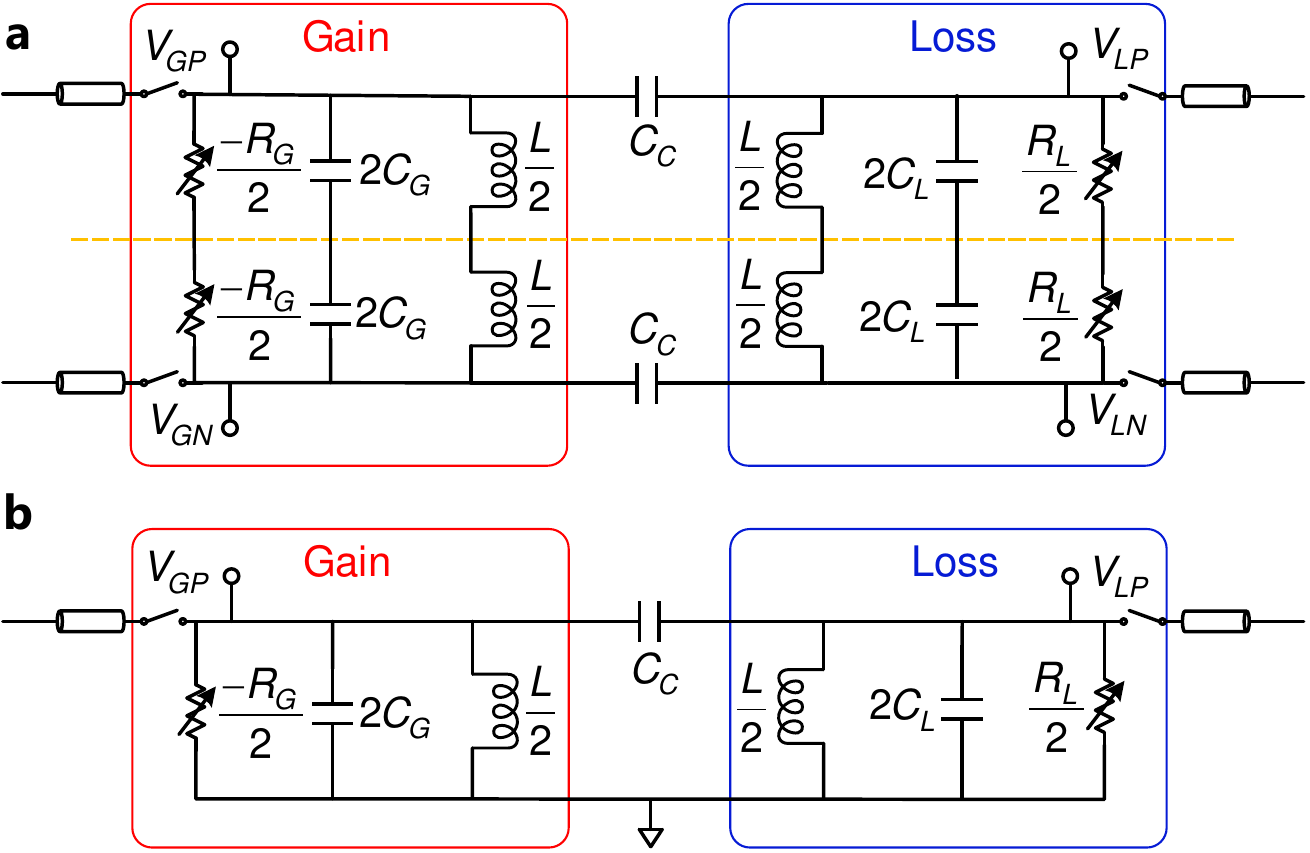}
    \caption{ \textbf{Transformation of differential architecture to equivalent single-ended architecture.} \textbf{a}. The symmetric representation of PT-symmetric electronic system. The yellow dash line is a symmetric axis, also as known as virtual ground. \textbf{b}. The equivalent single-ended circuit schematic of PT-symmetric electronic system. }
    \label{fig:fig_3}
\end{suppfigure}

A common way to analyze differential circuits is to convert them into single-ended equivalents. 
As the Supplementary Figure~\ref{fig:fig_3}a shows, by using a symmetrical axis\footnote{The horizontal dashed yellow line shown in Supplementary Figure~\ref{fig:fig_3}a.}, the differential circuit can be divided into two equal parts.
Either of them is an equivalent single-ended representation of the differential one, and can be used to derive mathematical expression of the PT-symmetric system. 
Therefore, the equivalent single-ended representation in Supplementary Figure~\ref{fig:fig_3}b is used as the simplified model for analysis through our paper. The difference between these two circuit topologies is that in differential architecture, each signal is represented by the amplitude difference between the differential wires; while in single-ended architecture, each signal is transmitted by only one wire and all the terminal signals are referenced to a common ground.
For example, in our system, the voltage $V_G$ ($V_L$) at the terminal of each RLC resonator is represented by a pair of differential signals, e.g., $V_{GP}$ and $V_{GN}$ for $V_G$, $V_{LP}$ and $V_{LN}$ for $V_L$. 
Note that in single-ended architecture, gain $-R_G$, loss $R_L$ and inductor $L$ will be half, but the capacitor $C_G$ ($C_L$) will be double. 
The PT symmetry condition is satisfied by setting $R_G\approx R_L=R$, $L_G\approx L_L=L$, and $C_G\approx C_L=C$.

\begin{suppfigure}[!t]
 \centering
    \includegraphics[width=0.75\linewidth]{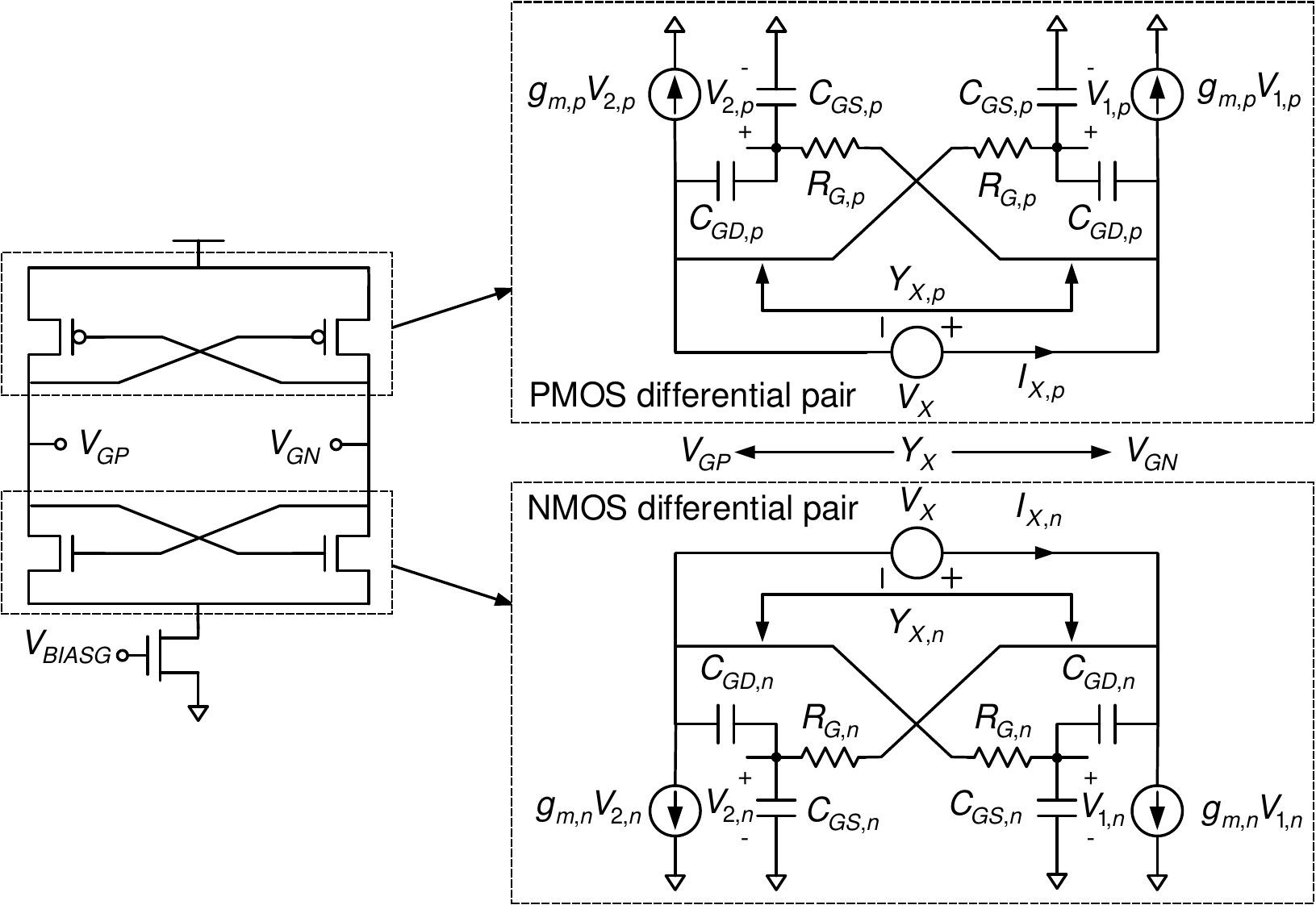}
    \caption{ \textbf{Small signal model of cross-coupled differential pair.} This model considers parasitic effect (gate resistor $R_G$, gate-drain capacitance $C_{GD}$, gate-source capacitance $C_{GS}$) when the circuit operates at high frequency. }
    \label{fig:fig_4}
\end{suppfigure}

\subsection{Analysis of Cross-coupled Differential Pair}
\label{sec: si_1_2}

A comprehensive analysis of $-R_{G0}$ can be obtained through the small signal model~\cite{S11} of XDP shown in Supplementary Figure~\ref{fig:fig_4}, where for NMOS differential pair we have
\begin{equation}
Y_{X,n}(S)= \frac{R_{G,n}C_{GS,n}C_{GD,n}S^2 + [C_{GS,n}+(4+g_{m,n}R_{G,n})C_{GD,n}]-g_{m,n} }{2[R_{G,n}(C_{GS,n}+C_{GD,n})+1]  },
    \label{eqn:nmos}
\end{equation}
and for PMOS differential pair,
\begin{equation}
Y_{X,p}(S)= \frac{R_{G,p}C_{GS,p}C_{GD,p}S^2 + [C_{GS,p}+(4+g_{m,p}R_{G,p})C_{GD,p}]-g_{m,p} }{2[R_{G,p}(C_{GS,p}+C_{GD,p})+1]  }.
    \label{eqn:pmos}
\end{equation}
Hence,
\begin{equation}
Re\{Y_{X,n}\}=\frac{-g_{m,n}+R_{G,n}\omega^2 [C_{GS,n}^2+C_{GD,n}(4+g_{m,n} R_{G,n})  (C_{GS,n}+C_{GD,n} )] }{   2[R_{G,n}^2 (C_{GS,n}+C_{GD,n} )^2 \omega^2+1] },
\label{eqn:nmos1}
    \end{equation}
\begin{equation}
Re\{Y_{X,p}\}=\frac{-g_{m,p}+R_{G,p}\omega^2 [C_{GS,p}^2+C_{GD,p}(4+g_{m,p} R_{G,p})  (C_{GS,p}+C_{GD,p} )] }{   2[R_{G,p}^2 (C_{GS,p}+C_{GD,p} )^2 \omega^2+1] }.
\label{eqn:pmos1}
    \end{equation}
Here, the subscript $n$ ($p$) denotes NMOS (PMOS) differential pair. 
If $R_{G,n}^2 (C_{GS,n}+C_{GD,n})^2 \omega^2 \ll 1$ and $R_{G,p}^2 (C_{GS,p}+C_{GD,p})^2 \omega^2 \ll 1$, then 
\begin{equation}
Re\{Y_{X,n}\} \approx -\frac{g_{m,n}}{2} + R_{G,n}\omega^2\cdot\frac{C_{GS,n}^2+C_{GD,n}(4+g_{m,n} R_{G,n})  (C_{GS,n}+C_{GD,n} )}{2},
\label{eqn:nmos2}
    \end{equation}
\begin{equation}
Re\{Y_{X,p}\} \approx -\frac{g_{m,p}}{2} + R_{G,p}\omega^2\cdot\frac{C_{GS,p}^2+C_{GD,p}(4+g_{m,p} R_{G,p})  (C_{GS,p}+C_{GD,p})}{2}.
\label{eqn:pmos2}
    \end{equation}
Combing Eq.~\eqref{eqn:nmos2} and Eq.~\eqref{eqn:pmos2}, one can obtain
\begin{equation}
Re\{Y_{X}\}=Re\{Y_{X,n}\}+Re\{Y_{X,p}\}=-(g_{m,n}+g_{m,p})/{2}+f(\omega).
\label{eqn:negr}
    \end{equation}
Here, $f(\omega)$ is a frequency-dependent term. 
In large signal domain,  $Re\{Y_X\}$ also dependents on the amplitude of oscillation voltage between the two terminals of the differential pair~\cite{S12}. 
A reasonable assumption is that when frequency $\omega$ is low and the XDP operates in the small signal domain, $Re\{Y_X\}$ can be expressed as       
\begin{equation}
Re\{Y_{X}\}={-(g_{m,n}+g_{m,p})}/{2}.
\label{eqn:negr1}
    \end{equation}
Therefore, the negative resistance $-R_{G0}$ can be obtained as
\begin{equation}
-R_{G0}=-{2}/{(g_{m,n}+g_{m,p})},
\label{eqn:negr2}
    \end{equation}
which is the reciprocal of summation of small signal transconductance of NMOS differential pair and PMOS differential pair. 
For theoretical analysis, we consider our system as a linear system with the assumption of small signal condition.
Supplementary Figure~\ref{fig:fig_voltage} shows the dependence of the negative resistance on the bias voltage.
The results are obtained from high-fidelity post-layout circuit simulations, where process, voltage, and temperature (PVT) variations are carefully considered.
The error bar at each point is the potential variation range of the negative resistance caused by PVT variations at the same bias voltage.
Note that for analog circuit design in mature technology (such as CMOS 130 nm), the simulated results from high-fidelity post-layout simulator (i.e., Cadence Spectre) often match well with the measured results, and thus can be used to verify the functionalities of the designed chip.

\begin{suppfigure}[!t]
 \centering
    \includegraphics[width=0.5\linewidth]{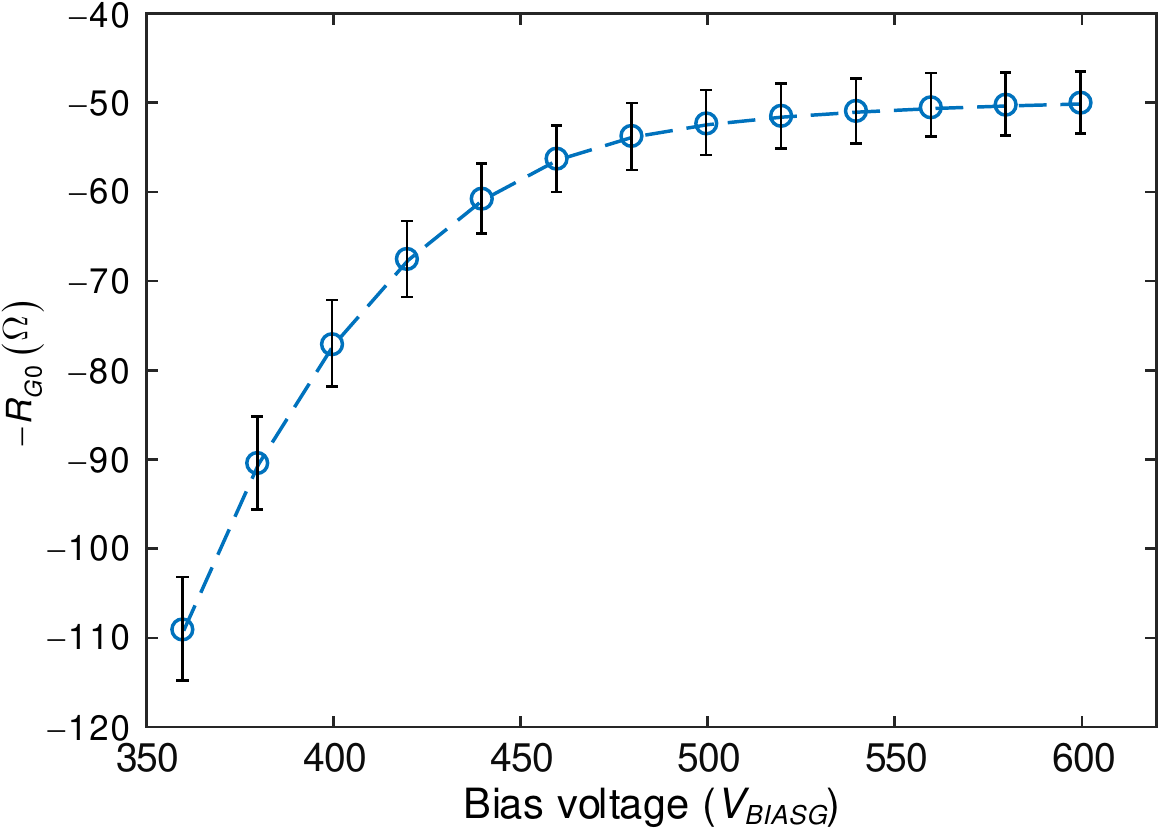}
    \caption{ \textbf{Tuning gain with the bias voltage.} The results are obtained from the thorough post-layout simulation with the consideration of PVT variations.}
    \label{fig:fig_voltage}
\end{suppfigure}

%% file: Scattering_analysis.tex
\section{Scattering Properties}
\label{sec:scat}
\subsection{Single-port Scattering}
\label{sec: si_2_1}

    \begin{suppfigure}[!t]
 \centering
    \includegraphics[width=0.60\linewidth]{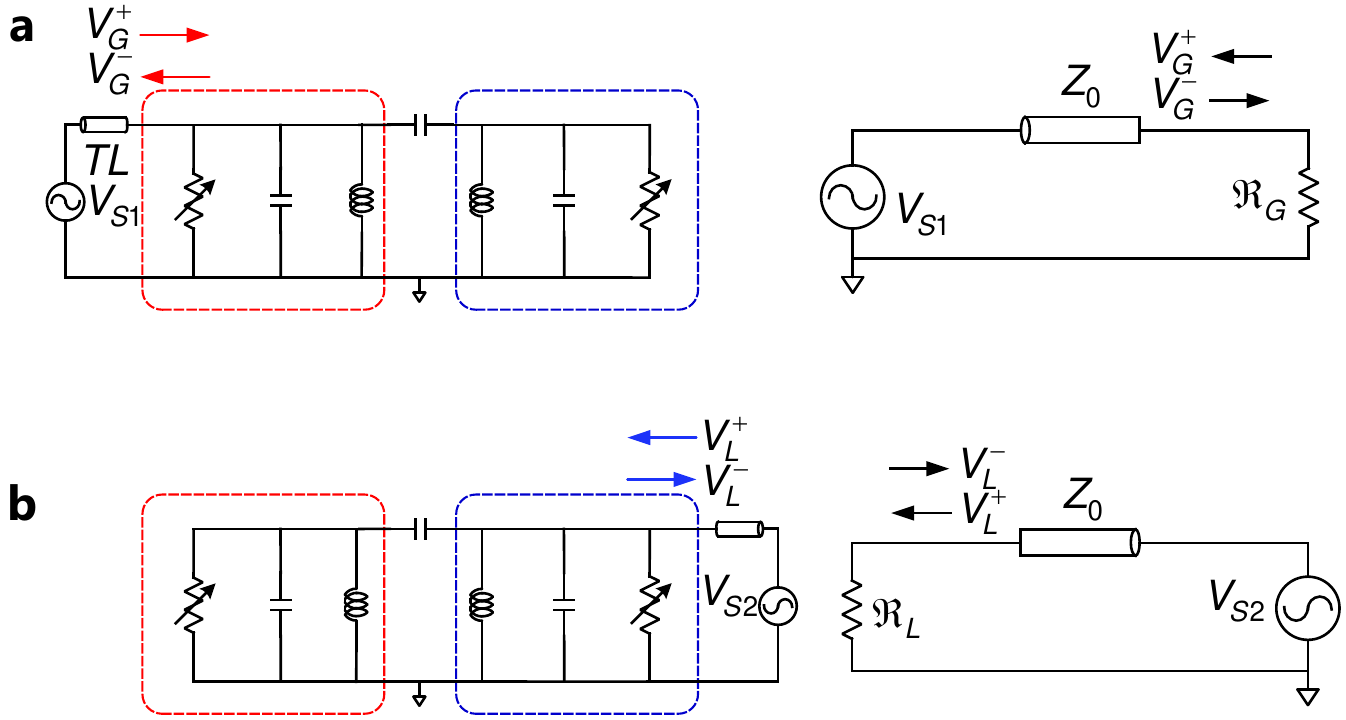}
    \caption{ \textbf{Simplified model for single-port scattering associated with the fully integrated PT-symmetric electronic system.} \textbf{a}. TL is attached to the gain side of the system. \textbf{b}. TL is connected to the loss side of the system.}
    \label{fig:fig_5}
\end{suppfigure}

    \begin{suppfigure}[!t]
 \centering
    \includegraphics[width=0.55\linewidth]{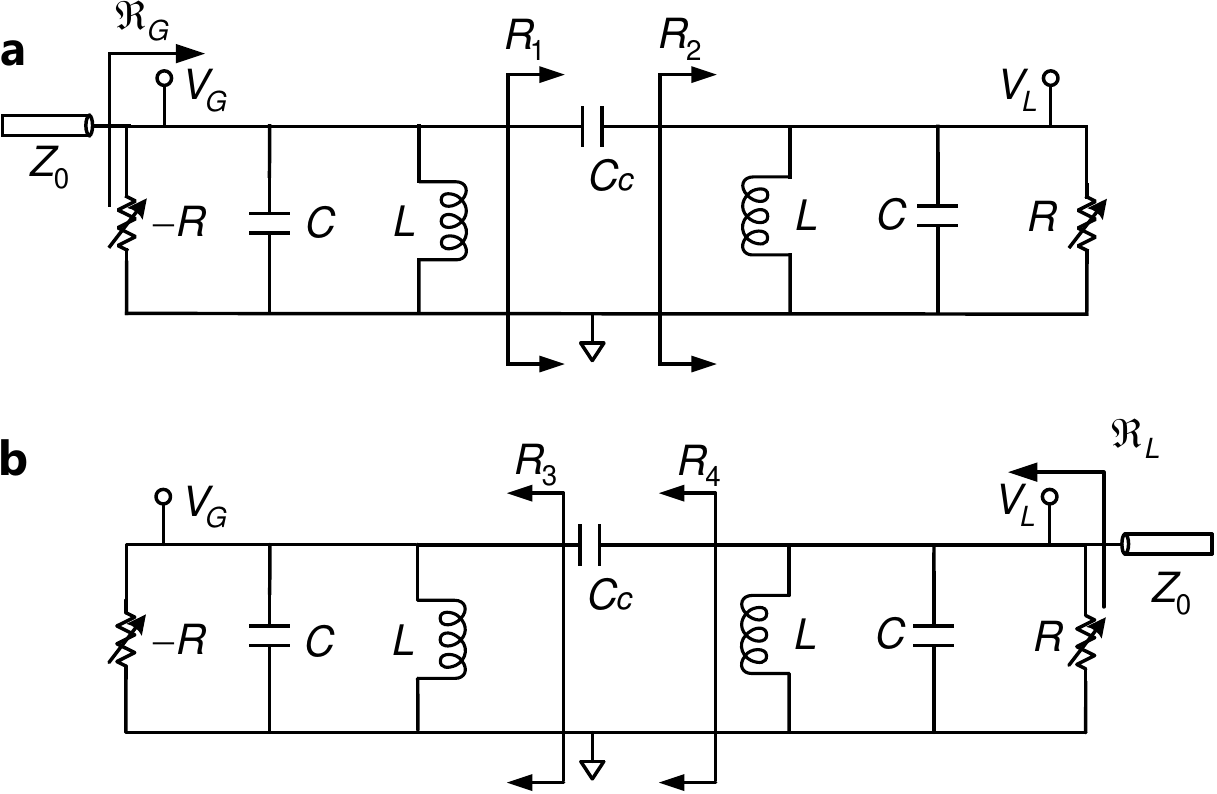}
    \caption{ \textbf{Circuit models to calculate $\mathcal{R}_G$ and $\mathcal{R}_L$.} \textbf{a}. Circuit model for calculating $\mathcal{R}_G$. \textbf{b}. Circuit model for calculating $\mathcal{R}_L$.}
    \label{fig:fig_6}
\end{suppfigure}

The theory of linear PT-symmetric systems has shown that single-port scattering fulfills generalized unitary relationship, that is gain side reflection $r_G$ and loss side reflection $r_L$ satisfy $r_G\cdot r_L=1$. 
We first derive the theory of single-port scattering for our system from the circuit perspective. 
Supplementary Figure~\ref{fig:fig_5} represents the conceptual illustration of single-port scattering and its equivalent circuit model. 
In scattering theory, the reflection coefficients are defined as the ratio of reflected wave and incident wave, that is $r_G=V_G^-/V_G^+$ and $r_L=V_L^-/V_L^+$. 
In circuit theory, the gain side reflection $r_G$ can be readily obtained as
\begin{equation}
r_G=\frac{\mathcal{R}_G(\omega^{'}) - Z_0}{\mathcal{R}_G(\omega^{'}) + Z_0}.
\label{eqn:resg}
    \end{equation}
 While for the loss side, the reflection $r_L$ can be directly written as  
\begin{equation}
r_L=\frac{\mathcal{R}_L(\omega^{'}) - Z_0}{\mathcal{R}_L(\omega^{'}) + Z_0}.
\label{eqn:resl}
    \end{equation}
Here, $\mathcal{R}_G$ represents the equivalent impedance of the system seen from left to right (shown in Supplementary Figure~\ref{fig:fig_6}a); $\mathcal{R}_L$ represents the impedance of the system seen from right to the left (shown in Supplementary Figure~\ref{fig:fig_6}b); and $Z_0$ is the characteristic resistor of TL. 
Let $\gamma=\sqrt{(L/C)}/R$, $c=C_C/C$, $\omega=\omega^{'} \sqrt{LC}$, and resort to the Supplementary Figure~\ref{fig:fig_6}a, then   
\begin{equation}
\mathcal{R}_G=R\cdot \frac{-\omega^2\gamma^2+j[\omega \gamma-(1+c)\omega^3\gamma ]   }{ (1-\omega^2)^2 - 2\omega^2c(1-\omega^2)+\omega^2\gamma^2}.
\label{eqn:resg1}
    \end{equation}
Similarly, we resort to the Supplementary Figure~\ref{fig:fig_6}b to calculate $\mathcal{R}_L$ which is expressed as
\begin{equation}
\mathcal{R}_L=R\cdot \frac{\omega^2\gamma^2+j[\omega \gamma-(1+c)\omega^3\gamma ]   }{ (1-\omega^2)^2 - 2\omega^2c(1-\omega^2)+\omega^2\gamma^2}.
\label{eqn:resl1}
    \end{equation}
Let $ E =R\omega^2\gamma^2;~
              F =R[\omega\gamma-(1+c)\omega^3\gamma];~
              G=(1-\omega^2)^2-2\omega^2c(1-\omega^2)+\omega^2\gamma^2$,
then 
\begin{equation}
 \mathcal{R}_G =({-E+jF})/{G}; ~~~ \mathcal{R}_L=({E+jF})/{G}.
    \label{eqn:nor1}
\end{equation}   
By combining Eq.~\eqref{eqn:resg}, Eq.~\eqref{eqn:resl}, and Eq.~\eqref{eqn:nor1}, we can obtain 
\begin{equation}
              ||r_G||\cdot||r_L|| =1; ~~~
              \phi_G+\phi_L=\pi.
    \label{eqn:nor2}
\end{equation}

    \begin{suppfigure}[!t]
 \centering
    \includegraphics[width=0.45\linewidth]{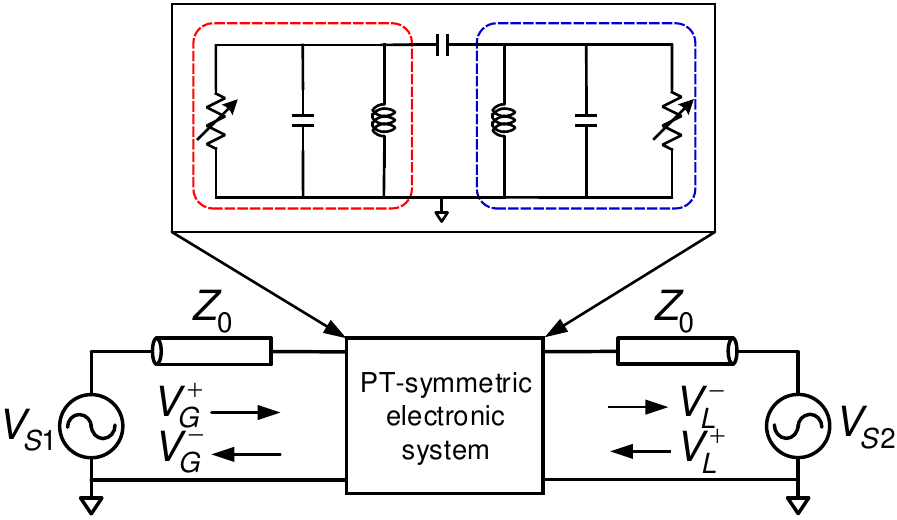}
    \caption{ \textbf{Circuit models to calculate the total output coefficient.} The upper is the PT-symmetric electronic system. Below is the circuit model.}
    \label{fig:fig_two_port}
\end{suppfigure}

\subsection{Two-port Scattering}
\label{sec: si_2_2}

The theory of linear PT-symmetric systems has also shown that two-port scattering exhibits a simultaneous coherent perfect absorber (CPA-) -amplifier property at a special frequency (Janus frequency).
We then derive the two-port scattering theory for our system from the circuit perspective. The two-port scattering can be considered as two-port TL model shown in Supplementary Figure~\ref{fig:fig_two_port}. 
We can write the basic voltage law of system by using $S$-parameter model~\cite{S21}:
\begin{equation}
\left\{
             \begin{array}{lr}
              \dfrac{ V_G^-}{\sqrt{Z_0}} =S_{11}\cdot \dfrac{ V_G^+}{\sqrt{Z_0}} + S_{12}\cdot \dfrac{ V_L^+}{\sqrt{Z_0}}; &  \\

              \dfrac{ V_L^-}{\sqrt{Z_0}}=S_{21}\cdot \dfrac{ V_G^+}{\sqrt{Z_0}} + S_{22}\cdot \dfrac{ V_L^+}{\sqrt{Z_0}}.
             \end{array}
\right.
    \label{eqn:S_0}
\end{equation}

\noindent Here, in our system,
\begin{equation}      
S=1/(A-iB)\cdot \left[                
  \begin{array}{cc}   
    -iD &  2\omega c\eta \\  
    2\omega c\eta & iC \\  
  \end{array}
\right] .
    \label{eqn:S_1}
\end{equation}
And, 
\begin{equation}
\left\{
             \begin{array}{lr}
              A = 2\eta\Omega ; &  \\
            B = \Omega^2-\eta^2-\omega^2c^2 + \gamma^2; &  \\
            C = (\gamma-\eta)^2 + \Omega^2-\omega^2c^2;\\
            D = (\gamma+\eta)^2 + \Omega^2-\omega^2c^2;\\
            \Omega = \omega(1+c)-1/\omega;\\
            \gamma = \sqrt{L/C}/R;\\
                  \eta = \sqrt{L/C}/Z_0;\\
                   c = C_C/C;\\
            \omega = \omega^{'} \sqrt{LC}.
             \end{array}
\right.
    \label{eqn:P_0}
\end{equation}
We can transform Eq.~\eqref{eqn:S_0} into the following equation~\cite{S21}:
\begin{equation}      
  \left[\begin{array}{c}   
    V_L^-  \\  
    V_L^+ \\  
  \end{array} \right] = \mathcal{M}\cdot \left[                
  \begin{array}{c}   
    V_G^-  \\  
    V_G^+ \\  
  \end{array}
\right].
    \label{eqn:M_0}
\end{equation}
Here,
\begin{equation}      
 \mathcal{M}=\dfrac{1}{2\omega c\eta}\cdot \left[                
  \begin{array}{cc}   
    A+iB &  iC \\  
    -iD &  A-iB \\  
  \end{array}
\right].
    \label{eqn:M_1}
\end{equation}
Note that $det( \mathcal{M})=1$.
Therefore,
\begin{equation}      
S=\dfrac{1}{\mathcal{M}_{22}} \cdot \left[                
  \begin{array}{cc}   
    - \mathcal{M}_{21} &  1 \\  
   1 &  \mathcal{M}_{12} \\  
  \end{array}
\right].
    \label{eqn:S_2}
\end{equation}

Generally, the reflection and transmission coefficients for the gain ($G$) and loss ($L$) incidence in terms of the transfer matrix elements as~\cite{S21}
\begin{equation}      
r_G =-\dfrac{\mathcal{M}_{21}}{\mathcal{M}_{22}},~~~r_L =-\dfrac{\mathcal{M}_{12}}{\mathcal{M}_{22}},~~~ t_G=t_L= \dfrac{1}{\mathcal{M}_{22}}.
    \label{eqn:coeff}
\end{equation}
It can be derived that 
\begin{equation}
\left\{
             \begin{array}{lr}
              ||r_G||\cdot||r_L|| =\sqrt{||( \mathcal{M}_{21}\cdot\mathcal{M}_{12}) /(\mathcal{M}_{22}\cdot\mathcal{M}_{22})||}=\sqrt{||T-1||}; &  \\

              \phi_G+\phi_L=\pi.
             \end{array}
\right.
\label{eqn:coeff1}
\end{equation}
Here, transmittance $T=t_G\cdot t_L$.
In the single-port scattering case, the transmittance $T=0$, as $\mathcal{M}_{22}\rightarrow \infty$ when $\eta \rightarrow 0$.
In other words, $||r_G||\cdot||r_L|| =1$.
Therefore, Eq.~\eqref{eqn:nor2} is a special case of Eq.~\eqref{eqn:coeff1}.

Using the scattering matrix, one can derive the conditions that our PT-symmetric system can simultaneously act either as an amplifier or as a perfect absorber~\cite{S21,S24,S25,chen2021non,wang2021non}. 
For a laser oscillator without an injected signal, the boundary condition satisfies $V_G^+=V_L^+=0$, which indicates $\mathcal{M}_{22}(\omega)=0$ in Eq.~\eqref{eqn:S_2}.
For a perfect absorber, the boundary condition satisfies $V_G^-=V_L^-=0$, which implies $det(S)=0$ in Eq.~\eqref{eqn:S_2}.
Therefore, $\mathcal{M}_{11}(\omega)=(1+\mathcal{M}_{12}\mathcal{M}_{21})/\mathcal{M}_{22}=0$, and the amplitudes of the incident waves must satisfy the condition $V_L^+=\mathcal{M}_{21}(\omega)V_G^+$. 
For the PT-symmetric structure, the matrix elements of $\mathcal{M}$ in Eq.~\eqref{eqn:M_1} satisfy the relationship $\mathcal{M}_{11}(\omega)=\mathcal{M}_{22}^{*} (\omega^{*})$. 
Thus, a real $\omega=\omega_J$ (Janus frequency) exists, that satisfies the amplifier/laser condition simultaneously with the absorber condition ($\mathcal{M}_{11} (\omega_J)=\mathcal{M}_{22}(\omega_J)=0$)~\cite{S21,S24}. 
Hence the two-port PT-symmetric system can behave simultaneously as a perfect absorber and as an amplifier.

This property can be explored using an overall output coefficient $\Theta$ defined as~\cite{S21,S24}
\begin{equation}      
\Theta= ({|V_G^-|^2+|V_L^-|^2})/({|V_G^+|^2+|V_L^+|^2}). 
    \label{eqn:theta}
\end{equation}
Note that in the case of a single-port scattering set-up discussed earlier in this section, the $\Theta$-function collapses to the gain/loss side reflectances.
Let ${V_L^+}/{V_G^+}$ be a generic ratio, then the perfect amplifier coefficient~\cite{S21} is obtained as:
\begin{equation}      
\Theta_{amp}(\omega)= \dfrac{ \big|\frac{V_L^+}{V_G^+}\mathcal{M}_{12}(\omega) +1 \big|^2  +  \big|\frac{V_L^+}{V_G^+}-\mathcal{M}_{21}(\omega) \big|^2}{\big(1+\frac{|V_L^+|^2}{|V_G^+|^2}\big)\big|\mathcal{M}_{22}(\omega)\big|^2} .
    \label{eqn:theta_1}
\end{equation}
At the singularity frequency point $\omega = \omega_J$, the 
$\Theta(\omega)$-function diverges as $\omega \rightarrow \omega_J$ and the circuit acts as an amplifier/laser. 
If on the other hand, we assume
that $V_L^+=\mathcal{M}_{21}(\omega)V_G^+$ (perfect adsorption condition), we can obtain~\cite{S21}
\begin{equation}      
\Theta_{abs}(\omega_J)= ({|\mathcal{M}_{22}(\omega_J) \mathcal{M}_{11}(\omega_J)|^2  })/({1+|\mathcal{M}_{21}(\omega_J)|^2| \mathcal{M}_{11}(\omega_J)|^2 })=0.
    \label{eqn:theta_2}
\end{equation}

%% file: Measurement_scat.tex
\section{Measurement Theory of Scattering Properties}
\subsection{Measurement Theory of Single-port Scattering}
\label{sec:S_scat_th}

In Section~\ref{sec:scat}, we derived the theoretical formula of single-port scattering coefficient. However, the theoretical formula cannot be used to calculate the coefficients for simulation and measurement.
Therefore, we need to find feasible method for practical measurement. We take the Supplementary Figure~\ref{fig:fig_5} for example.
It is obvious to know that $V_G = V_G^++V_G^-$.
Here, $V_G$ is the node voltage of gain side; $V_G^+$ is the incident wave, and $V_G^-$ is the reflected wave. 
We rewrite the reflection of gain side $r_G$ in Eq.~\eqref{eqn:resg} as   
\begin{equation}
r_G = \dfrac{\mathcal{R}_G(\omega)-Z_0}{\mathcal{R}_G(\omega)+Z_0} = \dfrac{V_G^-}{V_G^+ }=\beta_Ge^{j\phi},
    \label{eqn:refg1}
\end{equation} 
where, $\beta$ is the module of reflection coefficient, and $\phi$ is the phase of reflection coefficient. 
Therefore, $V_G=V_G^++V_G^-=V_G^+\cdot(1+\beta_G e^{j\phi} )=V_G^+\cdot(1+\beta_G  \cos(\phi)+\beta_G \sin(\phi)$), and 
\begin{equation}
\dfrac{V_G}{V_G^+} = 1+\beta_G  \cos(\phi)+\beta_G \sin(\phi).
    \label{eqn:refg2}
\end{equation} 
Here, $V_G^+=V_1/2$. 
$V_1$ is the voltage of $V_{S1}$. 

On the other hand, we also can let 
\begin{equation}
\dfrac{V_G}{V_G^+} = \alpha e^{j\xi}=\alpha\cos(\xi) +j\alpha\sin(\xi).
    \label{eqn:refg3}
\end{equation} 
Here, $V_G$ is the node voltage of gain side; $\xi$ is the phase difference between $V_G$ and source voltage $V_{S1}$; $\alpha$ is the amplitude ratio between $V_G$ and incident wave voltage $V_G^+$. 
Comparing Eq.~\eqref{eqn:refg2} and Eq.~\eqref{eqn:refg3}, we can obtain
\begin{equation}
\left\{
             \begin{array}{lr}
              \beta_G  \sin(\phi) =\alpha\sin(\xi); &  \\
              1+\beta_G  \cos(\phi)=\alpha\cos(\xi).
             \end{array}
\right.
    \label{eqn:refg4}
\end{equation} 
Then 
\begin{equation}
\left\{
             \begin{array}{lr}
              \beta_G  =\sqrt{1+\alpha^2-2\alpha\cos(\xi)}; &  \\

             \phi=\arctan((\alpha \sin(\xi))/(\beta_G\cos(\xi)-1)).
             \end{array}
\right.
    \label{eqn:refg5}
\end{equation} 
Therefore, if we measure the amplitude of $V_G$ and the phase difference $\xi$ between $V_G$ and source voltage $V_{S1}$, we can get the experiment results of reflection coefficient $r_G$ of gain side. 
Similarly, $r_L$ of loss side can also be obtained.

\subsection{Measurement Theory of Two-port Scattering}
From Supplementary Figure~\ref{fig:fig_two_port} and Eq.~\eqref{eqn:refg1}, we can easily get
\begin{equation}
\left\{
             \begin{array}{lr}
              V_G^- = V_G^+ \cdot \beta_G e^{j\phi_1}; &  \\

             V_L^- = V_L^+ \cdot \beta_L e^{j\phi_2}.
             \end{array}
\right.
    \label{eqn:two_P_1}
\end{equation} 
Then, the formula for measurement is
\begin{equation}      
\Theta= \dfrac{|V_G^-|^2+|V_L^-|^2}{|V_G^+|^2+|V_L^+|^2}=\dfrac{|V_L^+ \cdot \beta_L e^{j\phi_2}|^2+|V_G^+ \cdot \beta_G e^{j\phi_1}|^2}{|V_G^+|^2+|V_L^+|^2}. 
    \label{eqn:theta_m}
\end{equation}
Here, $V_G^+=V_1/2$, and $V_L^+=V_2/2$. 
$V_1$ and $V_2$ are the amplitude of $V_{S1}$ and $V_{S2}$, respectively. 
Base on the measurement theory of \ref{sec:S_scat_th}, $\phi_1$, $\phi_2$, $\beta_L$ and $\beta_G$ could be easily calculated. 
Therefore, $\Theta$ can be obtained by plugging these values into Eq.~\eqref{eqn:theta_m}.

%% file: VCO_theory.tex
\section{Microwave Generation}
\label{sec: si_4}

\subsection{Frequency Tuning Range Comparison}
\label{sec:VCO}
Microwave generation is very important on diverse on-chip applications~\cite{cao2019neuadc_C,cao2019neuadc,pipelineadc,pipelineadc_J,cao_tc,LV,zhouNEWCAS,cao2021phd,caoedssc1,caoedssc2,caomwscas,caonewcas}.
We theoretically compare the bandwidth of microwave generation of the fully integrated PT-symmetric electronic system and traditional oscillators.
As derived from the Methods of the main text, 
    \begin{suppfigure}[!t]
 \centering
    \includegraphics[width=0.60\linewidth]{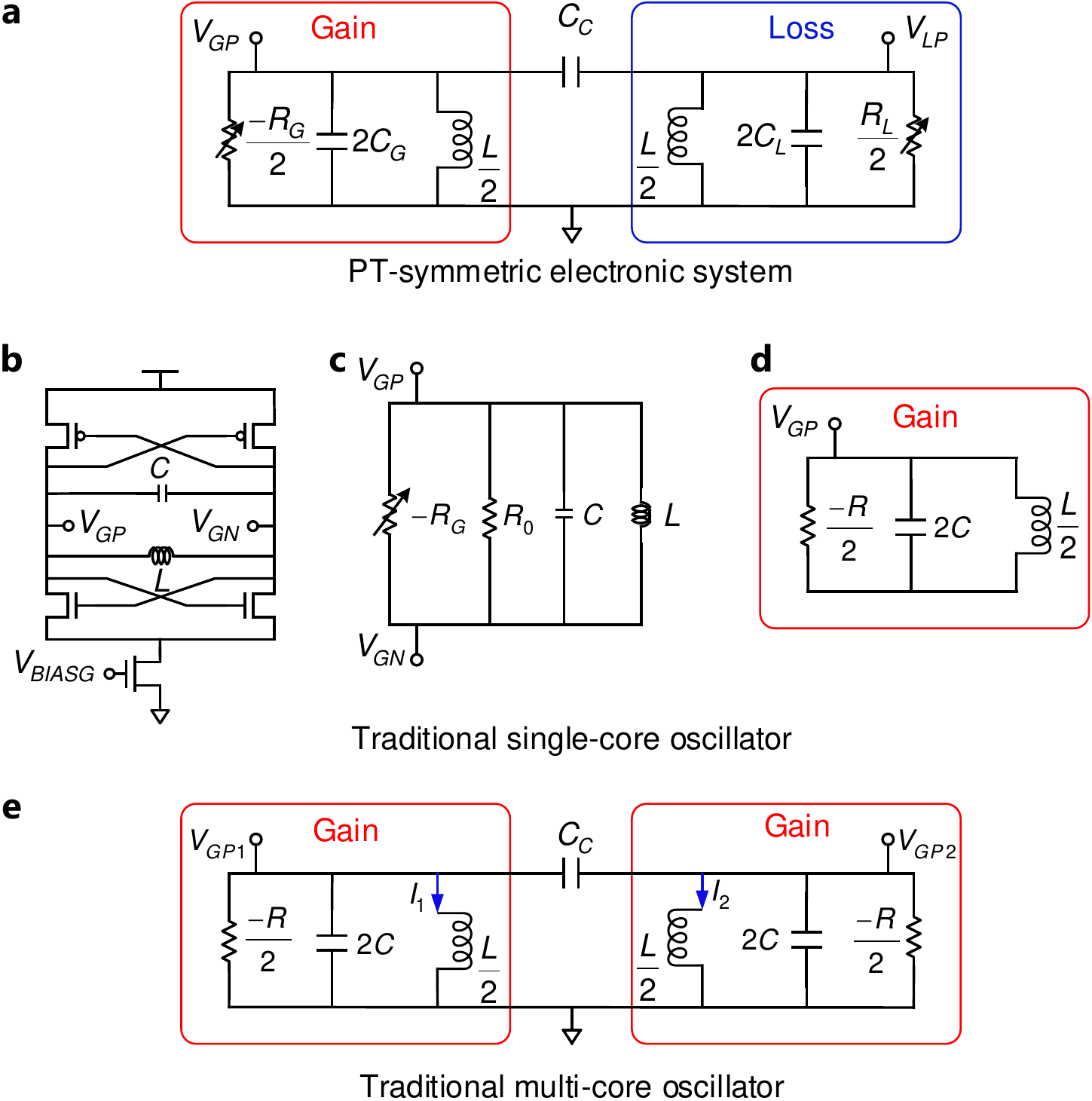}
    \caption{ \textbf{Illustration of three kinds of microwave generators.} \textbf{a}. The equivalent single-ended circuit schematic of the differential architecture of the fully integrated PT-symmetric electronic system. This figure is taken from Supplementary Figure~\ref{fig:fig_3}b. \textbf{b}. Schematic view of a traditional single-core oscillator. \textbf{c}. Equivalent differential model of the single-core oscillator. \textbf{d}. The equivalent single-ended circuit schematic of the single-core oscillator. \textbf{e}. The equivalent single-ended circuit schematic of multi-core oscillators. We use the multi-core oscillator built upon two capacitively coupled active RLC resonators as an example.}
    \label{fig:fig_7}
\end{suppfigure}
the fully integrated PT-symmetric electronic system (Supplementary Figure~\ref{fig:fig_7}a) has four normal mode frequencies,
\begin{equation}
\label{eq:mode}
{\omega}_{1,2}=\pm \frac{\sqrt{{\gamma}^2_{EP} - {\gamma}^2} + \sqrt{{\gamma}^2_{UP} - {\gamma}^2}}{2\sqrt{1+2c}}, ~~~ {\omega}_{3,4}=\pm \frac{\sqrt{{\gamma}^2_{EP}  - {\gamma}^2} - \sqrt{{\gamma}^2_{UP} - {\gamma}^2}}{2\sqrt{1+2c}},
\end{equation}
where, the breaking point (${\gamma}^{}_{EP}$) and the upper critical point (${\gamma}^{}_{UP}$) are identified as
\begin{equation}
\label{eq:point}
{\gamma}^{}_{EP}=\Big|1 - \sqrt{1+2c}\Big|, ~~~{\gamma}^{}_{UP}= 1 + \sqrt{1+2c}.
\end{equation}
The corresponding phase difference~\cite{S21} between the two RLC resonators can be expressed as
\begin{equation}
\label{eq:phi}
\phi_{1,3} = \frac{\pi}{2} - \tan^{-1}\left[\frac{1}{\gamma}\cdot(\frac{1}{{\omega}_{1,3}}-(1+c)\cdot{\omega}_{1,3})\right].
\end{equation}
Here, $\gamma$ is the gain-loss contrast tuning which is defined as $\gamma=\sqrt{L/C}/R$.

We then derive the theory for conventional single-core oscillators (Supplementary Figure~\ref{fig:fig_7}b).
Applying Kirchoff’s law on the equivalent circuit representation in Supplementary Figure~\ref{fig:fig_7}d yields the following expression:
\begin{equation}
\label{VIrela1_t}
\dfrac{V_{GP}}{-R/2}+\dfrac{V_{GP}}{i\omega^{'} L/2}+V_{GP}\cdot i\omega^{'} 2C=0.
\end{equation}
Here, $R=-R_{G}||R_0$ with $-R_{G}$ the tunable gain and $R_0$ the inherent loss of the resonator.
Using the same normalization methods presented before, that is ${\omega}_0={1}/{\sqrt{LC}}$, $\gamma=\sqrt{L/C}/R$, Eq.~\eqref{VIrela1_t} can be transferred into $\omega^2+i\gamma \omega -1 =0$ whose solutions are
\begin{equation}
\label{VIrela1_t_s}
\omega_{1,2}= \dfrac{-i\gamma \pm \sqrt{4-\gamma^2}}{2}.
\end{equation}
Eq.~\eqref{VIrela1_t_s} suggests that the oscillation happening in a single-core oscillator mainly goes through two phases: start-up phase and stable phase.
In the start-up phase, a small-signal gain $-R_{G}$ initially set slightly above the inherent loss $R_0$ is used to compensate for the loss so as to generate an oscillated microwave.
The oscillation frequency--the real part of the microwave--is in fact related to the amount of loss.
However, as the amplitude of the microwave exponentially grows, the small-signal gain $-R_{G}$ is degenerated in the large-signal domain due to the nonlinearity of the system, whose final value is equivalent to the loss $R_0$, leading to $\gamma\rightarrow0$.
The oscillation then steps into the stable phase, where the microwave's amplitude saturates at a fixed amplitude level and its oscillation frequency also becomes stable.
Such an oscillation frequency is independent of the gain-loss contrast and only determined by the natural frequency ($\omega_0=1/\sqrt{LC}$) of the resonator, i.e., 
\begin{equation}
\label{VIrela1_t_s_1}
\omega_{1,2}= 1.
\end{equation}
Note that in this stable phase, an oscillator generates stable sinusoidal waves for diverse on-chip applications.
Obviously, the stable oscillation frequency of conventional single-core oscillators can be tuned only by the capacitance $C$ or the inductance $L$.

\begin{suppfigure}[!t]
 \centering
    \includegraphics[width=0.7\linewidth]{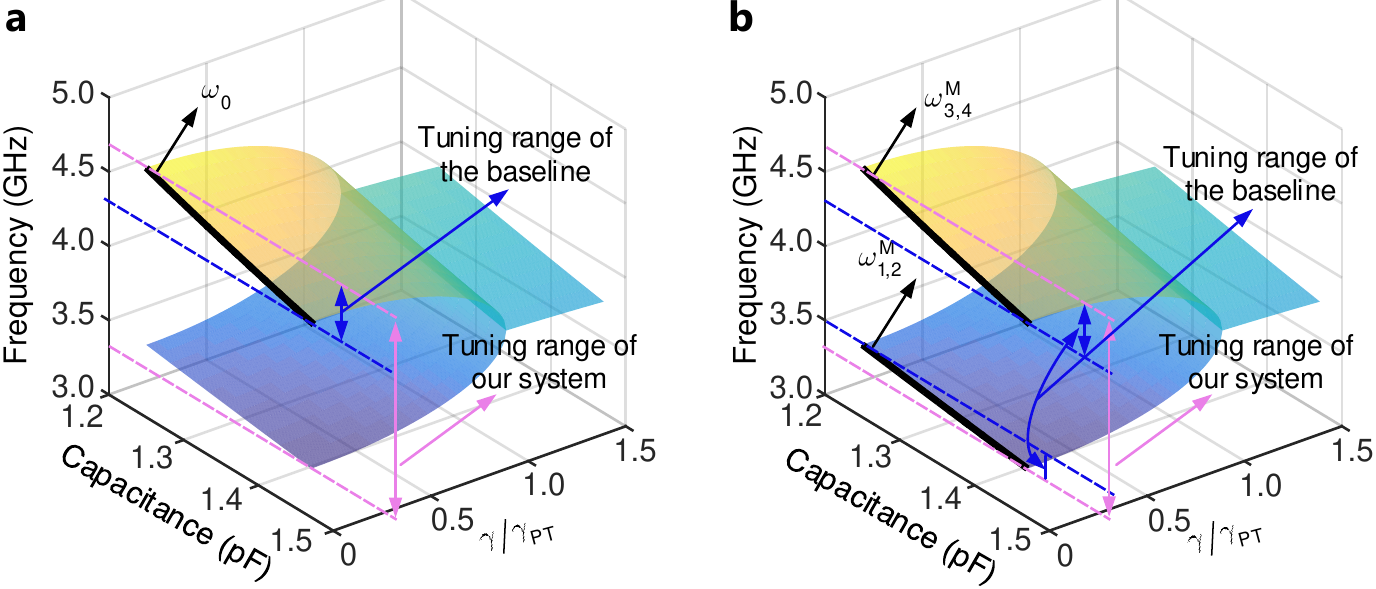}
    \caption{ \textbf{Theoretical comparisons of frequency tuning range between different oscillators}. \textbf{a}. Comparison between the single-core oscillator and our system. The black line indicates the tuning range of the single-core oscillator. \textbf{b}. Comparison between the multi-core oscillator and our system. The black line indicates the tuning range of the multi-core oscillator. }
    \label{fig:fig_com}
\end{suppfigure}

Multi-core VCOs are formed by coupling multiple identical single-core LC VCOs.
Here, we use a multi-core VCO built upon two coupled resonators as shown in Supplementary Figure~\ref{fig:fig_7}e as an example.
This oscillator with coupled-resonator structure without gain-loss contrast is used as another baseline of our system. 
A similar $I$-$V$ relations of the circuit can be obtained by using the Kirchoff’s law:
\begin{equation}
\left\{
             \begin{array}{lr}
              V_{GP1}=i{\omega}^{'}\dfrac{L}{2}\cdot I_1, ~~~I_1 - \dfrac{V_{GP1}}{R/2}+i{\omega}^{'}2C\cdot V_{GP1} + i{\omega}^{'}C_C\cdot(V_{GP1}-V_{GP2})=0; &  \\
&  \\
             V_{GP2}=i{\omega}^{'}\dfrac{L}{2}\cdot I_2, ~~~I_2 - \dfrac{V_{GP2}}{R/2}+i{\omega}^{'}2C\cdot V_{GP2} + i{\omega}^{'}C_C\cdot(V_{GP2}-V_{GP1})=0.
             \end{array}
\right.
    \label{eqn:VIrela1}
\end{equation} 
Using the same normalization methods as the single-core VCOs and considering $\gamma\rightarrow0$, the solutions are given by
\begin{equation}
\label{eq:mode_cou_1}
{\omega}^{\text{M}}_{1,2}=\pm \dfrac{1}{\sqrt{(1+2c)}}; ~~{\omega}^{\text{M}}_{3,4}= \pm 1.
\end{equation}

Comparing Eq.~\eqref{eq:mode}, Eq.~\eqref{VIrela1_t_s_1}, and Eq.~\eqref{eq:mode_cou_1}, it can be found that Eq.~\eqref{VIrela1_t_s_1} is a special form of $\omega_{1,2}$ in Eq.~\eqref{eq:mode} when $\gamma\rightarrow0$; Eq.~\eqref{eq:mode_cou_1} is a special form of Eq.~\eqref{eq:mode} when $\gamma\rightarrow0$.
The comparison shows that in addition to the inherent tuning freedoms preserved by $\omega_0$, ${\omega}_{1,2}$ and ${\omega}_{3,4}$ in Eq.~\eqref{eq:mode} also preserve an extra resistive tuning freedom, i.e., $\gamma=\sqrt{L/C}/R$.
Supplementary Figure~\ref{fig:fig_com} compares the theoretical frequency tuning range of three oscillators using the same tunable parameters, exhibiting a larger tuning range of our system.

\subsection{Phase Noise Comparison}
\label{sec: si_4_2}

Supplementary Figure~\ref{fig:fig_pn}a shows the passive resonator model for a conventional single-core oscillator whose phase noise (PN) model is illustrated in Supplementary Figure~\ref{fig:fig_pn}b.
The main noise sources come from resistor thermal noise ($I^{2}_{R_0}(\omega)=4kT/R_0,~\omega>0$) and transistor thermal noise (${I^{2}_{g_m}}(f)=4kTm g_m,~\omega>0$).
The classical PN formula of the conventional single-core oscillators~\cite{S35} is shown below
\begin{equation}
\label{eqn:pn}
\begin{split}
  &\ \mathcal{L}_{conv}(\triangle \omega) = 10 \cdot \log \Big[ \dfrac{\mathcal{P}_{sideband}(\omega+\triangle \omega,1Hz)}{\mathcal{P}_{carrier}} \Big] \\
  &\ = 10 \cdot \log \Bigg((1+m)\cdot \frac{4kTR_0}{V^2_{osc,conv}} \cdot\Big(\frac{\omega}{2Q_S\triangle \omega}\Big)^2\Bigg).  
\end{split}
\end{equation}
Here, $\mathcal{P}_{sideband}(\omega+\triangle \omega,1~Hz)$ represents the single sideband power of noise at a frequency offset of $\triangle \omega$ from the carrier with a
measurement bandwidth of $1~$Hz.
$\omega$ is the oscillation frequency.
$\triangle \omega$ is the frequency offset.
$k$ is Boltzmann's constant. 
$T$ is the absolute temperature. 
$R_0$ is the inherent resonator resistance. 
$m$ is a noise factor of the active device.
$V_{osc,conv}$ is the oscillation amplitude.
$Q_{S}$ is the quality factor of the resonator as shown in Supplementary Figure~\ref{fig:fig_pn}a, defined as $Q_{S}=\omega R_0C=R_0/(\omega L)$.

   \begin{suppfigure}[!t]
 \centering
    \includegraphics[width=0.9\linewidth]{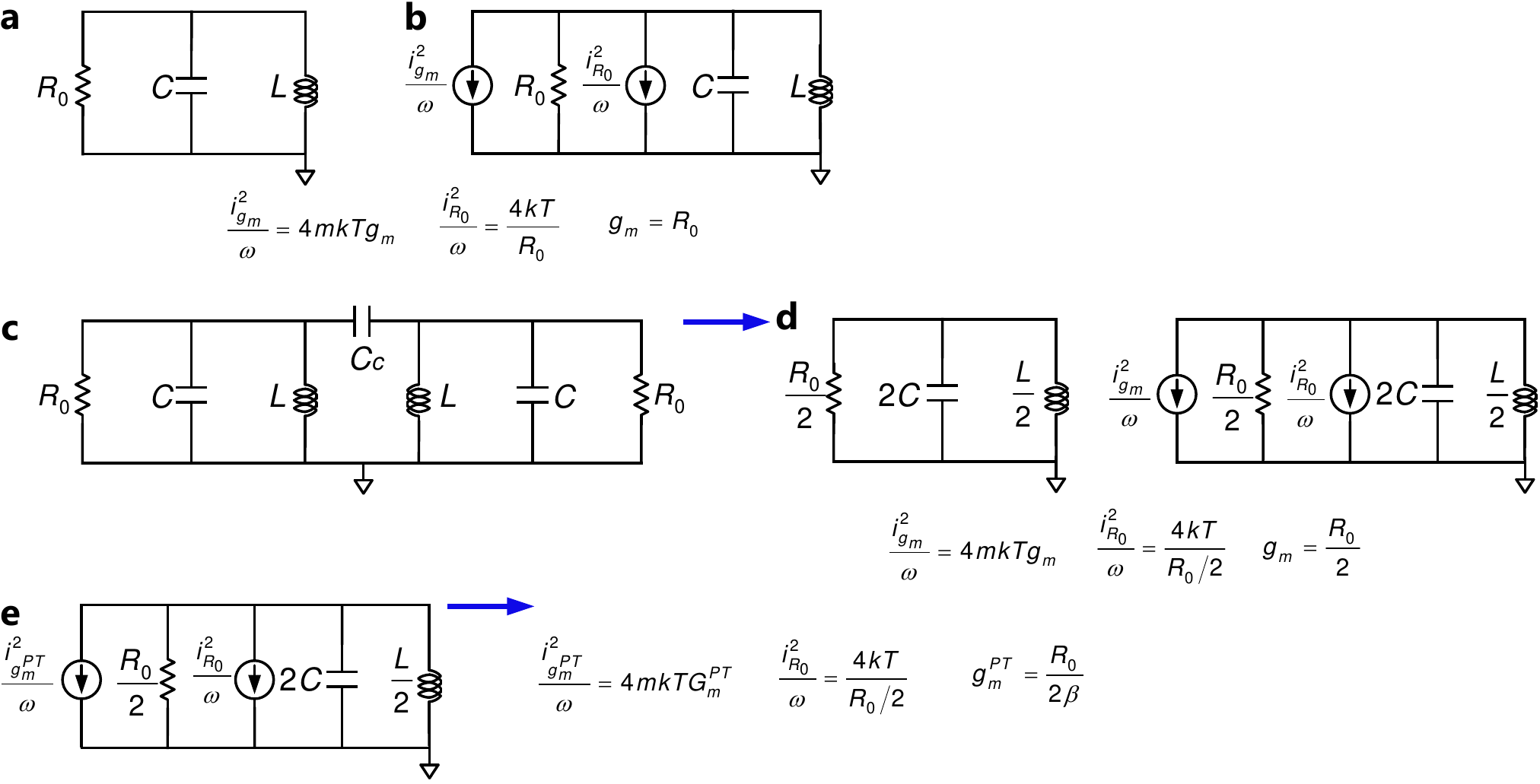}

        \vskip -6pt
    \caption{ \textbf{Phase noise models of different oscillators}. \textbf{a}. Passive resonator model of a conventional single-core oscillator. \textbf{b}. Phase noise model of the single-core oscillator. \textbf{c}. Passive resonator model of a conventional multi-core oscillator based on two coupled resonators. \textbf{d}. Equivalent resonator model and phase noise model for the multi-core oscillator.  \textbf{e}. Equivalent phase noise model of our system.}
    \label{fig:fig_pn}
    \vskip -6pt
\end{suppfigure}

It is well-known in the oscillator filed that multi-core oscillators built upon $N$ identically coupled resonators can lead to the PN reduction by $10\log_{10}N$ dB as compared to a single-core oscillator~\cite{c_vco_1,c_vco_2,c_vco_3,c_vco_4}.
A detailed theoretical analysis is proposed in a previous work~\cite{c_vco_1}.
We provide an intuitive understanding here by using a multi-core oscillator composed of two coupled resonators as an example.
Supplementary Figure~\ref{fig:fig_pn}c shows the passive resonator model for the multi-core oscillator.
One can imagine that the two coupled resonators can be equivalently considered as a single-core resonator with doubled capacitance, halved inductance and halved inherent resistor as shown in Supplementary Figure~\ref{fig:fig_pn}d.
Then, the oscillation frequency remains the same as the single-core oscillator.
According to Eq.~\eqref{eqn:pn}, PN is reduced by 3 dB in this case.
This case study indicates that although the noise sources of two coupled resonators double, the effective Q-factor ($Q_C$) of the system (Figure~\ref{fig:fig_pn}e) also doubles, i.e.,
\begin{equation}
\label{eqn:pn1}
\mathcal{L}_{c}(\triangle \omega) = \mathcal{L}_{conv}(\triangle \omega) - 3 =  10 \cdot \log_{10} \Bigg((1+m)\cdot \frac{8kTR}{V^2_{osc,conv}} \cdot\Big(\frac{\omega}{2\cdot 2Q_S\triangle \omega}\Big)^2\Bigg),
\end{equation}
where $Q_C=2Q_S$.
Our system built upon two coupled resonators also obey this rule.
However, with the unique gain-loss contrast tuning, our system achieves more PN reduction.
In conventional oscillators, the provided gain only demands to cancel the inherent loss.
However, in our system, the provided gain not only needs to compensate for the inherent loss, but also requires to balance the tunable loss.
Assuming the ratio between the provided gain and the inherent loss is $\beta$ ($\beta>1$), the PN of our system is expressed as
\begin{equation}
\label{eqn:pn_pt}
\mathcal{L}_{PT}(\triangle \omega) =  10 \cdot \log_{10} \Bigg((1+ \beta m)\cdot \frac{8kTR}{({\beta}^2 V_{osc,conv})^2} \cdot\Big(\frac{\omega}{2\cdot 2Q_{S}\triangle \omega}\Big)^2\Bigg).
\end{equation}
Here, the oscillation amplitude of our system increases to ${\beta}^2 \times$ as the current flowing into the resonator is quadratically proportional to the gain.
Note that in the saturation region, the resonator current $I_{D}$ is linear with the square of transconductance $g^{\text{PT}}_m$ based on the $I$-$V$ relationship of MOSFET: $g^{\text{PT}}_m={\partial i_{D}}/{\partial V_{GS}}=\mu_{n}C_{ox}(W/L)(V_{GS}-V_{th})=\sqrt{2\mu_{n}C_{ox}(W/L)I_D}$.
Comparing Eq.~\eqref{eqn:pn} and Eq.~\eqref{eqn:pn_pt}, we obtain
\begin{equation}
\label{eqn:com}
\mathcal{L}_{PT}(\triangle \omega) = \Big[\mathcal{L}_{c}(\triangle \omega)- 10\log_{10} \Big(\frac{\beta^4(1+m)}{1+\beta m}\Big)-3 \Big]<\mathcal{L}_{c}(\triangle \omega)-3.
\end{equation}
Eq.~\eqref{eqn:com} shows that the gain-loss contrast tuning of our system can further reduce PN by increasing the power of carrier.
Therefore, the PN improvement of our system is attributed to two facts: 1) the coupled-resonator structure of our system can enhance the effective Q-factor of the system, and 2) the gain-loss contrast tuning can increase the oscillation amplitude, decreasing the effect of noise.

%% file: Exper_Set.tex
\section{Experiments}

    \begin{suppfigure}[!t]
 \centering
    \includegraphics[width=0.60\linewidth]{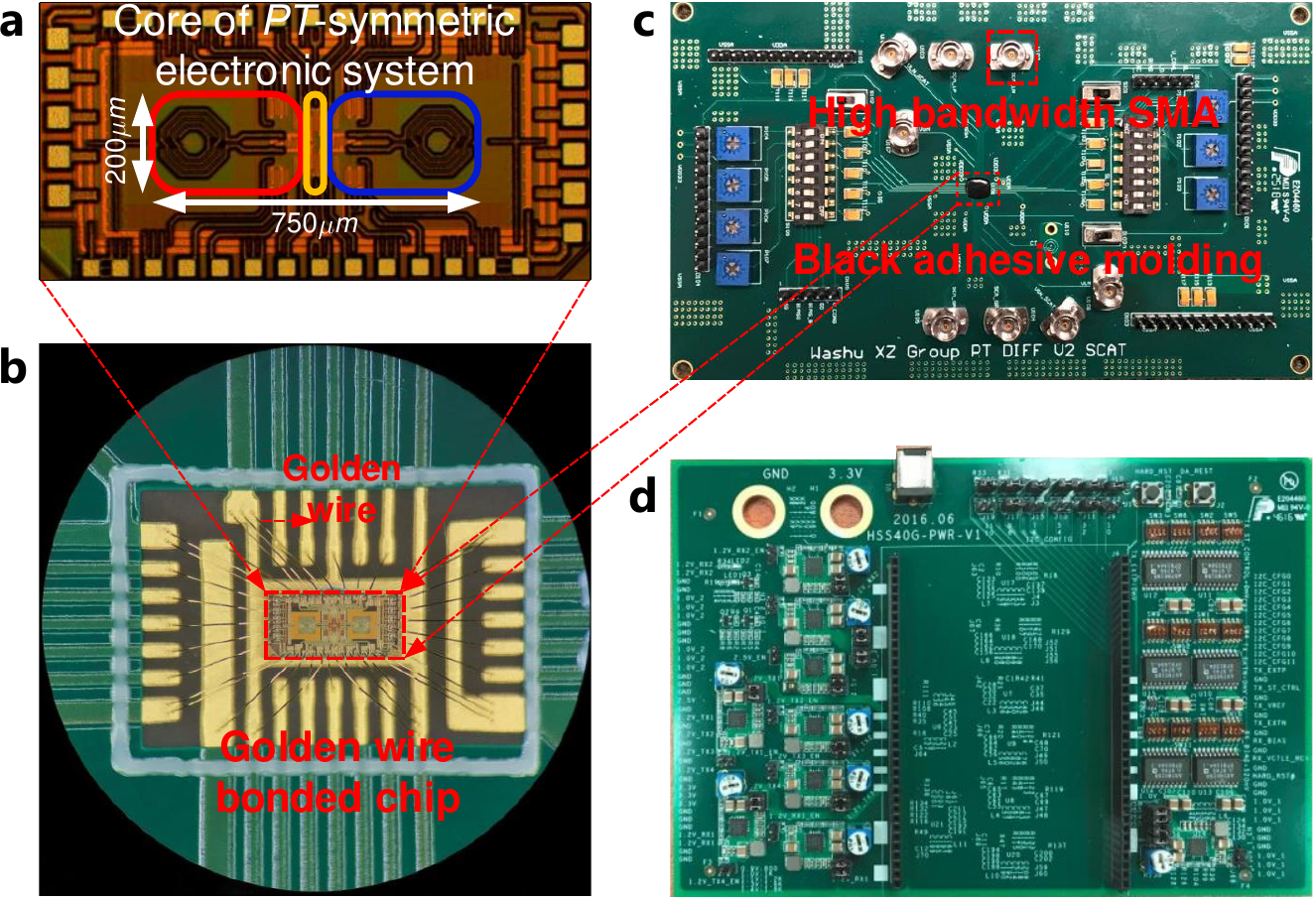}
    \caption{ \textbf{Experimental setups.} \textbf{a}. Chip die photo. \textbf{b}. Bonding diagram. \textbf{c}. Daughter PCB for controlling the biases of the IC components. \textbf{d}. Mother board for power supply.}
    \label{fig:fig_10}
\end{suppfigure}

    \begin{suppfigure}[!t]
 \centering
    \includegraphics[width=0.6\linewidth]{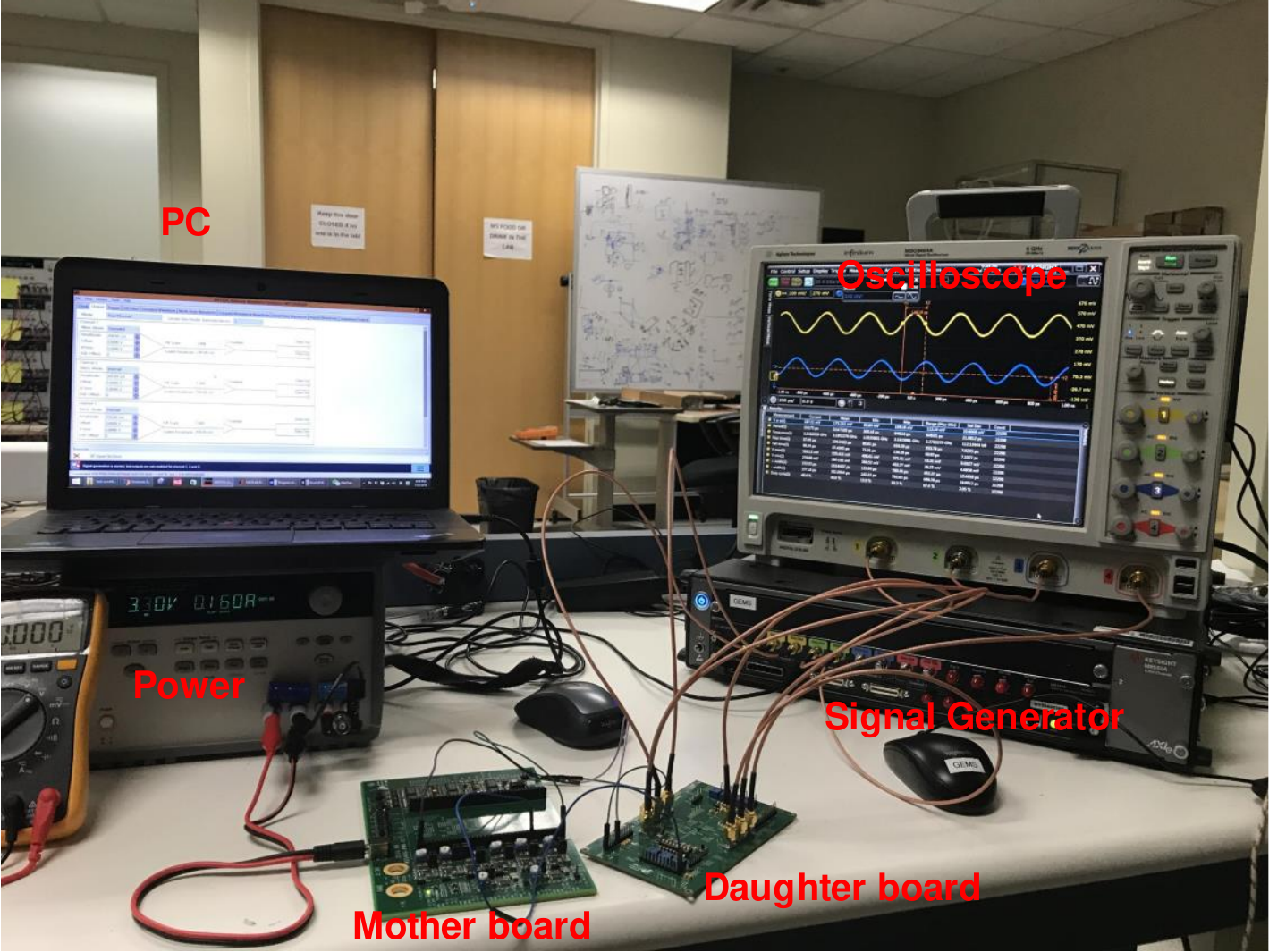}
    \caption{ \textbf{ Complete setup to test the system.} The test platform consists of a daughter board, a mother board, a power, a signal generator, an oscilloscope, and a PC.}
    \label{fig:fig_11}
\end{suppfigure}

\subsection{Experimental Setup}

We fabricated the chip with a $130$ nm CMOS technology. The chip die photo is shown in Supplementary Figure~\ref{fig:fig_10}a. The core area of the system is $200\mu m \times 750\mu m$. To test the chip, we designed two printed circuit boards (PCBs). 
One is a daughter board (Supplementary Figure~\ref{fig:fig_10}c) and the other one is a mother board (Supplementary Figure~\ref{fig:fig_10}d). 
The system chip was bonded on the daughter board by gold wires (Supplementary Figure~\ref{fig:fig_10}b). The daughter board provides all the control signals and high-speed inputs (outputs) for the chip, such as gain (loss) bias voltage, varactor bias voltage. 
All the high-speed input/output terminals of the chip are accessed by high bandwidth surface mount ahead (SMA) on the daughter board. 
The mother board is used as power supply for the daughter board.
A complete setup is shown in Supplementary Figure~\ref{fig:fig_11}. 
Our experimental setup consists of a bonded chip in a daughterboard, a motherboard, a power supply, a mixed signal oscilloscope (MSO, Agilent 9404A), an arbitrary wave generator (AWG, KEYSIGHT M8195A) and a personal computer (PC).
The MSO has four pairs of differential channels, and its highest sampling rate is $20~GSa/s$.
The AWG has four pairs of differential channels, each pair of which can generate arbitrary waves up to $50~$GHz with independently varying phase.

\subsection{Experiment And Simulation Procedures}
\label{sec: si_5_2}
In the phase transition experiments, the outputs of two RLC resonators were connected to the MSO.
To test the PT-symmetry spontaneous breaking, we used the zig-zagging method to make either eigen-frequency dominant~\cite{S21,S23}.

\begin{suppfigure}[ht]
 \centering
    \includegraphics[width=0.7\linewidth]{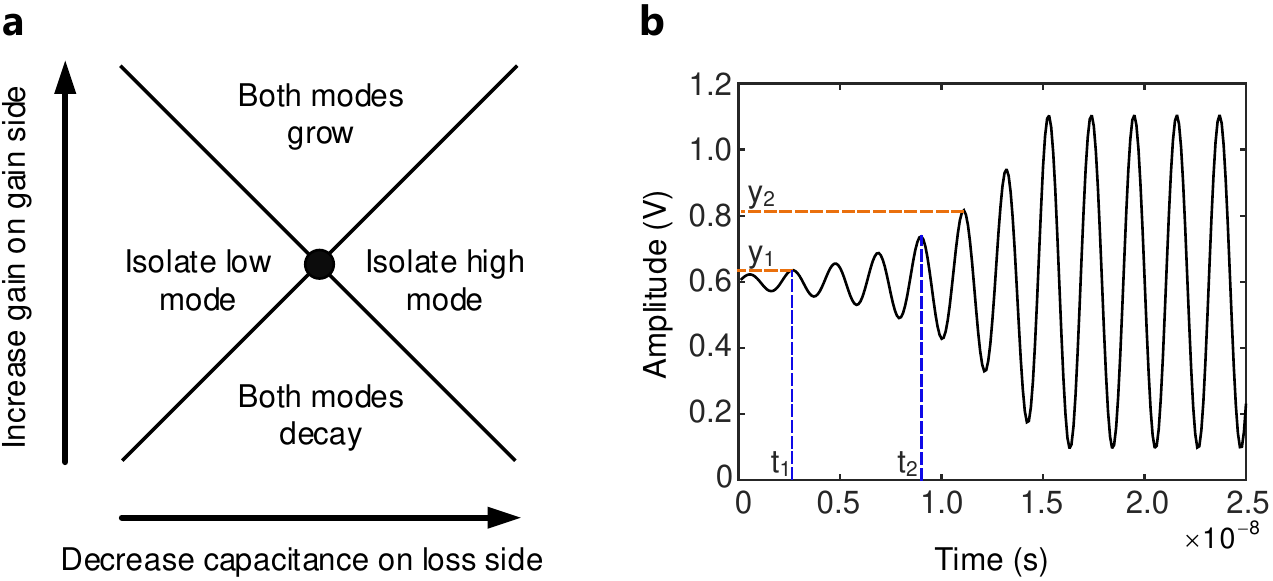}
        \vskip -6pt
    \caption{ \textbf{Measurement methodology to obtain the eigenfrequencies of our system~\cite{S23}}. \textbf{a}. The `X' plane used as the instructions to manually tune our system. \textbf{b}. A simplified example to show how to obtain the imaginary part of eigenfrequencies. }
    \label{fig:fig_measurement}
\end{suppfigure}

The method is originally proposed in~\cite{S23}, which can be best introduced using Supplementary Figure~\ref{fig:fig_measurement}\textbf{a}.
The horizontal axis represents the capacitance difference $\triangle C=C_L-C_G$ between the two resonators.
``Moving rightward on the figure indicates decreasing $\triangle C$.
The vertical axis is the gain. 
Upward movements indicate increasing gain.
The dot at the center of the X represents a point where the gain and loss are exactly balanced, but the capacitance is imbalanced by a small amount $\triangle C>0$.
The goal is to attain a stable dimer configuration just barely below that center dot. 
Then the slightest changes can cause a marginal instability in one, the other, or both modes, allowing each mode to be observed individually in a state of gain-loss balance.
An algorithm to find this balance point by zig-zagging along one of the bottom boundary lines of the X follows:
1) Reduce gain until all modes decay by ending just inside a border of the X; 
2) Change $\triangle C$ a bit in whichever direction doesn't immediately cause instability;
3) Increase gain until something oscillates;
4) Change $\triangle C$ enough to kill the oscillation, then a bit further;
5) Repeat 3) and 4) until tiny capacitance changes cause a switching from the high frequency to low frequency zone, with only a tiny ``dead zone'' in between.
Note that as $\gamma \rightarrow{\gamma}^{}_{EP}$, $\triangle C\rightarrow0$.
Beyond ${\gamma}^{}_{EP}$, $\triangle C$ was held fixed at its asymptotic value.''
Note that machine learning methods~\cite{cao2022aaai,cao2022dac} may be used to achieve an automated manner to adjust the gain-loss contrast for the system.
We will explore this method in our future work.

In the exact phase, mode frequencies were directly observed by balancing gain-loss and slightly unbalancing the capacitance, then correcting for the imbalance.
For each mode, once the system was brought to a state of marginal oscillation, oscilloscope waveform capture recorded $V_G (t)$ and $V_L (t)$, the voltage data at each side of the system. 
These data were analyzed for real frequency and amplitude.
This process described above forced the imaginary part of the frequency to be zero, and so the imaginary frequency data was automatically recorded as zero.
In the broken phase, the capacitance trim is kept fixed at its asymptotic value, and the gain trim is set to a bit higher than center dot. 
The exponential growth of transient data obtained in Figure~\ref{fig:fig_measurement}\textbf{b} then directly gives us the imaginary component: $\omega_{\text{Im}}=(\ln (y_2-V_{\text{cm}}) -\ln (y_1 -V_{\text{cm}}))/(t_2-t_1)$.
Here, $V_{\text{cm}}$ is the common mode voltage, which is set to be $V_{\text{DD}}/2=0.6V$ in our design.
Note that only a piece of the transient curve as shown in Figure~\ref{fig:fig_measurement}\textbf{b} is used to calculate the imaginary part.
As the amplitude increases, the gain will be degenerated due to the nonlinearity of the system, which can lead to the computation errors.

In the single-port scattering experiments, the system was biased in the exact phase.  
Then, a sinusoidal signal with varied frequency was applied into the system.
Note that the signal power was chosen to set the system in the linear region.
The incident wave $V^+_{G}$ ($V^+_{L}$) and the reflected wave $V^-_{G}$ ($V^-_{L}$) were extracted from the voltages at either side of the TL, from which the scattering coefficients $r_{G} ={V^-_{G}}/{V^+_{G}}$ and $r_{L} ={V^-_{L}}/{V^+_{L}}$ were calculated. 
In the two-port scattering simulations, the AWG sourced sinusoidal signals with varying frequencies or phase into the chip through TL.
Then signals on both terminals of the TL were sent into the MSO such that the incident wave and reflected wave could be captured. 
Theoretically, the ideal case of $\Theta_{abs}=0$ and $\Theta_{amp}=\infty$ can only occur when the gain and loss are perfectly balanced. 
In our system, small imbalance of RLC components existed in the two RLC circuits due to minor fabrication error, which could not be completely compensated by the external tuning. 
Such a tiny imbalance resulted in a large deviation of theoretical $\Theta_{abs}$ and $\Theta_{amp}$ of our system from the ideally balanced condition (see Supplementary Figure~\ref{fig:fig_two_scat}a), and experimental difficulty in measuring $\Theta_{abs}$.
Therefore, we performed SPICE simulation with special scanning techniques~\cite{S23} to obtain the corresponding results with $\Theta<0$ in Supplementary Figure~\ref{fig:fig_two_scat}b.
When the PT-symmetric dimmer acts as a perfect absorber, the condition $V_L^+=\mathcal{M}_{21}(\omega)V_G^+$ must be satisfied. In the simulation, we let $V_L^+=Ae^{i\phi^{'}}(\omega)V_G^+$. 
Each lower data point near the absorption point in Supplementary Figure~\ref{fig:fig_two_scat}b was found by fixing frequency and scanning through values of $\phi^{'}=90^{\circ}$ in tightly spaced increments, then recording the minimum $\Theta$ value. 
Within these $\phi^{'}$ scans, an iterative process of measurement and resetting was used at each step, to ensure that $A$ and $\phi^{'}$ were within a small tolerance level of the theoretically specified values.
The portion of the bottom (blue) curve in Supplementary Figure~\ref{fig:fig_two_scat}b near the minimum is an example of one of these high-precision scans.

In the nonreciprocal experiments, the AWG fed sinusoidal signals with varying frequencies into the system through the gain (loss) side TL.
Then both the incident wave on the input terminal of the gain (loss) side TL and the reflected wave on the output terminal of loss (gain) side TL could be captured by MSO.

%% file: Su_re.tex
\section{Supplementary Results}
\label{sec:results}
\subsection{Comparisons of Microwave Generation}
\label{sec:VCO_comp}

    \begin{suppfigure}[!t]
 \centering
    \includegraphics[width=1.0\linewidth]{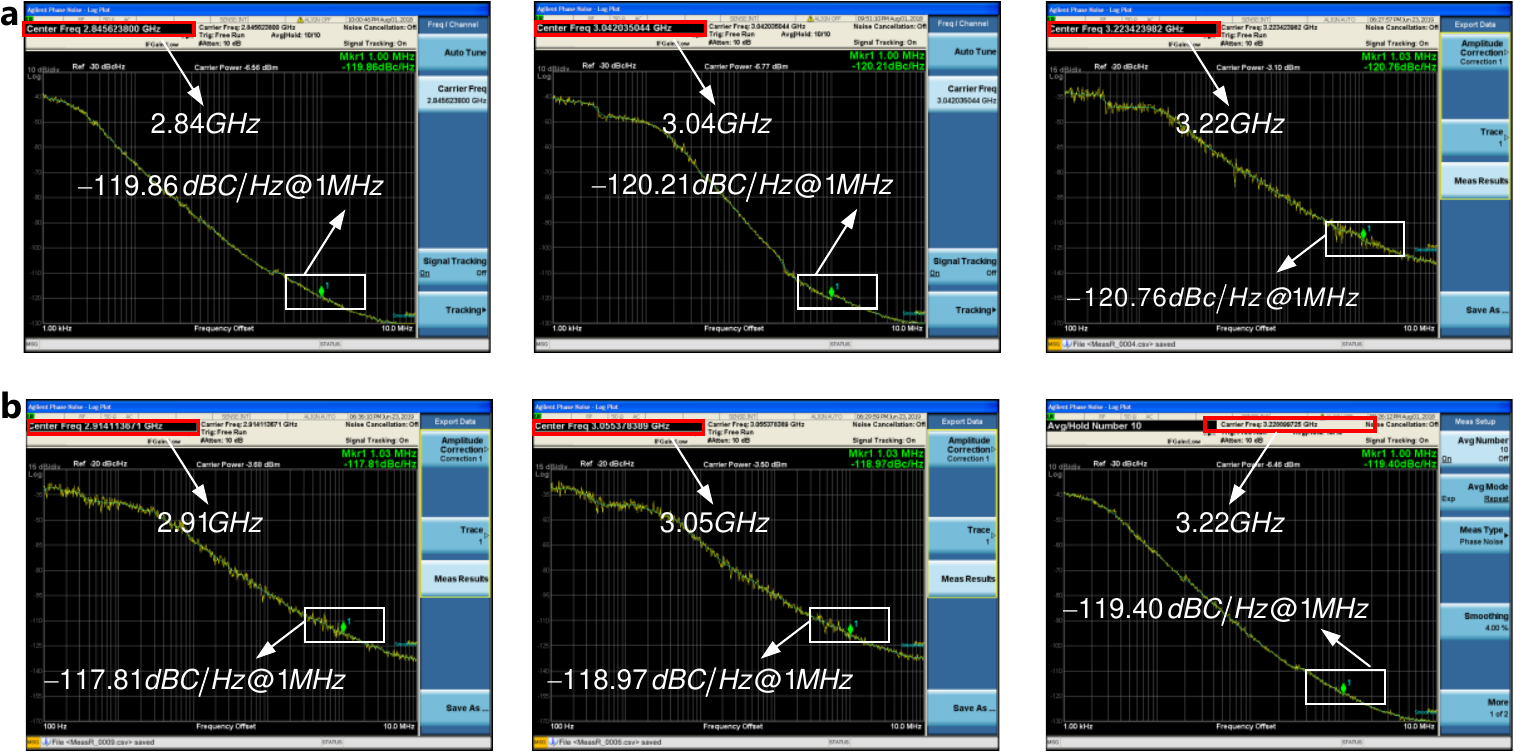}
    \caption{ \textbf{Comparison of phase noise performance of the microwave generation between the fully integrated PT-symmetric electronic system and the baseline conventional oscillator with a single-resonator structure.} \textbf{a}. Phase noise of our system at different frequencies: $2.84$ GHz (low frequency), $3.04$ GHz (medium frequency), $3.22$ GHz (high frequency). \textbf{b}. Phase noise of the baseline oscillator at different frequencies: $2.91$ GHz (low frequency), $3.05$ GHz (medium frequency), $3.22$ GHz (high frequency).}
    \label{fig:fig_9}
\end{suppfigure}

In our system, a conventional oscillator (Supplementary Figure~\ref{fig:fig_7}b) can be obtained by turning off SW1 to decouple the two RLC resonators in the fully integrated PT-symmetric electronic system shown in Supplementary Figure~\ref{fig:fig_1}. 
We compare the phase noise performance of the baseline oscillator and our PT-symmetric system at different frequencies (low, medium and high frequency in each individual system) in Supplementary Figure~\ref{fig:fig_9}.
The experimental results show that the fully integrated PT-symmetric electronic system generally has better phase noise performance in the tuning range than the conventional oscillator.
Table~\ref{tb:com_2} summarizes the comparison.

\input{table_1}

    \begin{suppfigure}[!t]
 \centering
    \includegraphics[width=0.50\linewidth]{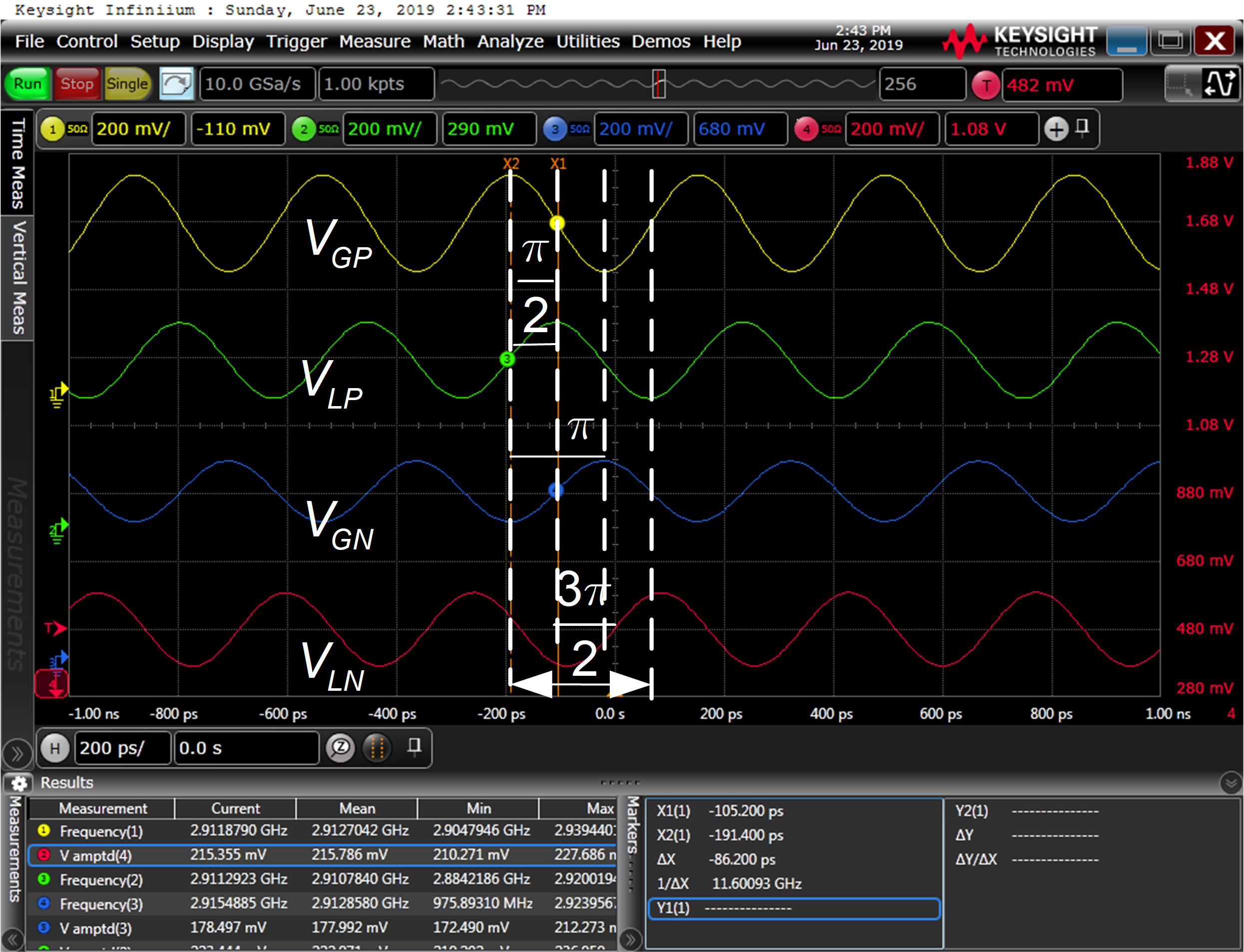}
    \caption{ \textbf{Example of quadrature microwave generation enabled by the PT-symmetric electronic system.} The system is biased around the exceptional point.}
    \label{fig:fig_8}
\end{suppfigure}

We then further examine the Eq.~\eqref{eq:mode} and Eq.~\eqref{eq:phi} in Section~\ref{sec:VCO} at the coalescence frequency $\omega_1=\omega_3$. 
We find that $\phi_{1,3}=\pi/2$ if $\omega_1=\omega_3=1/\sqrt{(1+c)}$.
This indicates that by carefully choosing design parameters at EP, the phase difference between two sides is $\pi/2$, then we can achieve quadrature microwave generation~\cite{S13,S14,S15,S16}. 
Note that we use a differential architecture to design the system, therefore the phase for $V_{GP}$, $V_{GN}$, $V_{LP}$, $V_{LN}$ is 0, $\pi$, $\pi/2$, $3\pi/2$. 
Supplementary Figure~\ref{fig:fig_8} shows the experimental results of the quadrature microwave generation enabled by our system.

\subsection{Scattering  Results}
\label{sec:scatt_re}
    \begin{suppfigure}[!t]
 \centering
    \includegraphics[width=0.80\linewidth]{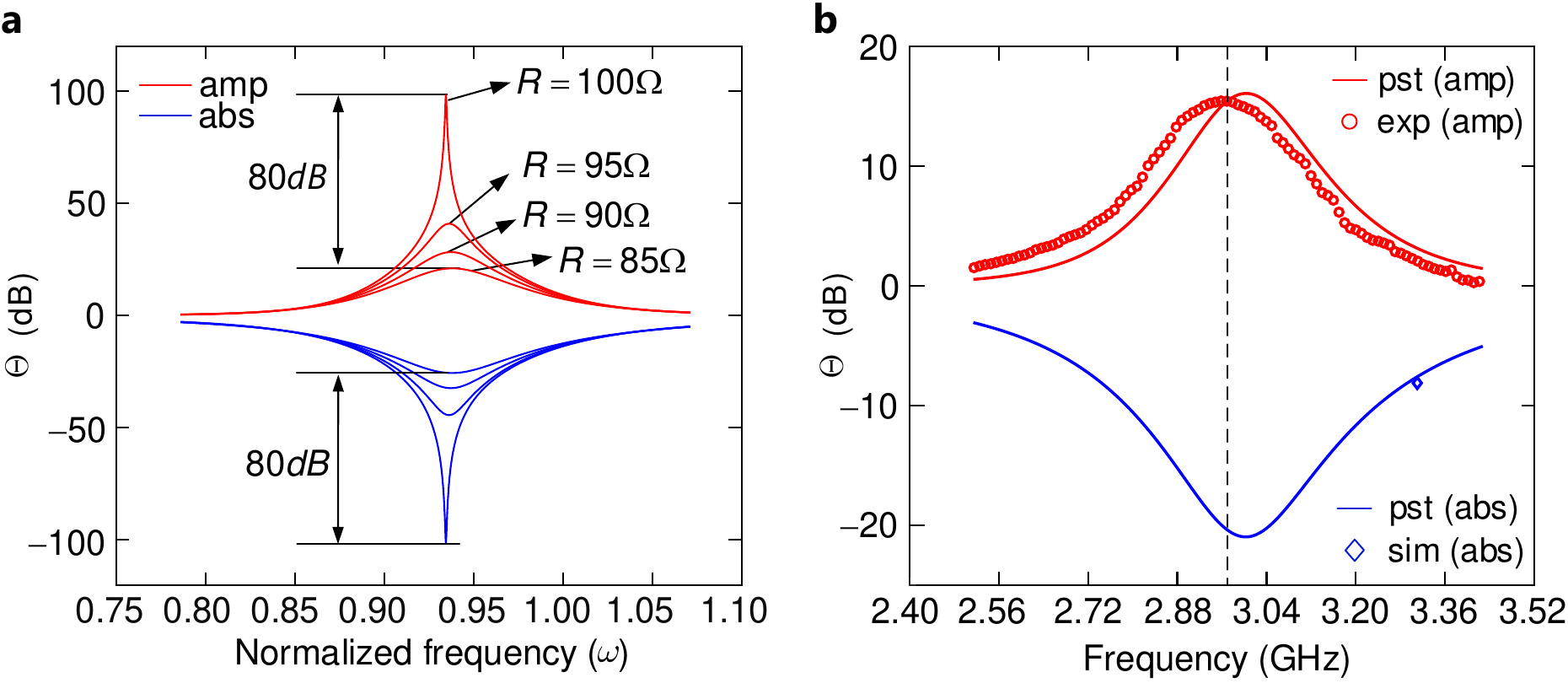}
    \caption{ \textbf{Simulations and experiments of two-port scattering property.} \textbf{a}. Theoretical deviation of $\Theta_{amp}$ and $\Theta_{abs}$ under different gain (loss) value R. \textbf{b}. Measured results.}
    \label{fig:fig_two_scat}
\end{suppfigure}

Theoretically, the $\omega_J$ is uniquely determined by the tuning parameter $\gamma=\sqrt{(L/C)}/R$ when the system is perfectly balanced. 
A small variation of gain/loss value $R$ in $\gamma$ will cause the significant deviation of $\Theta_{amp}$ and $\Theta_{abs}$ from ideal value ($\Theta_{amp}=\infty$, $\Theta_{abs}=0$). 
Supplementary Figure~\ref{fig:fig_two_scat}a shows the theoretical simulation of several groups $\Theta_{amp}$ and $\Theta_{abs}$ under different variation of $R$.
$R=100~\Omega$ can be consider as the case to achieve ideal $\Theta_{amp}$ and $\Theta_{abs}$. 
Even if there is only $15~\Omega$ deviation, the deviation of $\Theta_{amp}$  and $\Theta_{abs}$ is up to $80$ dB.
Considering the fabrication process leads to the imbalance of CMOS components between the two RLC resonators which cannot be completely compensated by the external tuning, the practical deviation becomes worse. 
Supplementary Figure~\ref{fig:fig_two_scat}b demonstrates a measured result (red dots) of two-port scattering property when the system suffers from small fabrication mismatches.

\subsection{Non-reciprocal Microwave Transmissions}
\label{sec:non_reci}

    \begin{suppfigure}[!t]
 \centering
   \includegraphics[width=0.8\linewidth]{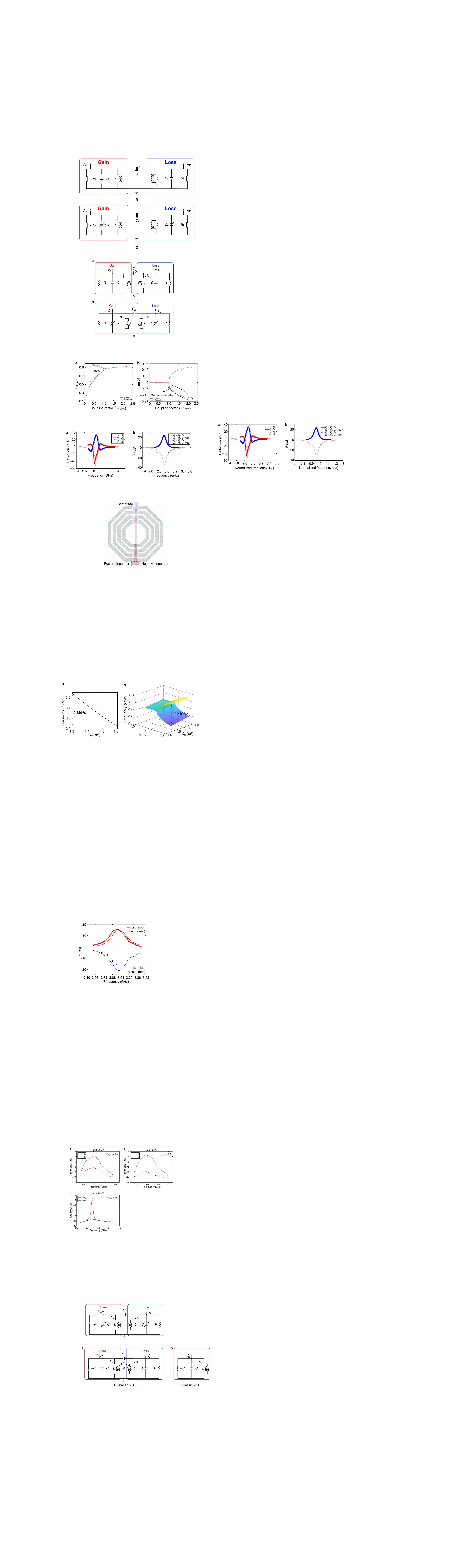}
    \caption{ \textbf{Supplementary experimental results for non-reciprocal transmission.}  \textbf{a}. Nonreciprocal transmission is observed in the exact phase ($\gamma/\gamma_{EP}=0.83$) with two peaks, where the forward transmission is up to $-10.5$ dB while the backward transmission is $1.0$ dB. \textbf{b}. Nonreciprocal transmission is observed in the exact phase ($\gamma/\gamma_{EP} =0.9$) with two peaks, where the forward transmission is up to $-14.5$ dB while the backward transmission is $1.9$ dB. \textbf{c}. Nonreciprocal transmission is observed in the broken phase ($\gamma/\gamma_{EP}=2$) with one peak, where the forward transmission is up to $-18.7$ dB while the backward transmission is $2.1$ dB.}
    \label{fig:fig_12}
\end{suppfigure}

\input{table_2}

Extra experimental results of non-reciprocal transmission are shown in Supplementary Figure~\ref{fig:fig_12} which together with the Figure~4a-e (main text) show the non-reciprocal trend of the isolation in Figure~4f (main text).
Our system shows strong isolation among a wide bandwidth in the microwave domain. 
Our system also requires a lower power threshold and shows interesting insertion gain due to enhanced nonlinearity enabled by PT-symmetry as compared with traditional nonlinearity-induced isolation (see Supplementary Table~\ref{tb:com_4}).

%% file: table_1.tex
\begin{supptable*}[!tb]
\footnotesize
\centering
\caption{Comparisons of microwave generation between the fully integrated PT-symmetric electronic system and the baseline traditional oscillator.}
\vspace{-0.2cm}
\begin{tabular}{|c|c|c|}
\hline
Works                                                                     & \begin{tabular}[c]{@{}c@{}}Baseline   \\ Oscillator\end{tabular}                        & \begin{tabular}[c]{@{}c@{}} Our PT-symmetric   \\ System\end{tabular}                                             \\ \hline
Technology (nm)                                                         & $130$                                                                              & $130$                                                                                                   \\ \hline
Supply (V)                                                                & $1.2$                                                                              & $1.2$                                                                                                   \\ \hline
Power (mW)                                                                & $4.31$                                                                             & $4.31$                                                                                                  \\ \hline
Area ($\text{mm}^2$)                                                                & $0.15$                                                                             & $0.15$                                                                                                  \\ \hline
$f_{\min}$ (GHz)                                                              & $2.93$                                                                             & $2.63$                                                                                                  \\ \hline
$f_{\max}$ (GHz)                                                                & $3.23 $                                                                            & $3.20 $                                                                                                 \\ \hline
\begin{tabular}[c]{@{}c@{}}LC Tuning \\ parameter\end{tabular}               & \begin{tabular}[c]{@{}c@{}}$L = 1.85$ nH;  \\   $C \in {[1.35, 1.55]}$ pF\end{tabular} & \begin{tabular}[c]{@{}c@{}}$L= 1.85$ nH; $C_C=500$ fF;  \\ $C \in {[1.35, 1.55]}$ pF. \end{tabular} \\ \hline
R tuning & N/A &$R \in {[80, 260]}~\Omega$ \\ \hline
FTR (\%)                                                                  & $9.70$                                                                             & $20.17$                                                                                                 \\ \hline
\begin{tabular}[c]{@{}c@{}}PN(dBc/Hz@1MHz ) \\ (Average)\end{tabular}     & $118.72$                                                                           & $120.28$                                                                                                \\ \hline
\end{tabular}
\label{tb:com_2}
\end{supptable*}

%% file: table_2.tex
\begin{supptable*}[!t]
\footnotesize
\centering
\caption{Non-reciprocity comparisons between our system and state-of-the-art isolators based on nonlinearity.}
\vspace{-0.2cm}
\begin{tabular}{|c|c|c|c|c|}
\hline
Works             & \begin{tabular}[c]{@{}c@{}}Our \\  System\end{tabular}      & \begin{tabular}[c]{@{}c@{}}PRL '13\\  \cite{S31}\end{tabular} & \begin{tabular}[c]{@{}c@{}}Nature Electron '19 \\ \cite{S32}\end{tabular}  & \begin{tabular}[c]{@{}c@{}}Nature Electron '20 \\ \cite{S32_1}\end{tabular}\\ \hline
Power threshold  & $-21$ dBm             & $9$ dBm                                                  & $17$ dBm     & $-20$ dBm                                       \\ \hline
Isolation         & $20$ dB                & \textless $5$ dB                                         & $35$ dB    & 10 dB                                             \\ \hline
Bandwidth         & $[2.75\sim3.10]$ GHz & $[38\sim40]$ KHz                                     & $[700\sim800]$ MHz    & $200$ MHz                              \\ \hline
Insertion gain?   & $5$ dB                 & No                                                     & No                     & No                               \\ \hline
Fully integrated? & Yes                  & No                                                     & No                    & No                                \\ \hline
\end{tabular}
\label{tb:com_4}
\end{supptable*}

%% file: Versatile_implemen.tex
\section{Versatile Fully Integrated PT-symmetric Electronic System}
\label{sec: si_7}

    \begin{suppfigure}[!t]
 \centering
    \includegraphics[width=0.43\linewidth]{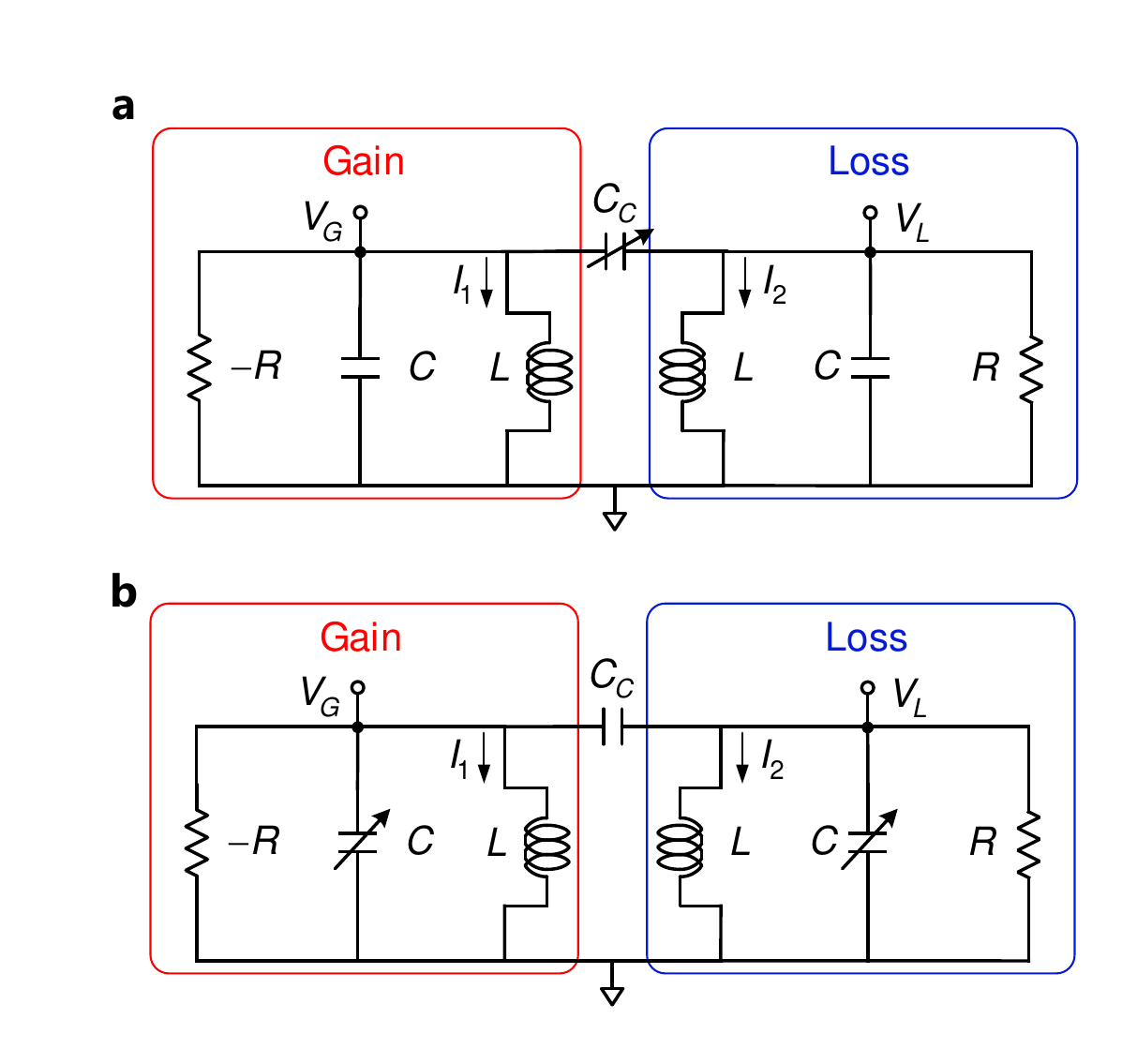}
    \caption{ \textbf{Versatile architectures to implement fully integrated PT-symmetric electronic system.} a. Coupling-tuning architecture. In this architecture, only coupling capacitance $C_C$ is adjustable, all other components are fixed. b. Capacitive tuning architecture. In this architecture, only capacitance in the RLC resonators is adjustable, all other components are fixed.}
    \label{fig:fig_13}
\end{suppfigure}

    \begin{suppfigure}[!t]
 \centering
    \includegraphics[width=0.68\linewidth]{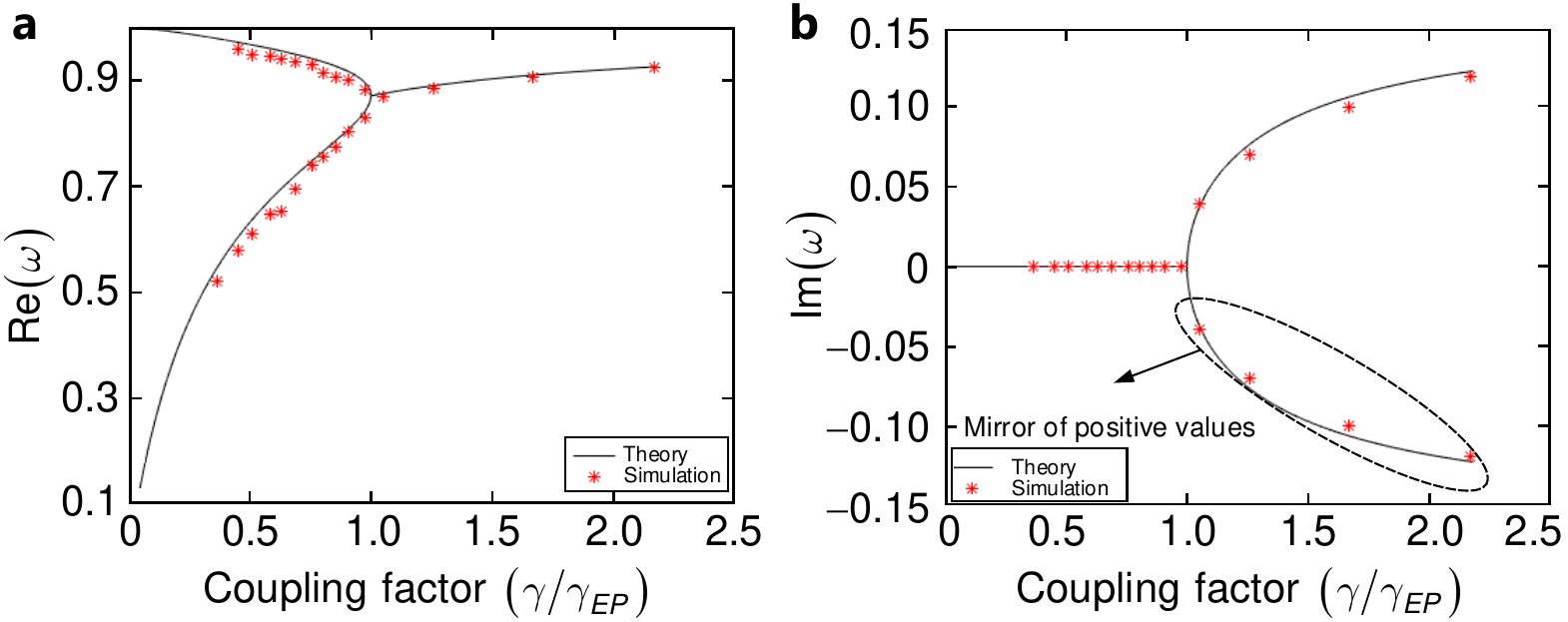}
    \caption{ \textbf{Phase transition of the system based on coupling-tuning architecture.} \textbf{a}. Real part of the eigenfrequencies. \textbf{b}. Imaginary part of the eigenfrequencies. The part below the zero axis is the symmetrical part of the positive one. Star symbols are simulations while lines are theoretical prediction.}
    \label{fig:fig_14}
\end{suppfigure}

    \begin{suppfigure}[!t]
 \centering
    \includegraphics[width=0.68\linewidth]{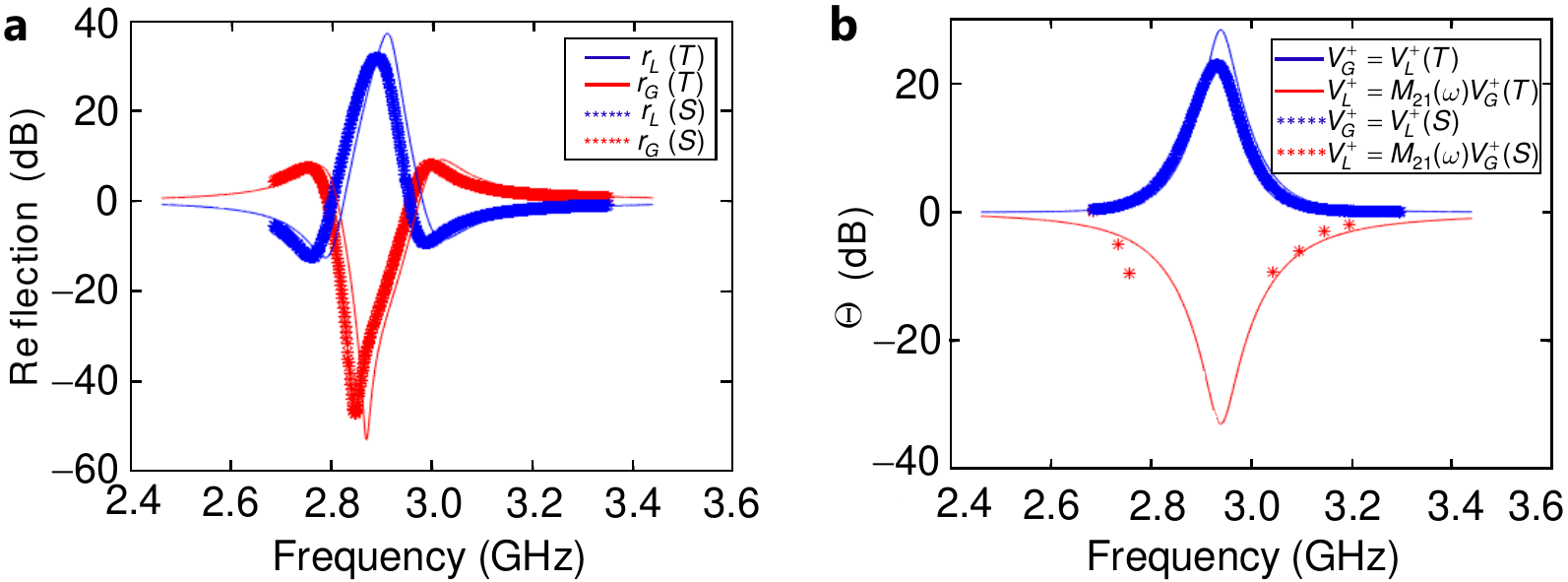}
    \caption{ \textbf{Scattering properties of the system based on coupling-tuning architecture.} \textbf{a}. Single-port scattering. \textbf{b}. Two-port scattering~\cite{S21}. Star symbols are simulation (S) results while lines are theoretical (T) prediction.}
    \label{fig:fig_15}
\end{suppfigure}

In addition to the proposed architecture based on the gain (loss) tuning, it is straightforward for us to implement other architectures for the system by leveraging the flexible tuning mechanisms of IC. 
We propose two variants of the system by using $130$ nm CMOS technology and show them in Supplementary Figure~\ref{fig:fig_13}.
The first variant is based on coupling-tuning architecture (Supplementary Figure~\ref{fig:fig_13}a). 
In this architecture, all the components are fixed except for the coupling capacitance $C_C$. 
The $C_C$ can be realized by switched-capacitor arrays or varactors. 
The second variant uses capacitance-tuning mechanism (Supplementary Figure~\ref{fig:fig_13}b).
In this architecture, all the components are fixed except for the RLC resonator's capacitance $C_G$ ($C_L$). The $C_G$ ($C_L$) can also be realized in the same way as $C_C$.
Although they are different in tuning mechanisms, the theory of the PT-symmetry spontaneous breaking keeps the same. 
Based on Eq.~\eqref{eq:point}, in the coupling-tuning architecture, $\gamma_{EP}$ and $\gamma_{UP}$ are evolving with the capacitance ratio $c$ while $\gamma$ is fixed. 
In the capacitance-tuning architecture, all $\gamma$, $\gamma_{EP}$ and $\gamma_{UP}$ are evolving with the RLC resonator's capacitance $C$, but $\gamma_{EP}$ and $\gamma_{UP}$ change faster.

We simulate the first variant of the system by using $130$ nm CMOS technology and show the corresponding results in Supplementary Figure~\ref{fig:fig_14} and Figure~\ref{fig:fig_15}. 
The design parameters of this variant are $L= 3.50$ nH, $C_C\in [0.3, 1.3]$ pF, $C = 0.4$ pF, and $R= 270~\Omega$.
All the simulation results are matched with theoretical predictions, demonstrating that ICs can provide versatile structures to study PT-symmetric electronics.

%% file: Extended_dis.tex
\section{Extended Discussions}
\subsection{Discussion on Periodic PT-symmetric Electronic Structures}
\label{sec: si_8_1}

PT-symmetric periodic structures, near the spontaneous PT symmetry breaking point, can act as unidirectional invisible media. In this regime, the reflection from one end is diminished while it is enhanced from the other.
In electronics, the unidirectional invisibility has been studied by using diverse board-level PT-symmetric systems~\cite{un_invi_1,un_invi_2,P_PT}.

\begin{suppfigure}[!ht]
 \centering
    \includegraphics[width=0.9\linewidth]{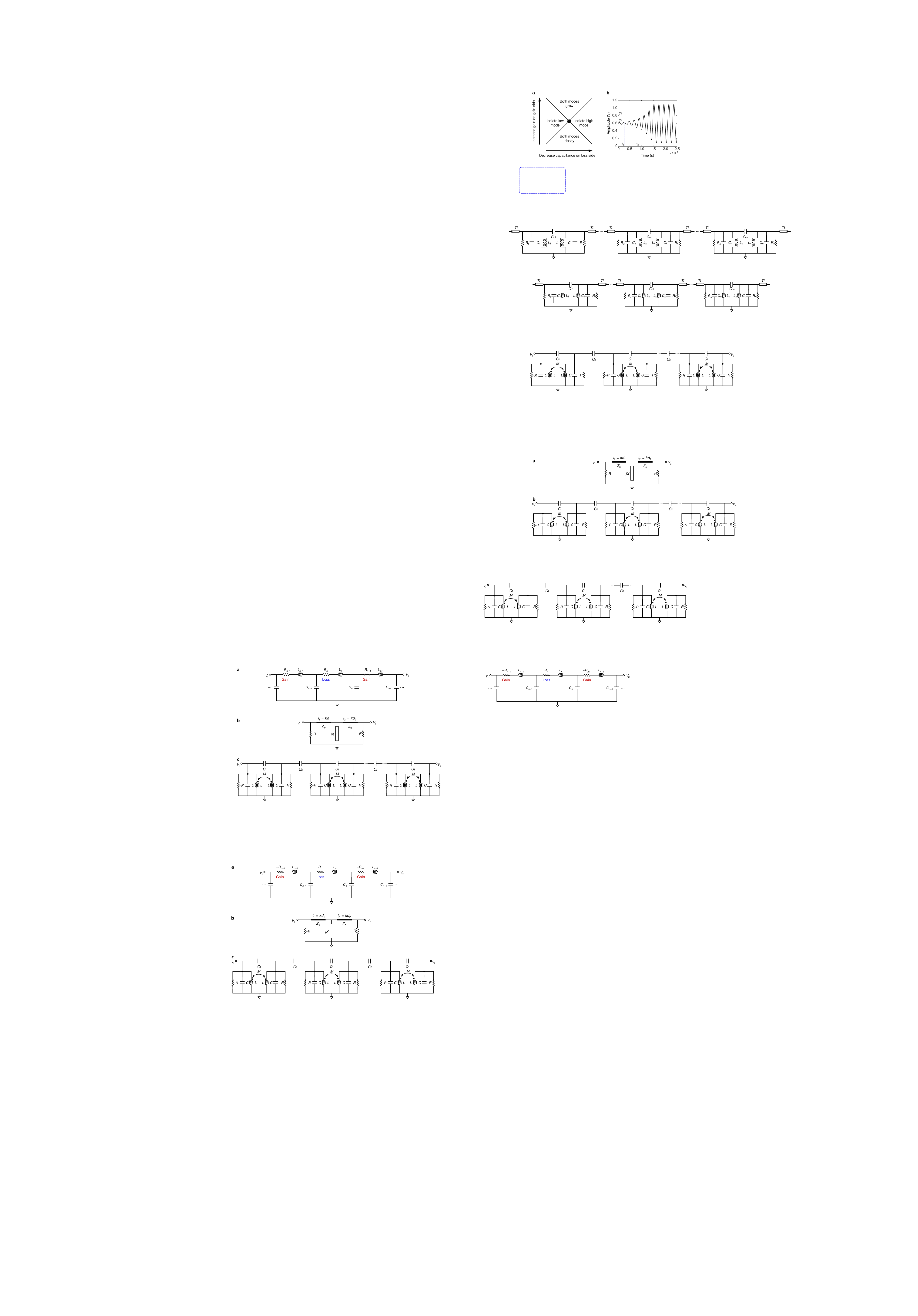}
    \caption{ \textbf{Circuit structure for unidirectional invisibility based on PT-symmetric electronic circuits}.  \textbf{a}. Circuit structure for unidirectional invisibility based on a periodic PT-symmetric transmission line circuit~\cite{P_PT}. \textbf{b}. The circuit schematic used in the previous work~\cite{un_invi_1}.}
    \label{fig:unidirecionality}
\end{suppfigure}

Supplementary Figure~\ref{fig:unidirecionality}a shows a circuit schematic of a PT-symmetric periodic structure based on a transmission line model~\cite{P_PT}, where the resistor $R_n$ is distributed according to the configuration of PT symmetry composed of a gain ($-R$) and loss ($+R$) sequence.
Theoretical analysis suggests that such a structure shows PT symmetry phase transition from real to complex eigenvalues as a function of resistance $R$.
It can be used as a counterpart of PT-symmetric Bragg periodic structures in the electronic domain to study the unidirectional invisibility around the exceptional point.

The system in reference~\cite{un_invi_1} is composed of lumped elements and transmission lines as shown in Supplementary Figure~\ref{fig:unidirecionality}b.
The two parallel resistors are separated by two transmission lines of which the electric lengths are $l_1=kd_1$, $l_2=kd_2$ and the characteristic impedance is $Z_0$, in which $k$ is the wave number and $d_{1,2}$ is the physical lengths of the transmission lines.
Furthermore, the resistance of reactance component which consists the capacitor $C$ or the inductor $L$ is $X = 1/\omega C$ or $X = \omega L$ between the two transmission lines.
Based on the scattering matrix method, the circuit can exhibit an ideal unidirectional performance at the spontaneous PT-symmetry breaking point by tuning the transmission lines between the lumped elements. 
Additionally, the resistance of the reactance component can alter the bandwidth of the
unidirectional invisibility flexibly.

The system in reference~\cite{un_invi_2} has the exactly same structure as our dimer.
An interesting result of two-port scattering in this paper
is that at specific $\omega$ values, the transmittance becomes $t = 1$, while at the same time one of the reflectances vanishes. 
Hence, the scattering for this direction of incidence is flux conserving and the structure is unidirectionally transparent. 
Periodic repetition of the PT-symmetric unit will result in the creation of unidirectionally transparent frequency bands.
We recommend these circuit structures for the study of unidirectional invisibility in the electronic domain.
With proper optimization and design techniques, all these circuits can be implemented on IC.

\subsection{Discussion on Topological PT-symmetric Electronics}
\label{sec: si_8_2}

\begin{suppfigure}[!t]
 \centering
    \includegraphics[width=0.9\linewidth]{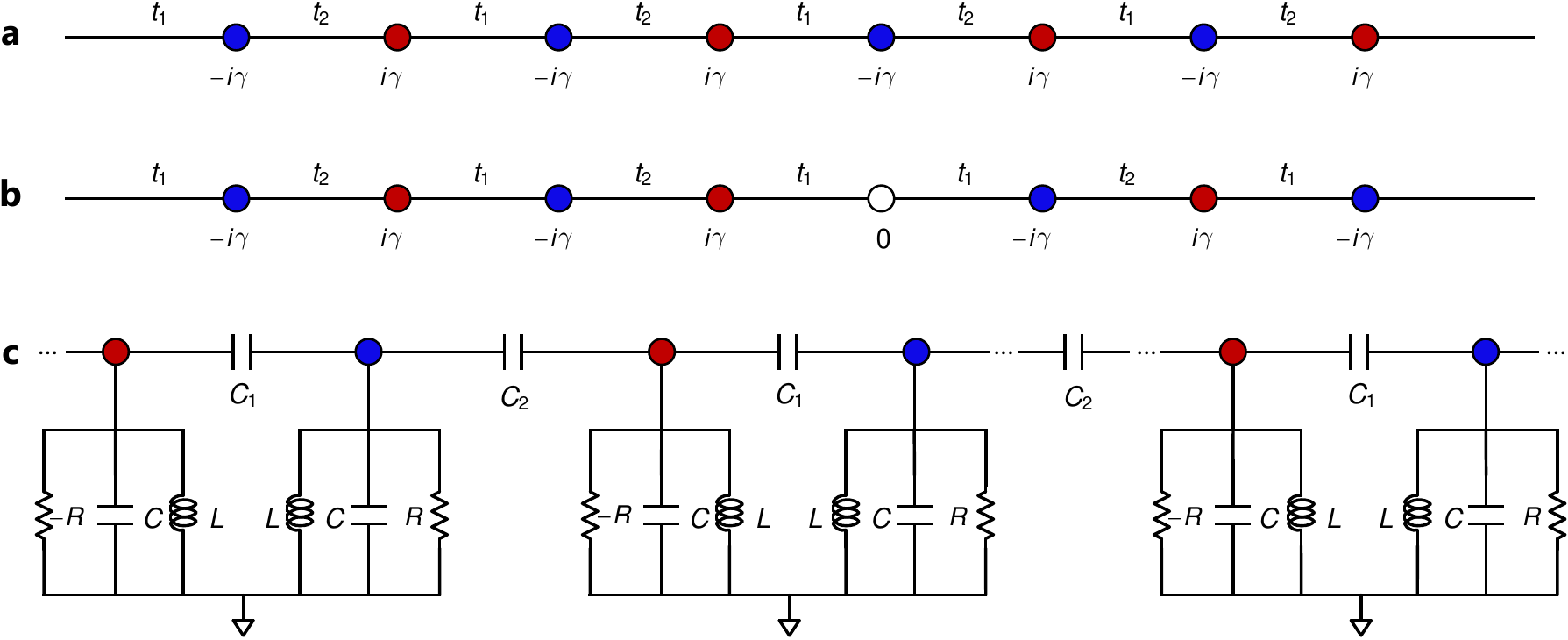}
    \caption{\textbf{Theoretical hopping model and circuit diagram for topological electronics, which is modified from ref~\cite{topo_ele}}.  \textbf{a}. Bulk model with two sites per unit cell (red and blue), with alternating hoppings $t_1$ and $t_2$ and on-site gain$/$loss terms $\pm i\gamma$. \textbf{b}. Insertion of the PT-symmetric defect (white empty circle). \textbf{c}. Circuit diagram of the experimental implementation. The hoppings are realized by capacitors $C_1$, $C_2$, the on-site gain and loss by resistive elements $-R$, $R$. The inductor $L$ or capacitor $C$ tunes the resonance frequency of the circuit.}
    \label{fig:topo}
\end{suppfigure}

Topological properties experience an intriguing degree of diversification when they are combined with PT symmetry.
Therefore, there have been considerable efforts devoted to studying topological insulators under the context of PT symmetry.
Here, we would like to give some discussions about studying topological effects with non-Hermitian topological electronic circuits.

Prior arts~\cite{topo_ele,topo,topo_ele_1,topo_ele_2} have used PT-symmetric electronic circuits to demonstrate various topological effects, such as topological defect engineering, topological insulating phase, and topological wireless power transfer.
So far, these experiments have been focused on low-frequency platform, i.e., printed circuit board.
A general 1-D PT-symmetric Su-Schrieffer-Heeger (SSH) tight-binding model is illustrated in Supplementary Figure~\ref{fig:topo}\textbf{a}, which is based on a chain with alternating hopping $t_1$ and $t_2$, and an alternating on-site gain and loss term $\pm i\gamma$.
The non-Hermitian SSH model is represented by the admittance matrix, also termed circuit Laplacian $J(\omega)$~\cite{topo_circuit} of the circuit.
The detailed theoretical analysis of such non-Hermitian SSH models can be found in prior works~\cite{topo_ele,topo_circuit}.
Supplementary Figure~\ref{fig:topo}\textbf{c} shows the corresponding circuit design of the 1-D topological chain, where the hoppings between non-Hermitian cells are represented by capacitors $C_1$, $C_2$, and the onsite gain and loss are realized by resistive elements.
In order to design a manifestly PT-symmetric topological midgap state in a waveguide system, a defect site can be inserted into chain as shown in Supplementary Figure~\ref{fig:topo}\textbf{b}.
Realize such a chain in CMOS is conceptually straightforward.
However, there are two concerns we want to discuss.
First, the parasitics are universal in IC designs.
Therefore, the parasitics between the connection of two adjacent non-Hermitian units should be minimized.
Second, nonlinearities are also common in ICs.
These non-idealities should be included into the theoretical analysis of the 1-D topological chain.

%% file: main_full.bbl
\begin{thebibliography}{10}
\providecommand{\url}[1]{#1}
\csname url@samestyle\endcsname
\providecommand{\newblock}{\relax}
\providecommand{\bibinfo}[2]{#2}
\providecommand{\BIBentrySTDinterwordspacing}{\spaceskip=0pt\relax}
\providecommand{\BIBentryALTinterwordstretchfactor}{4}
\providecommand{\BIBentryALTinterwordspacing}{\spaceskip=\fontdimen2\font plus
\BIBentryALTinterwordstretchfactor\fontdimen3\font minus
  \fontdimen4\font\relax}
\providecommand{\BIBforeignlanguage}[2]{{%
\expandafter\ifx\csname l@#1\endcsname\relax
\typeout{** WARNING: IEEEtran.bst: No hyphenation pattern has been}%
\typeout{** loaded for the language `#1'. Using the pattern for}%
\typeout{** the default language instead.}%
\else
\language=\csname l@#1\endcsname
\fi
#2}}
\providecommand{\BIBdecl}{\relax}
\BIBdecl

\bibitem{wang_EIT}
C.~Wang, X.~Jiang, G.~Zhao, M.~Zhang, C.~W. Hsu, B.~Peng, A.~D. Stone,
  L.~Jiang, and L.~Yang, ``Electromagnetically induced transparency at a chiral
  exceptional point,'' \emph{Nat. Phys.}, vol.~16, no.~3, pp. 334--340, Mar
  2020.

\bibitem{wang_EIT_1}
C.~Wang, X.~Jiang, W.~R. Sweeney, C.~W. Hsu, Y.~Liu, G.~Zhao, B.~Peng,
  M.~Zhang, L.~Jiang, A.~D. Stone, and L.~Yang, ``Induced transparency by
  interference or polarization,'' \emph{Proc. Natl Acad. Sci.}, vol. 118,
  no.~3, p. e2012982118, Jan. 2021.

\bibitem{CPA1}
C.~Wang, W.~R. Sweeney, A.~D. Stone, and L.~Yang, ``Coherent perfect absorption
  at an exceptional point,'' \emph{Science}, vol. 373, no. 6560, pp.
  1261--1265, 2021.

\bibitem{CPA2}
Y.~D. Chong, L.~Ge, H.~Cao, and A.~D. Stone, ``Coherent perfect absorbers:
  Time-reversed lasers,'' \emph{Phys. Rev. Lett.}, vol. 105, p. 053901, 2010.

\bibitem{CPA3}
Y.~D. Chong, L.~Ge, and A.~D. Stone, ``{$\mathcal{P}\mathcal{T}$-Symmetry
  Breaking and Laser-Absorber Modes in Optical Scattering Systems},''
  \emph{Phys. Rev. Lett.}, vol. 106, p. 093902, 2011.

\bibitem{CPA4}
W.~Wan, Y.~Chong, L.~Ge, H.~Noh, A.~D. Stone, and H.~Cao, ``{Time-Reversed
  Lasing and Interferometric Control of Absorption},'' \emph{Science}, vol.
  331, no. 6019, pp. 889--892, 2011.

\bibitem{topo_r_0}
H.~Zhao, X.~Qiao, T.~Wu, B.~Midya, S.~Longhi, and L.~Feng, ``Non-hermitian
  topological light steering,'' \emph{Science}, vol. 365, no. 6458, pp.
  1163--1166, 2019.

\bibitem{single_mode1}
L.~Feng, Z.~J. Wong, R.-M. Ma, Y.~Wang, and X.~Zhang, ``Single-mode laser by
  parity-time symmetry breaking,'' \emph{Science}, vol. 346, no. 6212, pp.
  972--975, 2014.

\bibitem{single_mode2}
H.~Hodaei, M.-A. Miri, M.~Heinrich, D.~N. Christodoulides, and M.~Khajavikhan,
  ``Parity-time{\textendash}symmetric microring lasers,'' \emph{Science}, vol.
  346, no. 6212, pp. 975--978, 2014.

\bibitem{single_mode3}
Z.~J. Wong, Y.-L. Xu, J.~Kim, K.~O'Brien, Y.~Wang, L.~Feng, and X.~Zhang,
  ``Lasing and anti-lasing in a single cavity,'' \emph{Nat. Photon.}, vol.~10,
  pp. 796--801, 2016.

\bibitem{sensor1}
W.~Chen, S.~Kaya~{\"O}zdemir, G.~Zhao, J.~Wiersig, and L.~Yang, ``Exceptional
  points enhance sensing in an optical microcavity,'' \emph{Nature}, vol. 548,
  pp. 192--196, 2017.

\bibitem{sensor2}
H.~Hodaei, A.~U. Hassan, S.~Wittek, H.~Garcia-Gracia, R.~El-Ganainy, D.~N.
  Christodoulides, and M.~Khajavikhan, ``Enhanced sensitivity at higher-order
  exceptional points,'' \emph{Nature}, vol. 548, pp. 187--191, 2017.

\bibitem{weijian_sensor}
W.~Chen, J.~Zhang, B.~Peng, \c{S}ahin Kaya~\"{O}zdemir, X.~Fan, and L.~Yang,
  ``{Parity-time-symmetric whispering-gallery mode nanoparticle sensor
  [Invited]},'' \emph{Photon. Res.}, vol.~6, pp. A23--A30, 2018.

\bibitem{nonre2}
B.~Peng, S.~K. {\"O}zdemir, F.~Lei, F.~Monifi, M.~Gianfreda, G.~L. Long,
  S.~Fan, F.~Nori, C.~M. Bender, and L.~Yang, ``Parity-time-symmetric
  whispering-gallery microcavities,'' \emph{Nat. Phys.}, vol.~10, pp. 394--398,
  2014.

\bibitem{nonre4}
L.~Chang, X.~Jiang, S.~Hua, C.~Yang, J.~Wen, L.~Jiang, G.~Li, G.~Wang, and
  M.~Xiao, ``Parity-time symmetry and variable optical isolation in
  active-passive-coupled microresonators,'' \emph{Nat. Photon.}, vol.~8, pp.
  524--529, 2014.

\bibitem{opmech1}
X.~Xu, Y.~Liu, C.~Sun, and Y.~Li, ``Mechanical $\mathcal{PT}$ symmetry in
  coupled optomechanical systems,'' \emph{Phys. Rev. A.}, vol.~92, p. 013852,
  2015.

\bibitem{opmech2}
X.~Xu, Y.~Li, A.~Chen, and Y.~Liu, ``Nonreciprocal conversion between microwave
  and optical photons in electro-optomechanical systems,'' \emph{Phys. Rev.
  A.}, vol.~93, p. 023827, 2016.

\bibitem{oeo1}
J.~Zhang and J.~Yao, ``Parity-time{\textendash}symmetric optoelectronic
  oscillator,'' \emph{Sci. Adv.}, vol.~4, no.~6, 2018.

\bibitem{aco1}
R.~Fleury, D.~Sounas, and A.~Al{\`u}, ``An invisible acoustic sensor based on
  parity-time symmetry,'' \emph{Nat. Commun.}, vol.~6, p. 5905, 2015.

\bibitem{aco2}
X.~Zhu, H.~Ramezani, C.~Shi, J.~Zhu, and X.~Zhang,
  ``$\mathcal{P}\mathcal{T}$-symmetric acoustics,'' \emph{Phys. Rev. X.},
  vol.~4, p. 031042, 2014.

\bibitem{NE_nonreci}
L.~Shao, W.~Mao, S.~Maity, N.~Sinclair, Y.~Hu, L.~Yang, and M.~Lon{\v{c}}ar,
  ``Non-reciprocal transmission of microwave acoustic waves in nonlinear
  parity--time symmetric resonators,'' \emph{Nat. Electron.}, vol.~3, no.~5,
  pp. 267--272, May 2020.

\bibitem{PT_E}
J.~Schindler, Z.~Lin, J.~M. Lee, H.~Ramezani, F.~M. Ellis, and T.~Kottos,
  ``$\mathcal{PT}$-symmetric electronics,'' \emph{Journal of Physics A:
  Mathematical and Theoretical.}, vol.~45, no.~44, p. 444029, 2012.

\bibitem{LRC_1}
J.~Schindler, A.~Li, M.~C. Zheng, F.~M. Ellis, and T.~Kottos, ``Experimental
  study of active {LRC} circuits with $\mathcal{PT}$ symmetries,'' \emph{Phys.
  Rev. A.}, vol.~84, p. 040101, 2011.

\bibitem{Dual_beh}
Z.~Lin, J.~Schindler, F.~M. Ellis, and T.~Kottos, ``Experimental observation of
  the dual behavior of $\mathcal{PT}$-symmetric scattering,'' \emph{Phys. Rev.
  A.}, vol.~85, p. 050101, 2012.

\bibitem{wireless_p}
S.~Assawaworrarit, X.~Yu, and S.~Fan, ``Robust wireless power transfer using a
  nonlinear parity-time-symmetric circuit,'' \emph{Nature}, vol. 546, pp.
  387--390, 2017.

\bibitem{NE_wireless}
S.~Assawaworrarit and S.~Fan, ``Robust and efficient wireless power transfer
  using a switch-mode implementation of a nonlinear parity--time symmetric
  circuit,'' \emph{Nat. Electron.}, vol.~3, no.~5, pp. 273--279, May 2020.

\bibitem{NE_sensor}
Z.~Dong, Z.~Li, F.~Yang, C.-W. Qiu, and J.~S. Ho, ``Sensitive readout of
  implantable microsensors using a wireless system locked to an exceptional
  point,'' \emph{Nat. Electron.}, vol.~2, no.~8, pp. 335--342, Aug 2019.

\bibitem{topo_ele}
A.~Stegmaier, S.~Imhof, T.~Helbig, T.~Hofmann, C.~H. Lee, M.~Kremer,
  A.~Fritzsche, T.~Feichtner, S.~Klembt, S.~H\"ofling, I.~Boettcher, I.~C.
  Fulga, L.~Ma, O.~G. Schmidt, M.~Greiter, T.~Kiessling, A.~Szameit, and
  R.~Thomale, ``Topological defect engineering and $\mathcal{P}\mathcal{T}$
  symmetry in non-hermitian electrical circuits,'' \emph{Phys. Rev. Lett.},
  vol. 126, p. 215302, May 2021.

\bibitem{topo_ele_1}
S.~Liu, S.~Ma, C.~Yang, L.~Zhang, W.~Gao, Y.~J. Xiang, T.~J. Cui, and S.~Zhang,
  ``{Gain- and Loss-Induced Topological Insulating Phase in a Non-Hermitian
  Electrical Circuit},'' \emph{Phys. Rev. Applied}, vol.~13, p. 014047, Jan
  2020.

\bibitem{topo_ele_2}
L.~Zhang, Y.~Yang, Z.~Jiang, Q.~Chen, Q.~Yan, Z.~Wu, B.~Zhang, J.~Huangfu, and
  H.~Chen, ``Demonstration of topological wireless power transfer,''
  \emph{Science Bulletin}, vol.~66, no.~10, pp. 974--980, 2021.

\bibitem{PTX}
P.-Y. Chen, M.~Sakhdari, M.~Hajizadegan, Q.~Cui, M.~M.-C. Cheng, R.~El-Ganainy,
  and A.~Al{\`u}, ``Generalized parity-time symmetry condition for enhanced
  sensor telemetry,'' \emph{Nat. Electron.}, vol.~1, no.~5, pp. 297--304, 2018.

\bibitem{3d_IC}
Y.~Liu, J.~Zhang, and L.-M. Peng, ``Three-dimensional integration of plasmonics
  and nanoelectronics,'' \emph{Nat. Electron.}, vol.~1, no.~12, pp. 644--651,
  Dec 2018.

\bibitem{con_non}
A.~Nagulu, N.~Reiskarimian, and H.~Krishnaswamy, ``Non-reciprocal electronics
  based on temporal modulation,'' \emph{Nat. Electron.}, vol.~3, no.~5, pp.
  241--250, May 2020.

\bibitem{mag_book}
D.~M. Pozar, \emph{Microwave Engineering, 4th edn}.\hskip 1em plus 0.5em minus
  0.4em\relax John Wiley \& Sons, New York, 2012.

\bibitem{WJ}
W.~Chen, D.~Leykam, Y.~Chong, and L.~Yang, ``Nonreciprocity in synthetic
  photonic materials with nonlinearity,'' \emph{MRS Bulletin}, vol.~43, p.
  443–451, 2018.

\bibitem{wave_gene}
J.~Wang, H.~Shen, L.~Fan, R.~Wu, B.~Niu, L.~T. Varghese, Y.~Xuan, D.~E. Leaird,
  X.~Wang, F.~Gan, A.~M. Weiner, and M.~Qi, ``Reconfigurable radio-frequency
  arbitrary waveforms synthesized in a silicon photonic chip,'' \emph{Nat.
  Commun.}, vol.~6, no.~1, p. 5957, Jan 2015.

\bibitem{fre_modu}
G.~J. Schneider, J.~A. Murakowski, C.~A. Schuetz, S.~Shi, and D.~W. Prather,
  ``Radiofrequency signal-generation system with over seven octaves of
  continuous tuning,'' \emph{Nat. Photon.}, vol.~7, no.~2, pp. 118--122, Feb
  2013.

\bibitem{non_bro}
D.~L. Sounas, J.~Soric, and A.~Al{\`u}, ``Broadband passive isolators based on
  coupled nonlinear resonances,'' \emph{Nat. Electron.}, vol.~1, pp. 113--119,
  2018.

\bibitem{XDP}
B.~Razavi, ``{A 300 {GHz} Fundamental Oscillator in 65 nm {CMOS} Technology},''
  \emph{IEEE J. Solid State Circ}, vol.~46, no.~4, pp. 894--903, 2011.

\bibitem{VCO}
Z.~Li and K.~K. O, ``A low-phase-noise and low-power multiband {CMOS}
  voltage-controlled oscillator,'' \emph{IEEE J. Solid State Circ}, vol.~40,
  no.~6, pp. 1296--1302, 2005.

\bibitem{VCO_IND}
Y.~Chen and K.~Mouthaan, ``{Wideband Varactorless $ LC $ VCO Using a Tunable
  Negative-Inductance Cell},'' \emph{IEEE Trans. Circuits Syst. I}, vol.~57,
  no.~10, pp. 2609--2617, 2010.

\bibitem{amp-depen}
T.~Djurhuus, V.~Krozer, J.~Vidkjaer, and T.~K. Johansen, ``Nonlinear analysis
  of a cross-coupled quadrature harmonic oscillator,'' \emph{IEEE Trans.
  Circuits Syst. I}, vol.~52, no.~11, pp. 2276--2285, 2005.

\bibitem{Asym_t}
N.~Bender, S.~Factor, J.~D. Bodyfelt, H.~Ramezani, D.~N. Christodoulides, F.~M.
  Ellis, and T.~Kottos, ``{Observation of Asymmetric Transport in Structures
  with Active Nonlinearities},'' \emph{Phys. Rev. Lett.}, vol. 110, p. 234101,
  2013.

\bibitem{gpc}
L.~Ge, Y.~D. Chong, and A.~D. Stone, ``Conservation relations and anisotropic
  transmission resonances in one-dimensional $\mathcal{PT}$-symmetric photonic
  heterostructures,'' \emph{Phys. Rev. A.}, vol.~85, p. 023802, Feb 2012.

\bibitem{ic_sens}
A.~A. {Helmy}, H.~{Jeon}, Y.~{Lo}, A.~J. {Larsson}, R.~{Kulkarni}, J.~{Kim},
  J.~{Silva-Martinez}, and K.~{Entesari}, ``{A Self-Sustained CMOS Microwave
  Chemical Sensor Using a Frequency Synthesizer},'' \emph{IEEE J. Solid State
  Circ}, vol.~47, no.~10, pp. 2467--2483, 2012.

\bibitem{uni_d0}
L.~Feng, Y.-L. Xu, W.~S. Fegadolli, M.-H. Lu, J.~E.~B. Oliveira, V.~R. Almeida,
  Y.-F. Chen, and A.~Scherer, ``Experimental demonstration of a unidirectional
  reflectionless parity-time metamaterial at optical frequencies,'' \emph{Nat.
  Mater.}, vol.~12, no.~2, pp. 108--113, Feb 2013.

\bibitem{uni_d1}
P.~Miao, Z.~Zhang, J.~Sun, W.~Walasik, S.~Longhi, N.~M. Litchinitser, and
  L.~Feng, ``Orbital angular momentum microlaser,'' \emph{Science}, vol. 353,
  no. 6298, pp. 464--467, 2016.

\bibitem{mag_1}
N.~Reiskarimian and H.~Krishnaswamy, ``Magnetic-free non-reciprocity based on
  staggered commutation,'' \emph{Nat. Commun.}, vol.~7, p. 11217, 2016.

\bibitem{Wang2012}
Z.~Wang, Z.~Wang, J.~Wang, B.~Zhang, J.~Huangfu, J.~D. Joannopoulos,
  M.~Soljacic, and L.~Ran, ``Gyrotropic response in the absence of a bias
  field,'' \emph{Proc. Natl Acad. Sci.}, vol. 109, no.~33, pp.
  13\,194--13\,197, Jul. 2012.

\bibitem{nonlinearreci}
J.~Zang, A.~Alvarez-Melcon, and J.~Gomez-Diaz, ``Nonreciprocal phased-array
  antennas,'' \emph{Phys. Rev. Applied}, vol.~12, p. 054008, Nov 2019.

\bibitem{cao2022fully}
W.~Cao, C.~Wang, W.~Chen, S.~Hu, H.~Wang, L.~Yang, and X.~Zhang, ``Fully
  integrated parity--time-symmetric electronics,'' \emph{Nature
  nanotechnology}, vol.~17, no.~3, pp. 262--268, 2022.

\bibitem{S21}
J.~Schindler, Z.~Lin, J.~M. Lee, H.~Ramezani, F.~M. Ellis, and T.~Kottos,
  ``{PT}-symmetric electronics,'' \emph{Journal of Physics A: Mathematical and
  Theoretical}, vol.~45, no.~44, p. 444029, oct 2012.

\bibitem{S22}
P.-Y. Chen, M.~Sakhdari, M.~Hajizadegan, Q.~Cui, M.~M.-C. Cheng, R.~El-Ganainy,
  and A.~Al{\`u}, ``{Generalized parity--time symmetry condition for enhanced
  sensor telemetry},'' \emph{Nature Electronics}, vol.~1, no.~5, May 2018.

\bibitem{mos_r1}
H.~{Jeon} and K.~W. {Kobayashi}, ``Linear voltage controlled variable resistor
  using body potential in soi process,'' \emph{IEEE Microwave and Wireless
  Components Letters}, vol.~26, no.~10, pp. 816--818, 2016.

\bibitem{mos_r2}
\BIBentryALTinterwordspacing
R.~S. SITARAM and W.~G. TOWNSEND, ``A voltage-controlled variable-resistance
  mosfet,'' \emph{International Journal of Electronics}, vol.~38, no.~2, pp.
  253--257, 1975. [Online]. Available:
  \url{https://doi.org/10.1080/00207217508920395}
\BIBentrySTDinterwordspacing

\bibitem{S11}
B.~{Razavi}, ``{A 300-GHz Fundamental Oscillator in 65-nm CMOS Technology},''
  \emph{IEEE Journal of Solid-State Circuits}, vol.~46, no.~4, pp. 894--903,
  2011.

\bibitem{S12}
T.~{Djurhuus}, V.~{Krozer}, J.~{Vidkjaer}, and T.~K. {Johansen}, ``Nonlinear
  analysis of a cross-coupled quadrature harmonic oscillator,'' \emph{IEEE
  Transactions on Circuits and Systems I: Regular Papers}, vol.~52, no.~11, pp.
  2276--2285, 2005.

\bibitem{S24}
Y.~D. Chong, L.~Ge, and A.~D. Stone, ``{$\mathcal{P}\mathcal{T}$-Symmetry
  Breaking and Laser-Absorber Modes in Optical Scattering Systems},''
  \emph{Phys. Rev. Lett.}, vol. 106, p. 093902, Mar 2011.

\bibitem{S25}
A.~Mostafazadeh, ``{Spectral Singularities of Complex Scattering Potentials and
  Infinite Reflection and Transmission Coefficients at Real Energies},''
  \emph{Phys. Rev. Lett.}, vol. 102, p. 220402, Jun 2009.

\bibitem{chen2021non}
W.~Chen, C.~Wang, B.~Peng, and L.~Yang, ``Non-hermitian physics and exceptional
  points in high-quality optical microresonators,'' in \emph{Ultra-high-q
  Optical Microcavities}.\hskip 1em plus 0.5em minus 0.4em\relax World
  Scientific, 2021, pp. 269--313.

\bibitem{wang2021non}
C.~Wang, Z.~Fu, and L.~Yang, ``Non-hermitian physics and engineering in silicon
  photonics,'' in \emph{Silicon Photonics IV}.\hskip 1em plus 0.5em minus
  0.4em\relax Springer, 2021, pp. 323--364.

\bibitem{cao2019neuadc_C}
W.~Cao, X.~He, A.~Chakrabarti, and X.~Zhang, ``{NeuADC: Neural Network-Inspired
  RRAM-Based Synthesizable Analog-to-Digital Conversion with Reconfigurable
  Quantization Support},'' in \emph{2019 Design, Automation Test in Europe
  Conference Exhibition (DATE)}, 2019, pp. 1477--1482.

\bibitem{cao2019neuadc}
W.~{Cao}, X.~{He}, A.~{Chakrabarti}, and X.~{Zhang}, ``{NeuADC: Neural
  Network-Inspired Synthesizable Analog-to-Digital Conversion},'' \emph{IEEE
  Transactions on Computer-Aided Design of Integrated Circuits and Systems
  (TCAD)}, vol.~39, no.~9, pp. 1841--1854, 2020.

\bibitem{pipelineadc}
W.~{Cao}, L.~{Ke}, A.~{Chakrabarti}, and X.~{Zhang}, ``{Neural Network-Inspired
  Analog-to-Digital Conversion to Achieve Super-Resolution with Low-Precision
  RRAM Devices},'' in \emph{2019 IEEE/ACM International Conference on
  Computer-Aided Design (ICCAD)}, 2019, pp. 1--7.

\bibitem{pipelineadc_J}
W.~Cao, L.~Ke, A.~Chakrabarti, and X.~Zhang, ``{Evaluating Neural
  Network-Inspired Analog-to-Digital Conversion With Low-Precision RRAM},''
  \emph{IEEE Transactions on Computer-Aided Design of Integrated Circuits and
  Systems (TCAD)}, vol.~40, no.~5, pp. 808--821, 2021.

\bibitem{cao_tc}
W.~Cao, Y.~Zhao, A.~Boloor, Y.~Han, X.~Zhang, and L.~Jiang, ``Neural-pim:
  Efficient processing-in-memory with neural approximation of peripherals,''
  \emph{IEEE Transactions on Computers}, pp. 1--1, 2021.

\bibitem{LV}
F.~Lv, X.~Zheng, F.~Zhao, J.~Wang, S.~Yue, {Ziqiang Wang}, W.~Cao, Y.~He,
  C.~Zhang, H.~Jiang, and Z.~Wang, ``A power scalable 2–10 gb/s pi-based
  clock data recovery for multilane applications,'' \emph{Microelectronics
  Journal}, vol.~82, pp. 36--45, 2018.

\bibitem{zhouNEWCAS}
N.~Zhou, L.~Wu, Z.~Wang, X.~Zheng, W.~Cao, C.~Zhang, F.~Li, and Z.~Wang, ``A
  28-gb/s transmitter with 3-tap ffe and t-coil enhanced terminal in 65-nm cmos
  technology,'' in \emph{2016 14th IEEE International New Circuits and Systems
  Conference (NEWCAS)}.\hskip 1em plus 0.5em minus 0.4em\relax IEEE, 2016, pp.
  1--4.

\bibitem{cao2021phd}
W.~Cao, ``Machine learning for analog/mixed-signal integrated circuit design
  automation,'' 2021.

\bibitem{caoedssc1}
W.~Cao, X.~Zheng, Z.~Wang, D.~Li, F.~Li, S.~Yue, and Z.~Wang, ``A 15gb/s
  wireline repeater in 65nm cmos technology,'' in \emph{2015 IEEE International
  Conference on Electron Devices and Solid-State Circuits (EDSSC)}.\hskip 1em
  plus 0.5em minus 0.4em\relax IEEE, 2015, pp. 590--593.

\bibitem{caoedssc2}
W.~Cao, Z.~Wang, D.~Li, F.~Li, and Z.~Wang, ``A 40gb/s adaptive equalizer with
  amplitude approaching technique in 65nm cmos,'' in \emph{2015 IEEE
  International Conference on Electron Devices and Solid-State Circuits
  (EDSSC)}.\hskip 1em plus 0.5em minus 0.4em\relax IEEE, 2015, pp. 451--454.

\bibitem{caomwscas}
W.~Cao, Z.~Wang, D.~Li, X.~Zheng, K.~Huang, S.~Yuan, F.~Li, and Z.~Wang, ``{A
  40Gb/s 39mW 3-tap adaptive closed-loop decision feedback equalizer in 65nm
  CMOS},'' in \emph{2015 IEEE 58th International Midwest Symposium on Circuits
  and Systems (MWSCAS)}, 2015, pp. 1--4.

\bibitem{caonewcas}
W.~Cao and et~al, ``{A 40Gb/s 27mW 3-tap closed-loop decision feedback
  equalizer in 65nm CMOS},'' in \emph{2015 IEEE 13th International New Circuits
  and Systems Conference (NEWCAS)}.\hskip 1em plus 0.5em minus 0.4em\relax
  IEEE, 2015, pp. 1--4.

\bibitem{S35}
B.~{Razavi}, ``A study of phase noise in cmos oscillators,'' \emph{IEEE Journal
  of Solid-State Circuits}, vol.~31, no.~3, pp. 331--343, 1996.

\bibitem{c_vco_1}
H.-C. Chang, X.~Cao, U.~Mishra, and R.~York, ``Phase noise in coupled
  oscillators: theory and experiment,'' \emph{IEEE Transactions on Microwave
  Theory and Techniques}, vol.~45, no.~5, pp. 604--615, 1997.

\bibitem{c_vco_2}
G.~Li, L.~Liu, Y.~Tang, and E.~Afshari, ``A low-phase-noise wide-tuning-range
  oscillator based on resonant mode switching,'' \emph{IEEE Journal of
  Solid-State Circuits}, vol.~47, no.~6, pp. 1295--1308, 2012.

\bibitem{c_vco_3}
S.~A.-R. Ahmadi-Mehr, M.~Tohidian, and R.~B. Staszewski, ``Analysis and design
  of a multi-core oscillator for ultra-low phase noise,'' \emph{IEEE
  Transactions on Circuits and Systems I: Regular Papers}, vol.~63, no.~4, pp.
  529--539, 2016.

\bibitem{c_vco_4}
A.~ElSayed and M.~Elmary, ``Low-phase-noise lc quadrature vco using coupled
  tank resonators in a ring structure,'' \emph{IEEE Journal of Solid-State
  Circuits}, vol.~36, no.~4, pp. 701--705, 2001.

\bibitem{S23}
Schindler and J.~Caulfield, ``{PT-Symmetric Electronics},'' \emph{Masters
  Theses, \url{https://wesscholar.wesleyan.edu/etd\_mas\_theses/42}}, vol.~42,
  2013.

\bibitem{cao2022aaai}
W.~Cao and et~al, ``Domain knowledge-based automated analog circuit design with
  deep reinforcement learning,'' \emph{arXiv preprint arXiv:2202.13185}, 2022.

\bibitem{cao2022dac}
W.~Cao, M.~Benosman, X.~Zhang, and R.~Ma, ``Domain knowledge-infused deep
  learning for automated analog/radio-frequency circuit parameter
  optimization,'' \emph{arXiv preprint arXiv:2204.12948}, 2022.

\bibitem{S13}
A.~{Mirzaei}, M.~E. {Heidari}, R.~{Bagheri}, S.~{Chehrazi}, and A.~A. {Abidi},
  ``{The Quadrature LC Oscillator: A Complete Portrait Based on Injection
  Locking},'' \emph{IEEE Journal of Solid-State Circuits}, vol.~42, no.~9, pp.
  1916--1932, 2007.

\bibitem{S14}
G.~{Cusmai}, M.~{Repossi}, G.~{Albasini}, A.~{Mazzanti}, and F.~{Svelto}, ``{A
  Magnetically Tuned Quadrature Oscillator},'' \emph{IEEE Journal of
  Solid-State Circuits}, vol.~42, no.~12, pp. 2870--2877, 2007.

\bibitem{S15}
J.~{Kim}, J.~{Kim}, G.~{Kim}, and D.~{Jeong}, ``{A Fully Integrated 0.13-
  $\mu$m CMOS 40-Gb/s Serial Link Transceiver},'' \emph{IEEE Journal of
  Solid-State Circuits}, vol.~44, no.~5, pp. 1510--1521, 2009.

\bibitem{S16}
D.~{Guermandi}, P.~{Tortori}, E.~{Franchi}, and A.~{Gnudi}, ``{A 0.83-2.5-GHz
  continuously tunable quadrature VCO},'' \emph{IEEE Journal of Solid-State
  Circuits}, vol.~40, no.~12, pp. 2620--2627, 2005.

\bibitem{S31}
N.~Bender, S.~Factor, J.~D. Bodyfelt, H.~Ramezani, D.~N. Christodoulides, F.~M.
  Ellis, and T.~Kottos, ``{Observation of Asymmetric Transport in Structures
  with Active Nonlinearities},'' \emph{Phys. Rev. Lett.}, vol. 110, p. 234101,
  Jun 2013.

\bibitem{S32}
D.~L. Sounas, J.~Soric, and A.~Al{\`u}, ``Broadband passive isolators based on
  coupled nonlinear resonances,'' \emph{Nature Electronics}, vol.~1, pp.
  113--119, 2018.

\bibitem{S32_1}
L.~Shao, W.~Mao, S.~Maity, N.~Sinclair, Y.~Hu, L.~Yang, and M.~Lon{\v{c}}ar,
  ``Non-reciprocal transmission of microwave acoustic waves in nonlinear
  parity--time symmetric resonators,'' \emph{Nature Electronics}, vol.~3,
  no.~5, pp. 267--272, May 2020.

\bibitem{un_invi_1}
B.~Lv, J.~Fu, B.~Wu, R.~Li, Q.~Zeng, X.~Yin, Q.~Wu, L.~Gao, W.~Chen, Z.~Wang,
  Z.~Liang, A.~Li, and R.~Ma, ``Unidirectional invisibility induced by
  parity-time symmetric circuit,'' \emph{Scientific Reports}, vol.~7, no.~1, p.
  40575, Jan 2017.

\bibitem{un_invi_2}
Z.~Lin, J.~Schindler, F.~M. Ellis, and T.~Kottos, ``Experimental observation of
  the dual behavior of $\mathcal{PT}$-symmetric scattering,'' \emph{Phys. Rev.
  A}, vol.~85, p. 050101, May 2012.

\bibitem{P_PT}
F.~R. Humire and E.~Lazo, ``$\mathcal{PT}$-symmetric direct electrical
  transmission lines: Localization behavior,'' \emph{Phys. Rev. E}, vol. 100,
  p. 022221, Aug 2019.

\bibitem{topo}
H.~Zhao, X.~Qiao, T.~Wu, B.~Midya, S.~Longhi, and L.~Feng, ``Non-hermitian
  topological light steering,'' \emph{Science}, vol. 365, no. 6458, pp.
  1163--1166, 2019.

\bibitem{topo_circuit}
C.~H. Lee, S.~Imhof, C.~Berger, F.~Bayer, J.~Brehm, L.~W. Molenkamp,
  T.~Kiessling, and R.~Thomale, ``Topolectrical circuits,''
  \emph{Communications Physics}, vol.~1, no.~1, p.~39, Jul 2018.

\end{thebibliography}


\begin{thebibliography}{10}
\providecommand{\url}[1]{#1}
\csname url@samestyle\endcsname
\providecommand{\newblock}{\relax}
\providecommand{\bibinfo}[2]{#2}
\providecommand{\BIBentrySTDinterwordspacing}{\spaceskip=0pt\relax}
\providecommand{\BIBentryALTinterwordstretchfactor}{4}
\providecommand{\BIBentryALTinterwordspacing}{\spaceskip=\fontdimen2\font plus
\BIBentryALTinterwordstretchfactor\fontdimen3\font minus
  \fontdimen4\font\relax}
\providecommand{\BIBforeignlanguage}[2]{{%
\expandafter\ifx\csname l@#1\endcsname\relax
\typeout{** WARNING: IEEEtran.bst: No hyphenation pattern has been}%
\typeout{** loaded for the language `#1'. Using the pattern for}%
\typeout{** the default language instead.}%
\else
\language=\csname l@#1\endcsname
\fi
#2}}
\providecommand{\BIBdecl}{\relax}
\BIBdecl

\bibitem{cao2022fully}
W.~Cao, C.~Wang, W.~Chen, S.~Hu, H.~Wang, L.~Yang, and X.~Zhang, ``Fully
  integrated parity--time-symmetric electronics,'' \emph{Nature
  nanotechnology}, vol.~17, no.~3, pp. 262--268, 2022.

\bibitem{S21}
J.~Schindler, Z.~Lin, J.~M. Lee, H.~Ramezani, F.~M. Ellis, and T.~Kottos,
  ``{PT}-symmetric electronics,'' \emph{Journal of Physics A: Mathematical and
  Theoretical}, vol.~45, no.~44, p. 444029, oct 2012.

\bibitem{S22}
P.-Y. Chen, M.~Sakhdari, M.~Hajizadegan, Q.~Cui, M.~M.-C. Cheng, R.~El-Ganainy,
  and A.~Al{\`u}, ``{Generalized parity--time symmetry condition for enhanced
  sensor telemetry},'' \emph{Nature Electronics}, vol.~1, no.~5, May 2018.

\bibitem{mos_r1}
H.~{Jeon} and K.~W. {Kobayashi}, ``Linear voltage controlled variable resistor
  using body potential in soi process,'' \emph{IEEE Microwave and Wireless
  Components Letters}, vol.~26, no.~10, pp. 816--818, 2016.

\bibitem{mos_r2}
\BIBentryALTinterwordspacing
R.~S. SITARAM and W.~G. TOWNSEND, ``A voltage-controlled variable-resistance
  mosfet,'' \emph{International Journal of Electronics}, vol.~38, no.~2, pp.
  253--257, 1975. [Online]. Available:
  \url{https://doi.org/10.1080/00207217508920395}
\BIBentrySTDinterwordspacing

\bibitem{S11}
B.~{Razavi}, ``{A 300-GHz Fundamental Oscillator in 65-nm CMOS Technology},''
  \emph{IEEE Journal of Solid-State Circuits}, vol.~46, no.~4, pp. 894--903,
  2011.

\bibitem{S12}
T.~{Djurhuus}, V.~{Krozer}, J.~{Vidkjaer}, and T.~K. {Johansen}, ``Nonlinear
  analysis of a cross-coupled quadrature harmonic oscillator,'' \emph{IEEE
  Transactions on Circuits and Systems I: Regular Papers}, vol.~52, no.~11, pp.
  2276--2285, 2005.

\bibitem{S24}
Y.~D. Chong, L.~Ge, and A.~D. Stone, ``{$\mathcal{P}\mathcal{T}$-Symmetry
  Breaking and Laser-Absorber Modes in Optical Scattering Systems},''
  \emph{Phys. Rev. Lett.}, vol. 106, p. 093902, Mar 2011.

\bibitem{S25}
A.~Mostafazadeh, ``{Spectral Singularities of Complex Scattering Potentials and
  Infinite Reflection and Transmission Coefficients at Real Energies},''
  \emph{Phys. Rev. Lett.}, vol. 102, p. 220402, Jun 2009.

\bibitem{chen2021non}
W.~Chen, C.~Wang, B.~Peng, and L.~Yang, ``Non-hermitian physics and exceptional
  points in high-quality optical microresonators,'' in \emph{Ultra-high-q
  Optical Microcavities}.\hskip 1em plus 0.5em minus 0.4em\relax World
  Scientific, 2021, pp. 269--313.

\bibitem{wang2021non}
C.~Wang, Z.~Fu, and L.~Yang, ``Non-hermitian physics and engineering in silicon
  photonics,'' in \emph{Silicon Photonics IV}.\hskip 1em plus 0.5em minus
  0.4em\relax Springer, 2021, pp. 323--364.

\bibitem{cao2019neuadc_C}
W.~Cao, X.~He, A.~Chakrabarti, and X.~Zhang, ``{NeuADC: Neural Network-Inspired
  RRAM-Based Synthesizable Analog-to-Digital Conversion with Reconfigurable
  Quantization Support},'' in \emph{2019 Design, Automation Test in Europe
  Conference Exhibition (DATE)}, 2019, pp. 1477--1482.

\bibitem{cao2019neuadc}
W.~{Cao}, X.~{He}, A.~{Chakrabarti}, and X.~{Zhang}, ``{NeuADC: Neural
  Network-Inspired Synthesizable Analog-to-Digital Conversion},'' \emph{IEEE
  Transactions on Computer-Aided Design of Integrated Circuits and Systems
  (TCAD)}, vol.~39, no.~9, pp. 1841--1854, 2020.

\bibitem{pipelineadc}
W.~{Cao}, L.~{Ke}, A.~{Chakrabarti}, and X.~{Zhang}, ``{Neural Network-Inspired
  Analog-to-Digital Conversion to Achieve Super-Resolution with Low-Precision
  RRAM Devices},'' in \emph{2019 IEEE/ACM International Conference on
  Computer-Aided Design (ICCAD)}, 2019, pp. 1--7.

\bibitem{pipelineadc_J}
W.~Cao, L.~Ke, A.~Chakrabarti, and X.~Zhang, ``{Evaluating Neural
  Network-Inspired Analog-to-Digital Conversion With Low-Precision RRAM},''
  \emph{IEEE Transactions on Computer-Aided Design of Integrated Circuits and
  Systems (TCAD)}, vol.~40, no.~5, pp. 808--821, 2021.

\bibitem{cao_tc}
W.~Cao, Y.~Zhao, A.~Boloor, Y.~Han, X.~Zhang, and L.~Jiang, ``Neural-pim:
  Efficient processing-in-memory with neural approximation of peripherals,''
  \emph{IEEE Transactions on Computers}, pp. 1--1, 2021.

\bibitem{LV}
F.~Lv, X.~Zheng, F.~Zhao, J.~Wang, S.~Yue, {Ziqiang Wang}, W.~Cao, Y.~He,
  C.~Zhang, H.~Jiang, and Z.~Wang, ``A power scalable 2–10 gb/s pi-based
  clock data recovery for multilane applications,'' \emph{Microelectronics
  Journal}, vol.~82, pp. 36--45, 2018.

\bibitem{zhouNEWCAS}
N.~Zhou, L.~Wu, Z.~Wang, X.~Zheng, W.~Cao, C.~Zhang, F.~Li, and Z.~Wang, ``A
  28-gb/s transmitter with 3-tap ffe and t-coil enhanced terminal in 65-nm cmos
  technology,'' in \emph{2016 14th IEEE International New Circuits and Systems
  Conference (NEWCAS)}.\hskip 1em plus 0.5em minus 0.4em\relax IEEE, 2016, pp.
  1--4.

\bibitem{cao2021phd}
W.~Cao, ``Machine learning for analog/mixed-signal integrated circuit design
  automation,'' 2021.

\bibitem{caoedssc1}
W.~Cao, X.~Zheng, Z.~Wang, D.~Li, F.~Li, S.~Yue, and Z.~Wang, ``A 15gb/s
  wireline repeater in 65nm cmos technology,'' in \emph{2015 IEEE International
  Conference on Electron Devices and Solid-State Circuits (EDSSC)}.\hskip 1em
  plus 0.5em minus 0.4em\relax IEEE, 2015, pp. 590--593.

\bibitem{caoedssc2}
W.~Cao, Z.~Wang, D.~Li, F.~Li, and Z.~Wang, ``A 40gb/s adaptive equalizer with
  amplitude approaching technique in 65nm cmos,'' in \emph{2015 IEEE
  International Conference on Electron Devices and Solid-State Circuits
  (EDSSC)}.\hskip 1em plus 0.5em minus 0.4em\relax IEEE, 2015, pp. 451--454.

\bibitem{caomwscas}
W.~Cao, Z.~Wang, D.~Li, X.~Zheng, K.~Huang, S.~Yuan, F.~Li, and Z.~Wang, ``{A
  40Gb/s 39mW 3-tap adaptive closed-loop decision feedback equalizer in 65nm
  CMOS},'' in \emph{2015 IEEE 58th International Midwest Symposium on Circuits
  and Systems (MWSCAS)}, 2015, pp. 1--4.

\bibitem{caonewcas}
W.~Cao and et~al, ``{A 40Gb/s 27mW 3-tap closed-loop decision feedback
  equalizer in 65nm CMOS},'' in \emph{2015 IEEE 13th International New Circuits
  and Systems Conference (NEWCAS)}.\hskip 1em plus 0.5em minus 0.4em\relax
  IEEE, 2015, pp. 1--4.

\bibitem{S35}
B.~{Razavi}, ``A study of phase noise in cmos oscillators,'' \emph{IEEE Journal
  of Solid-State Circuits}, vol.~31, no.~3, pp. 331--343, 1996.

\bibitem{c_vco_1}
H.-C. Chang, X.~Cao, U.~Mishra, and R.~York, ``Phase noise in coupled
  oscillators: theory and experiment,'' \emph{IEEE Transactions on Microwave
  Theory and Techniques}, vol.~45, no.~5, pp. 604--615, 1997.

\bibitem{c_vco_2}
G.~Li, L.~Liu, Y.~Tang, and E.~Afshari, ``A low-phase-noise wide-tuning-range
  oscillator based on resonant mode switching,'' \emph{IEEE Journal of
  Solid-State Circuits}, vol.~47, no.~6, pp. 1295--1308, 2012.

\bibitem{c_vco_3}
S.~A.-R. Ahmadi-Mehr, M.~Tohidian, and R.~B. Staszewski, ``Analysis and design
  of a multi-core oscillator for ultra-low phase noise,'' \emph{IEEE
  Transactions on Circuits and Systems I: Regular Papers}, vol.~63, no.~4, pp.
  529--539, 2016.

\bibitem{c_vco_4}
A.~ElSayed and M.~Elmary, ``Low-phase-noise lc quadrature vco using coupled
  tank resonators in a ring structure,'' \emph{IEEE Journal of Solid-State
  Circuits}, vol.~36, no.~4, pp. 701--705, 2001.

\bibitem{S23}
Schindler and J.~Caulfield, ``{PT-Symmetric Electronics},'' \emph{Masters
  Theses, \url{https://wesscholar.wesleyan.edu/etd\_mas\_theses/42}}, vol.~42,
  2013.

\bibitem{cao2022aaai}
W.~Cao and et~al, ``Domain knowledge-based automated analog circuit design with
  deep reinforcement learning,'' \emph{arXiv preprint arXiv:2202.13185}, 2022.

\bibitem{cao2022dac}
W.~Cao, M.~Benosman, X.~Zhang, and R.~Ma, ``Domain knowledge-infused deep
  learning for automated analog/radio-frequency circuit parameter
  optimization,'' \emph{arXiv preprint arXiv:2204.12948}, 2022.

\bibitem{S13}
A.~{Mirzaei}, M.~E. {Heidari}, R.~{Bagheri}, S.~{Chehrazi}, and A.~A. {Abidi},
  ``{The Quadrature LC Oscillator: A Complete Portrait Based on Injection
  Locking},'' \emph{IEEE Journal of Solid-State Circuits}, vol.~42, no.~9, pp.
  1916--1932, 2007.

\bibitem{S14}
G.~{Cusmai}, M.~{Repossi}, G.~{Albasini}, A.~{Mazzanti}, and F.~{Svelto}, ``{A
  Magnetically Tuned Quadrature Oscillator},'' \emph{IEEE Journal of
  Solid-State Circuits}, vol.~42, no.~12, pp. 2870--2877, 2007.

\bibitem{S15}
J.~{Kim}, J.~{Kim}, G.~{Kim}, and D.~{Jeong}, ``{A Fully Integrated 0.13-
  $\mu$m CMOS 40-Gb/s Serial Link Transceiver},'' \emph{IEEE Journal of
  Solid-State Circuits}, vol.~44, no.~5, pp. 1510--1521, 2009.

\bibitem{S16}
D.~{Guermandi}, P.~{Tortori}, E.~{Franchi}, and A.~{Gnudi}, ``{A 0.83-2.5-GHz
  continuously tunable quadrature VCO},'' \emph{IEEE Journal of Solid-State
  Circuits}, vol.~40, no.~12, pp. 2620--2627, 2005.

\bibitem{S31}
N.~Bender, S.~Factor, J.~D. Bodyfelt, H.~Ramezani, D.~N. Christodoulides, F.~M.
  Ellis, and T.~Kottos, ``{Observation of Asymmetric Transport in Structures
  with Active Nonlinearities},'' \emph{Phys. Rev. Lett.}, vol. 110, p. 234101,
  Jun 2013.

\bibitem{S32}
D.~L. Sounas, J.~Soric, and A.~Al{\`u}, ``Broadband passive isolators based on
  coupled nonlinear resonances,'' \emph{Nature Electronics}, vol.~1, pp.
  113--119, 2018.

\bibitem{S32_1}
L.~Shao, W.~Mao, S.~Maity, N.~Sinclair, Y.~Hu, L.~Yang, and M.~Lon{\v{c}}ar,
  ``Non-reciprocal transmission of microwave acoustic waves in nonlinear
  parity--time symmetric resonators,'' \emph{Nature Electronics}, vol.~3,
  no.~5, pp. 267--272, May 2020.

\bibitem{un_invi_1}
B.~Lv, J.~Fu, B.~Wu, R.~Li, Q.~Zeng, X.~Yin, Q.~Wu, L.~Gao, W.~Chen, Z.~Wang,
  Z.~Liang, A.~Li, and R.~Ma, ``Unidirectional invisibility induced by
  parity-time symmetric circuit,'' \emph{Scientific Reports}, vol.~7, no.~1, p.
  40575, Jan 2017.

\bibitem{un_invi_2}
Z.~Lin, J.~Schindler, F.~M. Ellis, and T.~Kottos, ``Experimental observation of
  the dual behavior of $\mathcal{PT}$-symmetric scattering,'' \emph{Phys. Rev.
  A}, vol.~85, p. 050101, May 2012.

\bibitem{P_PT}
F.~R. Humire and E.~Lazo, ``$\mathcal{PT}$-symmetric direct electrical
  transmission lines: Localization behavior,'' \emph{Phys. Rev. E}, vol. 100,
  p. 022221, Aug 2019.

\bibitem{topo_ele}
A.~Stegmaier, S.~Imhof, T.~Helbig, T.~Hofmann, C.~H. Lee, M.~Kremer,
  A.~Fritzsche, T.~Feichtner, S.~Klembt, S.~Höfling, I.~Boettcher, I.~C.
  Fulga, O.~G. Schmidt, M.~Greiter, T.~Kiessling, A.~Szameit, and R.~Thomale,
  ``Topological defect engineering and {PT}-symmetry in non-hermitian
  electrical circuits,'' \emph{Preprint at}, p.
  https://arxiv.org/abs/2011.14836, 2020.

\bibitem{topo}
H.~Zhao, X.~Qiao, T.~Wu, B.~Midya, S.~Longhi, and L.~Feng, ``Non-hermitian
  topological light steering,'' \emph{Science}, vol. 365, no. 6458, pp.
  1163--1166, 2019.

\bibitem{topo_ele_1}
S.~Liu, S.~Ma, C.~Yang, L.~Zhang, W.~Gao, Y.~J. Xiang, T.~J. Cui, and S.~Zhang,
  ``{Gain- and Loss-Induced Topological Insulating Phase in a Non-Hermitian
  Electrical Circuit},'' \emph{Phys. Rev. Applied}, vol.~13, p. 014047, Jan
  2020.

\bibitem{topo_ele_2}
L.~Zhang, Y.~Yang, Z.~Jiang, Q.~Chen, Q.~Yan, Z.~Wu, B.~Zhang, J.~Huangfu, and
  H.~Chen, ``Topological wireless power transfer,'' \emph{Preprint at}, p.
  https://arxiv.org/abs/2008.02592, 2020.

\bibitem{topo_circuit}
C.~H. Lee, S.~Imhof, C.~Berger, F.~Bayer, J.~Brehm, L.~W. Molenkamp,
  T.~Kiessling, and R.~Thomale, ``Topolectrical circuits,''
  \emph{Communications Physics}, vol.~1, no.~1, p.~39, Jul 2018.

\end{thebibliography}
